\documentclass[longauth]{aa}  

\usepackage{graphicx}
\usepackage{amsmath}
\usepackage{txfonts}
\usepackage{longtable}
\usepackage{natbib}
\bibpunct{(}{)}{;}{a}{}{,}

\begin{document}
\hypersetup{pdfauthor={Name}}
\title{A sub-Neptune planet around TOI-1695 discovered and characterized with SPIRou and TESS}

\author{
Flavien Kiefer\inst{\ref{inst:1},\ref{inst:2}}\thanks{flavien.kiefer@obspm.fr}, 
G. H\'ebrard\inst{\ref{inst:2},\ref{inst:3}}, 
E. Martioli\inst{\ref{inst:4},\ref{inst:2}}, 
E. Artigau\inst{\ref{inst:5}},  
R. Doyon\inst{\ref{inst:5}}, 
J.-F. Donati\inst{\ref{inst:6}}, 
C. Cadieux\inst{\ref{inst:5}}, 
A. Carmona\inst{\ref{inst:7}},
D. R. Ciardi\inst{\ref{inst:8}}, 
P. I. Cristofari\inst{\ref{inst:6}}, 
L. de\,Almeida\inst{\ref{inst:4}}, 
P. Figueira\inst{\ref{inst:11},\ref{inst:12}},
E. Gaidos\inst{\ref{inst:28}},
E. Gonzales\inst{\ref{inst:9}}, 
A. Lecavelier\inst{\ref{inst:2}}, 
K. G.\ Stassun\inst{\ref{inst:10}}, 
L. Arnold\inst{\ref{inst:13}},
B. Benneke\inst{\ref{inst:14}},
I. Boisse\inst{\ref{inst:15}},
X. Bonfils\inst{\ref{inst:7}},
N. J. Cook\inst{\ref{inst:5}},
P. Cort\'es-Zuleta\inst{\ref{inst:15}},
X. Delfosse\inst{\ref{inst:7}},
J. Dias\,do\,Nascimento\inst{\ref{inst:33,34}},
M. Fausnaugh\inst{\ref{inst:16}},
W. Fong\inst{\ref{inst:16}},
P. Fouqué\inst{\ref{inst:5}},
T. Forveille\inst{\ref{inst:7}},
J. Gomes da Silva\inst{\ref{inst:25}},
K. Hesse\inst{\ref{inst:16}},
\'A. K\'osp\'al\inst{\ref{inst:29},\ref{inst:30},\ref{inst:31},\ref{inst:32}},
H. Lewis\inst{\ref{inst:17}},
C.-F. Liu\inst{\ref{inst:27}},
J. H. C. Martins\inst{\ref{inst:25}},
M. Paegert\inst{\ref{inst:18}},
S. Seager\inst{\ref{inst:19},\ref{inst:20},\ref{inst:21}},
H. Shang\inst{\ref{inst:27}},
J. D. Twicken\inst{\ref{inst:22},\ref{inst:23}},
T. Vandal\inst{\ref{inst:5}},
S. Vinatier\inst{\ref{inst:1}},
T. Widemann\inst{\ref{inst:1},\ref{inst:26}},
J. N. Winn\inst{\ref{inst:24}}}

   \institute{
        \label{inst:1}LESIA, Observatoire de Paris, Université PSL, CNRS, Sorbonne Université, Université Paris Cité, 5 place Jules Janssen, 92195 Meudon, France
        \and
        \label{inst:2}Sorbonne Université, CNRS, UMR 7095, Institut d’Astrophysique de Paris, 98 bis bd Arago, 75014 Paris, France
        \and
        \label{inst:3}Observatoire de Haute Provence, St Michel l’Observatoire, France
         \and
        \label{inst:4}Laboratório Nacional de Astrofísica, Rua Estados Unidos 154, 37504-364, Itajubá - MG, Brazil
        \and
        \label{inst:5}Université de Montréal, Département de Physique, IREX, Montréal,
        QC H3C 3J7, Canada
        \and
        \label{inst:6}Institut de Recherche en Astrophysique et Planétologie, Université 
        de Toulouse, CNRS, IRAP/UMR 5277, 14 avenue Edouard Belin, 
        31400, Toulouse, France
        \and 
        \label{inst:7}Université Grenoble Alpes, CNRS, IPAG, 38000 Grenoble, France
        \and
        \label{inst:8}NASA Exoplanet Science Institute, Caltech/IPAC, Pasadena, CA 91125, USA
        \and
        \label{inst:9}Department of Astronomy and Astrophysics, University of California, Santa Cruz, CA 95064, USA
        \and
        \label{inst:10}Department of Physics and Astronomy, Vanderbilt University, Nashville, TN 37235, USA
        \and
        \label{inst:11}Instituto de Astrof\'isica e Ciencias do Espa\c co, Universidade do Porto, CAUP, Rua das Estrelas, P-4150-762 Porto, Portugal
        \and 
        \label{inst:12}European Southern Observatory, Alonso de Cordova, Vitacura, Santiago, Chile
        \and
        \label{inst:13}Canada-France-Hawaii Telescope, CNRS, 96743 Kamuela, Hawai'i, 
        USA
        \and
        \label{inst:14}Institut de Recherche sur les Exoplanètes, Département de Physique, Université de Montréal, 1375 Avenue Thérèse-Lavoie-Roux, Montreal, QC H2V 0B3, Canada
        \and
        \label{inst:15}Aix Marseille Univ, CNRS, CNES, LAM, Marseille, France
        \and
        \label{inst:16}Department of Physics and Kavli Institute for Astrophysics and Space Research, Massachusetts Institute of Technology, Cambridge, MA 02139, USA
        \and 
        \label{inst:17}Space Telescope Science Institute, 3700 San Martin Drive, Baltimore, MD, 21218, USA
        \and
        \label{inst:18}Center for Astrophysics, Harvard \& Smithsonian, 60 Garden St, Cambridge, MA 02138, USA
        \and
        \label{inst:19}MIT Kavli Institute for Astrophysics and Space Research, Massachusetts Institute of Technology, Cambridge, MA 02139, USA
        \and 
        \label{inst:20}Earth and Planetary Sciences, Massachusetts Institute of Technology, 77 Massachusetts Avenue, Cambridge, MA 02139, USA
        \and 
        \label{inst:21}Department of Aeronautics and Astronautics, MIT, 77 Massachusetts Avenue, Cambridge, MA 02139, USA
        \and
        \label{inst:22}NASA Ames Research Center, Moffett Field, CA 94035, USA
        \and
        \label{inst:23}SETI Institute, Mountain View, CA 94043, USA
        \and 
        \label{inst:24}Department of Astrophysical Sciences, Peyton Hall, 4 Ivy Lane, Princeton, NJ 08544, USA
        \and
        \label{inst:25}Instituto de Astrof\'isica e Ci\^{e}ncias do Espa\c{c}o, Universidade do Porto, CAUP, Rua das Estrelas, 4150-762, Porto, Portugal 
        \and
        \label{inst:26}UVSQ, Université Paris-Saclay, 78047 Guyancourt, France
        \and
        \label{inst:27}Institute of Astronomy and Astrophysics, Academia Sinica, Taipei 10617, Taiwan
        \and
        \label{inst:28}Department of Earth Sciences, University of Hawai'i at M\-anoa, Honolulu, Hawaii 96822 USA
        \and
        \label{inst:29}Konkoly Observatory, Research Centre for Astronomy and Earth Sciences, E\"otv\"os Lor\'and Research Network (ELKH), Konkoly-Thege Mikl\'os \'ut 15-17, H-1121 Budapest, Hungary
        \and
        \label{inst:30}CSFK, MTA Centre of Excellence, Budapest, Konkoly Thege Mikl\'os \'ut 15-17, H-1121, Hungary
        \and
        \label{inst:31}Max-Planck-Institute for Astronomy, K\"onigstuhl 17, D-69117 Heidelberg, Germany
        \and
        \label{inst:32}ELTE E\"otv\"os Lor\'and University, Institute of Physics, P\'azm\'any P\'eter s\'et\'any 1/A, H-1117 Budapest, Hungary
        \and
        \label{inst:33}Universidade Federal do Rio Grande do Norte (UFRN), 59078-970, Natal, RN, Brazil
        \and
        \label{inst:34}Center for Astrophysics | Harvard \& Smithsonian, 60 Garden Street, Cambridge, MA 02138, USA
        }

   \date{Received ; accepted }

 
  \abstract{
   TOI-1695 is a V-mag=13 M-dwarf star from the northern hemisphere at 45\,pc from the Sun, around which a 3.134-day periodic transit signal from a super-Earth candidate was identified in TESS photometry. With a transit depth of 1.3\,mmag, the radius of candidate TOI-1695.01 was estimated by the TESS pipeline to be 1.82 R$_\oplus$ with an equilibrium temperature of $\sim$620\,K. We successfully detect a reflex motion of the star and establish it is due to a planetary companion at an orbital period consistent with the photometric transit period thanks to a year-long radial-velocity monitoring of TOI-1695 by the SPIRou infrared spectropolarimeter. We use and compare different methods to reduce and analyse those data. We report a 5.5-$\sigma$ detection of the planetary signal, giving a mass of $5.5$$\pm$$1.0$\,M$_\oplus$ and a radius of 2.03$\pm$0.18\,R$_\oplus$. We derive a mean equilibrium planet temperature of 590$\pm$90\,K. The mean density of this small planet of 3.6$\pm$1.1\,g\,cm$^{-3}$ is similar (1.7--$\sigma$ lower) than that of the Earth. It leads to a non-negligible fraction of volatiles in its atmosphere with $f_{H,He}$=0.28$^{+0.46}_{-0.23}$\% or $f_\text{water}$=23$\pm$12\%. TOI-1695\,b is a new sub-Neptune planet at the border of the M-dwarf radius valley that can help test formation scenarios for super-Earth/sub-Neptune-like planets. 
  }


\titlerunning{TOI-1695\,b with TESS \& SPIRou}
\authorrunning{Kiefer et al.}
   \maketitle

\section{Introduction}

Over the next few years, new space- and ground-based observatories (e.g. the James Webb Space Telescope and ESO's extremely large telescope) will be paramount to extend the current boundaries of planetary science.
%
%
These will boost the exploration of the atmospheres of exoplanets to an unprecedented level of precision, and probe the gas content of super-Earth and terrestrial exoplanets. It is thus of key importance to identify new low-mass exoplanets amenable to future atmospheric characterisation. 

A central feature of small exoplanets ($R$$<$$4$\,R$_\oplus$) at small orbital separation ($P$$<$$100$\,days) is the radius valley \citep{Owen2013,Fulton2017,CloutierMenou2020} separating solid super-Earths from gaseous sub-Neptunes. One possible explanation for this dichotomy is runaway evaporation or grinding of the atmospheres of low-mass exoplanets by a strong XUV stellar radiation \citep{Owen2013}. Alternative explanations are core-powered mass loss \citep{Ginzburg2018}, impact erosion by planetesimals \citep{Shuvalov2009}, or formation of distinct rocky and non-rocky planet populations with delayed gas accretion \citep{Lee2014,Lee2021}. Around solar-like stars, photoevaporation and core-powered mass loss scenarios predict that the valley location shifts towards lower radius with lower stellar mass and larger orbital period \citep{VanEylen2018,Fulton2018,Martinez2019,Wu2019,Gupta2022}. 

Around low-mass stars, \citet{CloutierMenou2020} find that the slope of the valley around M dwarfs in the radius-period diagram is inverse to the valley slope around FGK stars, leading to increasing radius of the transition with orbital period. This relationship is more compatible with a gas-depleted planet formation scenario in which solid cores increase in size with increasing distance to the star \citep{Lee2014,Lopez2018}. Similarly than for solar-like stars, for planets around low-mass stars, photoevaporation and core-powered mass loss are also expected to shape a radius gap with a negative slope in $R_p--P$ diagram. However this slope is expected to be closer and closer to zero with decreasing stellar mass~\citep{Gupta2022}.

Here we characterize a candidate planet with a super-Earth/sub-Neptune size, that was detected and identified around the M2--type star TOI-1695, with the NASA Transiting Exoplanet Survey Satellite (TESS; \citealt{Ricker2015}). With an initially estimated radius of 1.8\,R$_\oplus$ and an orbital period of 3.134\,days, TOI-1695.01 lies in a radius/separation region where the question of formation and evolution of the atmosphere of super-Earth or sub-Neptune around low-mass star can be tackled. Investigating whether or not such a key planet harbors a significant atmosphere may help explain which scenario is dominantly shaping the radius valley depending on the host star mass. 

In this paper, we seek to first establish the planetary nature of TOI-1695.01, and to characterize its mass and radius by combining TESS photometric measurements with radial velocities (RV) variations observed with the infrared (0.98-2.45\,$\mu$m) SpectroPolarimetre InfraRouge (SPIRou; \citealt{Donati2020}) installed on the 3.6-m Canada France Hawaii Telescope (CFHT). 
 
In Section~\ref{sec:obs} we review the observations available for TOI-1695, including those from TESS and SPIRou. In Section~\ref{sec:carac} we characterize the host star using spectroscopy, spectrophotometry and spectropolarimetry. In Section~\ref{sec:planet} we characterize the planet properties through a joint fit of TESS and SPIRou data. In Section~\ref{sec:discussion} we discuss the results. We summarize and conclude in Section~\ref{sec:conclusion}.
 
\section{Observations}
\label{sec:obs}
\subsection{TESS lightcurves}
\label{sec:TESS}
\begin{figure}[hbt]    \centering
    \begin{picture}(89.3,90)
    \put(0,0){\includegraphics[width=89.3mm]{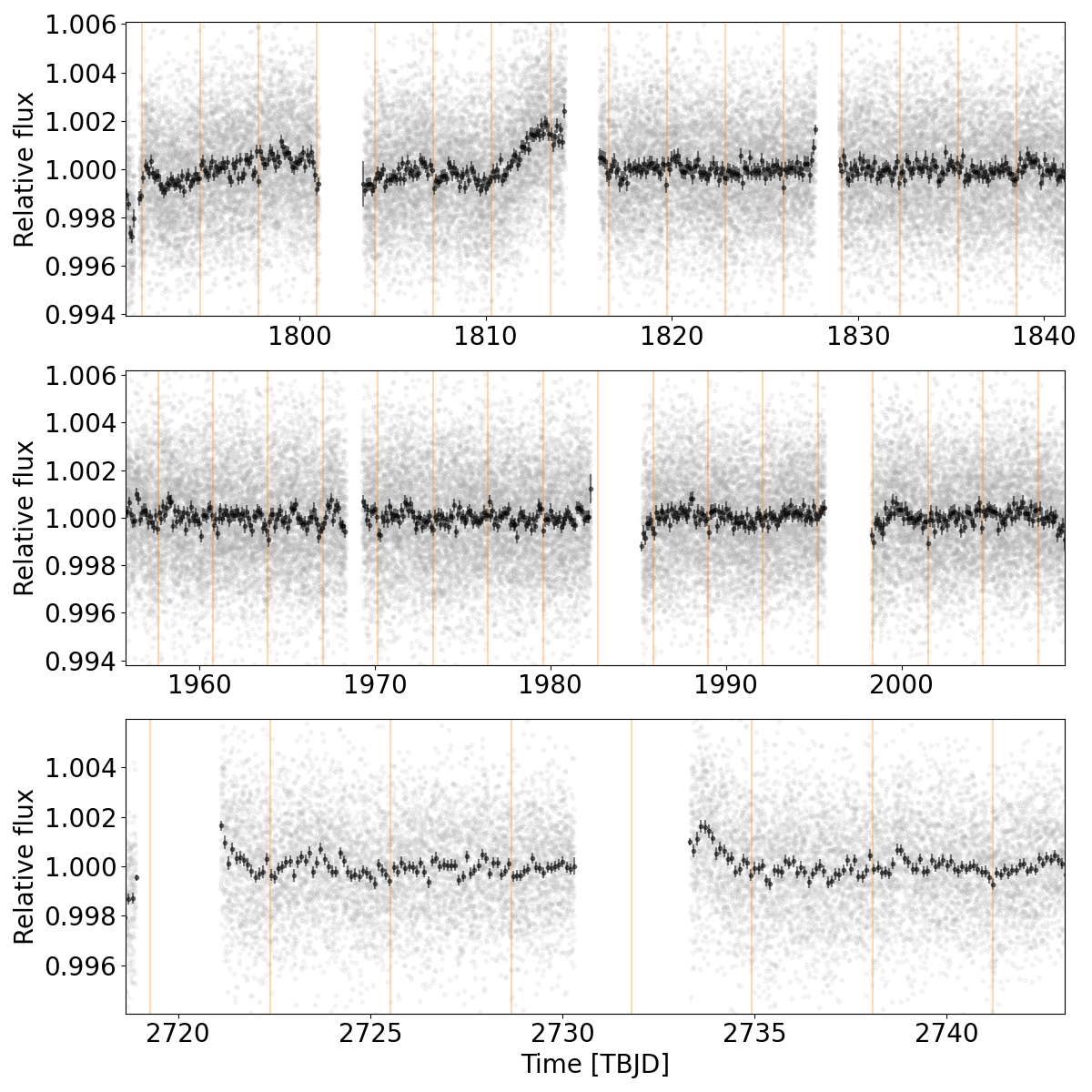}}
    \put(15,83){Sector 18}
    \put(50,83){Sector 19}
    \put(15,54){Sector 23}
    \put(55,54){Sector 24}
    \put(45,26){Sector 52}
    \end{picture}
    \caption{The full TESS lightcurve separated in sectors (gray dots) and binned with a 0.1-day timestep (black). The identified transit locations are shown with orange vertical lines. \label{fig:transit_detection}}
\end{figure}
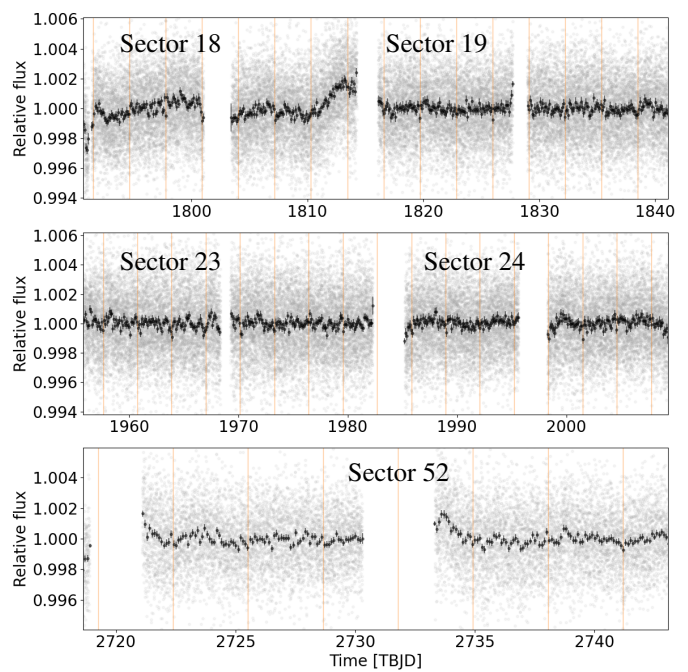

TOI-1695, also known as TIC-422756130.01 in the TESS Input Catalog (TIC), was covered by TESS sectors 18, 19, 24, 25 and 52 between November 2019 and June 2022 with 120-s cadence. In the data from these observing campaigns, a signal at a 3.13-day period is found using the Box-Least-Square (BLS) algorithm \citep{Kovacs2002} with 34 transits found, among which 32 were already identified in the data validation timeseries (DVT hereafter ; \citealt{Twicken2018,Li2019}) of sectors 18 to 25, allowing the event TOI-1695.01 to be identified. We retrieved and used in the rest of this study the Science Processing Operations Center Pipeline (SPOC; \citealt{Jenkins2016}) simple aperture photometry with presearch data conditioning (PDCSAP ; \citealt{Smith2012,Stumpe2014}). The full SPOC PDCSAP lightcurve with all identified transits is shown in Fig.~\ref{fig:transit_detection}.

We first performed a preliminary fit of the TESS PDCSAP data of TOI-1695 by a photometric \verb+batman+ model \citep{Kreidberg2015} with a third-degree polynomial modeling of the continuum of all individual transits. It has a transit depth of $\sim$1000\,ppm for a total transit duration of 1.2\,hour and a period of 3.134319$\pm$0.000026\,days. It was pointed out\footnote{\texttt{https://archive.stsci.edu/missions/tess/doc/\\tess\_drn/tess\_sector\_27\_drn38\_v02.pdf}} that the SPOC photometry pipeline was prone to over-estimate the background level and over-correct for background in crowded fields in the TESS primary mission (sectors 1-26). The pipeline was updated to prevent this in the extended mission (sectors 27+). We estimate that the combined transit search of the TOI\,1695 data for sectors 18, 19, 24, 25 and 52 would have over-estimated the transit depth (planet radius) by approximately 2.8\% (1.4\%). This is a minor correction compared to uncertainties ($\sim$8.9\%) resulting from the full analysis performed in Section~\ref{sec:planet}.


Using \verb+TRICERATOPS+ \citep{Giacalone2020}, we examined possible sources of false-positive (FP) transit detection in the TOI-1695 lightcurve, from the system itself and from nearby background systems within the TESS aperture. The full list of possible causes of an FP is given in Table\,1 of \citet{Giacalone2021}. We show the aperture used in Sector 18 and possible contaminating sources around TOI-1695 in Fig.~\ref{fig:aperture}. This source is located in a crowded region with 411 neighbors with T-mag$<$18 at less than 200\arcsec. At most 8 stars identified in the different TESS apertures could be contributing to the transit signal with TIC IDs 422756137, 422756132, 422756120, 629325853, 422756126, 422756145, 422756114, and 422756147. They have a flux ratio with TOI-1695 in TESS passband between 5.3 and 7.4 mag. We find an FP probability of transit from TOI-1695 itself of 0.18\% and an FP probability of a background star pollution of 0.10\%. These probabilities accounts for the contrast curve around TOI-1695 obtained in Section~\ref{sec:imaging}. According to the criteria defined by \citet{Giacalone2021}, TOI-1695\,b is thus a "validated planet". Nevertheless, complementary follow-up with RV, as performed in the present paper, is necessary to independently validate its planetary nature and measure its mass.

\begin{figure}
    \centering
    \includegraphics[width=89.3mm]{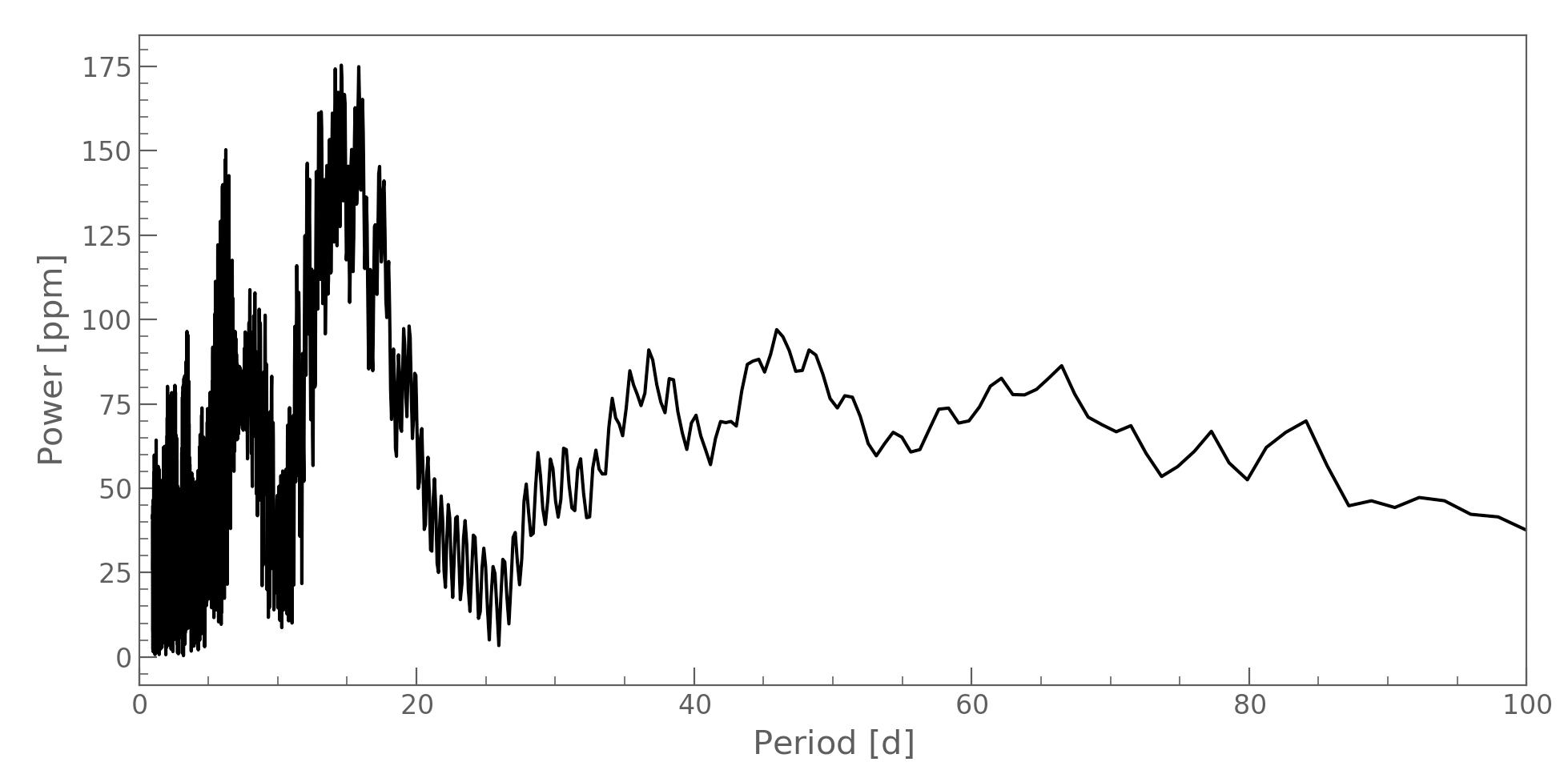}
    \caption{Periodogram of the TESS SPOC lightcurve with a peak at $\sim$15\,days for a 175\,ppm signal.}
    \label{fig:LC_periodogram}
\end{figure}

Beyond the transit signal, a 15-day modulation with an average amplitude of $\sim$175\,ppm seems to be present in the PDCSAP light curves. A periodogram of the TESS PDCSAP data produced with \verb+lightkurve+ \citep{Lightkurve2018} is shown in Fig.~\ref{fig:LC_periodogram}. This 15-day periodic signal is in fact mainly seen in the Sector 18 with an amplitude of $\sim$700\,ppm (Fig.~\ref{fig:transit_detection}). We discard effects from variable background stars, since given the sources identified above within TOI-1695 apertures, such amplitude of variations in the lightcurve would require those sources to have proper oscillations $>$10\% which is rare~\citep{McQuillan2012}. If not of instrumental origin, which we cannot fully exclude, this fluctuating signal could well be due to spots, in which case the 15-day period would be linked with the rotation period of the star. We discuss this possibility in more details and confront it to other observations in Section~\ref{sec:rotation_discussion}.

\subsection{SPIRou spectra}

\subsubsection{Observations}
SPIRou is a near-infrared high-resolution (980-2450\,nm; R=75,000) high-stability ($\sim$1\,m\,s$^{-1}$) velocimeter and spectropolarimeter installed at the Cassegrain focus of the 3.6-m CFHT at Maunakea \citep{Donati2020}. TOI-1695 was observed between Dec. 2020 and Jan. 2022 with SPIRou as part of the large program SPIRou Legacy Survey (SLS; id P42, PI: Jean-François Donati). It was observed at 45 observing epochs with 4 polarimetric exposures per epoch, totaling 180 SPIRou spectra. Table~\ref{tab:log} presents the observation log associated with each exposure. All SPIRou observations of TOI-1695 consist of polarization sequences, each split into 4 sub-exposures associated with different rhomb retarder configurations, from which we retrieve the corresponding Stokes I (unpolarised) and V (circularly polarized) spectra of the star at each epoch~\citep{Donati2020}. The sequence number 1-4 of any sub-exposure is also given in Table~\ref{tab:log}. Science data are acquired on fibers A and B fed by the 2 orthogonal states of the selected polarization, while fiber C is used for simultaneously recording the spectrum of the SPIRou Fabry-P\'erot RV reference,  as described in \citet{Hobson2021}. 

A calibration sequence, including flats, darks, and spectra of comparison lamps (Fabry-P\'erot, UNe), is performed every afternoon and morning, preceding and following each night of observation with SPIRou. A master-dark is constructed from a subset  of well characterized darks. The SPIRou detector software also archives ramp-files, that are detector images in ADU/s, during which creation, a correction for non-linearity is performed. Early-type stars (A-type) are observed nightly as telluric absorption standards. Bright inactive cool stars are also regularly observed as RV standards. These data are used to calibrate the measurements extracted from the SPIRou spectra. The detailed calibration sequence is described in Cook et al. (2022, submitted).

\subsubsection{APERO reduction}

The raw data were first reduced with the SPIRou reduction software, A PipelinE to Reduce Observations or APERO, with version 0.6.132 (Cook et al., submitted), or v6 hereafter. APERO first corrects detector effects, removes constant background thermal component, and identify bad pixels and cosmic ray impacts. It then calculates the position of the 49 of the 50 echelle spectral orders\footnote{APERO does not extract the bluest order due to low S/N.} and optimally extracts spectra from fibers A, B and C into 2D order-separated e2ds and 1D order-merged s1d spectra. A blaze function is derived from the flat-field exposures.

The wavelengths are calibrated on the spectrum collected on Fiber C. An absolute calibration of wavelengths with respect to the Solar System barycentric rest frame using the current Barycentric Earth RV (BERV) and the Barycentric Julian Date (BJD) of each exposure is performed by the code \verb+barycorrpy+ \citep{Kanodia2018,Wright2014}. 
Finally, for each exposure (\verb+e.fits+ file) APERO calculates the corresponding spectrum of the telluric transmission out of the whole collection of standard star observations carried out with SPIRou since 2018 using PCA algorithm \citep{Artigau2014} and divides it out, producing a telluric-corrected \verb+t.fits+ spectrum. APERO also calculates the Stokes V spectra using the method of \citet{Donati1997}, as described in detail in \citet{Martioli2020}.

A more advanced version of the APERO pipeline, version 0.7.194, was released during the writing of this work~\citep{Artigau2022,Cadieux2022}, hereafter called v7. We used both the v6 and v7 versions in our study, and compare them in Section~\ref{sec:planet}.

\subsubsection{RV derivation}
The RV were derived from the telluric corrected \verb+t.fits+ spectra using the line-by-line (LBL) algorithm~\citep{Artigau2022}. As in \citet{Bouchy2001}, LBL requires a template spectrum with S/N as large as possible, since a derivative of the template is used to determine the RV of each spectral line. For TOI-1695, with only 40 epochs, the combined spectrum produced by APERO reaches a S/N much lower than other bright standard stars monitored with SPIRou. Instead of TOI-1695, we thus used as template another target
 with a similar M2V spectral type, GL15A ($T_\text{eff}$=3600\,K; $\log g$=4.8\,cgs units; $[$M/H$]$=-0.16\,dex; \citealt{Passegger2019,Cristofari2022}). Table~\ref{tab:star_param} shows the stellar parameters that we find for TOI-1695. We combined 1040 spectra of GL15A with for each a S/N$\sim$350 at 1670\,nm leading to a template with S/N close to 10\,000. 

The second and third derivatives of each spectral line are used as a proxy for the full width at half maximum (FWHM) and bisector span (BIS) \citep{Artigau2022}. Their variations with respect to RV are studied in Section~\ref{sec:activity}.

Calibration drifts are derived for all RV measurement epochs using the simultaneous Fabry-P\'erot on fiber C. They lead to drift-corrected (DC hereafter) RV. An average zero-point correction (ZPC, hereafter) is also derived from the monitoring of RV standards \citep{Cadieux2022}. Our reference RVs in the rest of the paper will be those corrected from drift and zero-point, but we do a comparison of RV data reduction with or without drift and zero-point correction when analysing the planet signal in Section~\ref{sec:planet}.

All RV data, FWHM and BIS used for TOI-1695 and reduced with versions v6 and v7 are available in Tables~\ref{tab:rvs_v6} and~\ref{tab:rvs_v7}, respectively. Their variations are inspected in more details in Section~\ref{sec:activity}. In both cases, we ignored the measurement of epoch \#44 (JD$-$2459605) that has only a single sub-exposure. With the v6, we also ignored the measurements of epoch \#33 (JD$-$2459591) for which the drift calculation failed.

\subsection{Imaging}\label{sec:imaging}
As part of a standard process to validate transiting exoplanets and to assess the possible contamination of bound or unbound companions on the derived planetary radii \citep{ciardi2015}, TOI-1695 was observed with infrared high-resolution adaptive optics (AO) imaging at Keck Observatory with the NIRC2 instrument on Keck-II behind the natural guide star AO system \citep{wizinowich2000}.  The observations were made on 2020~May~28 UT in the standard 3-point dither pattern that is used with NIRC2 to avoid the lower left quadrant of the detector which is typically noisier than the other three quadrants. The dither pattern step size was $3\arcsec$ and was repeated twice, with each dither offset from the previous dither by $0.5\arcsec$. The camera was in the narrow-angle mode with a full field of view of $\sim10\arcsec$ and a pixel scale of approximately $0.01\arcsec$ per pixel. The observations were made in the narrow-band Br-$\gamma$ filter ($\lambda_o$=2.17$\mu$m; $\Delta\lambda$=0.03$\mu$m) with an integration time of 4\,s with one coadd per frame for a total of 36\,s on target.

The AO data were processed and analyzed with a custom set of IDL tools.  The science frames were flat-fielded and sky-subtracted.  The flat fields were generated from a median of dark subtracted flats taken on-sky, and the flats were normalized such that the median value of the flats is unity.  Sky frames were generated from the median average of the 9 dithered science frames; each science image was then sky-subtracted and flat-fielded.  The reduced science frames were combined into a single combined image using an intra-pixel interpolation that conserves flux, shifts the individual dithered frames by the appropriate fractional pixels, and median-coadds the frames.  The final resolution of the combined dithers was determined from the FWHM of the point spread function to 0.054$\arcsec$.  

The sensitivities of the final combined AO image were determined by injecting simulated sources azimuthally around the primary target every $20^\circ $ at separations of integer multiples of the central source's FWHM \citep{furlan2017}. The brightness of each injected source was scaled until standard aperture photometry detected it with $5\sigma $ significance. The resulting brightness of the injected sources relative to the target set the contrast limits at that injection location. The final $5\sigma $ limit at each separation was determined from the average of all of the determined limits at that separation and the uncertainty on the limit was set by the RMS dispersion of the azimuthal slices at a given radial distance.

\begin{figure}[hbt]
    \centering
    \includegraphics[width=89.3mm]{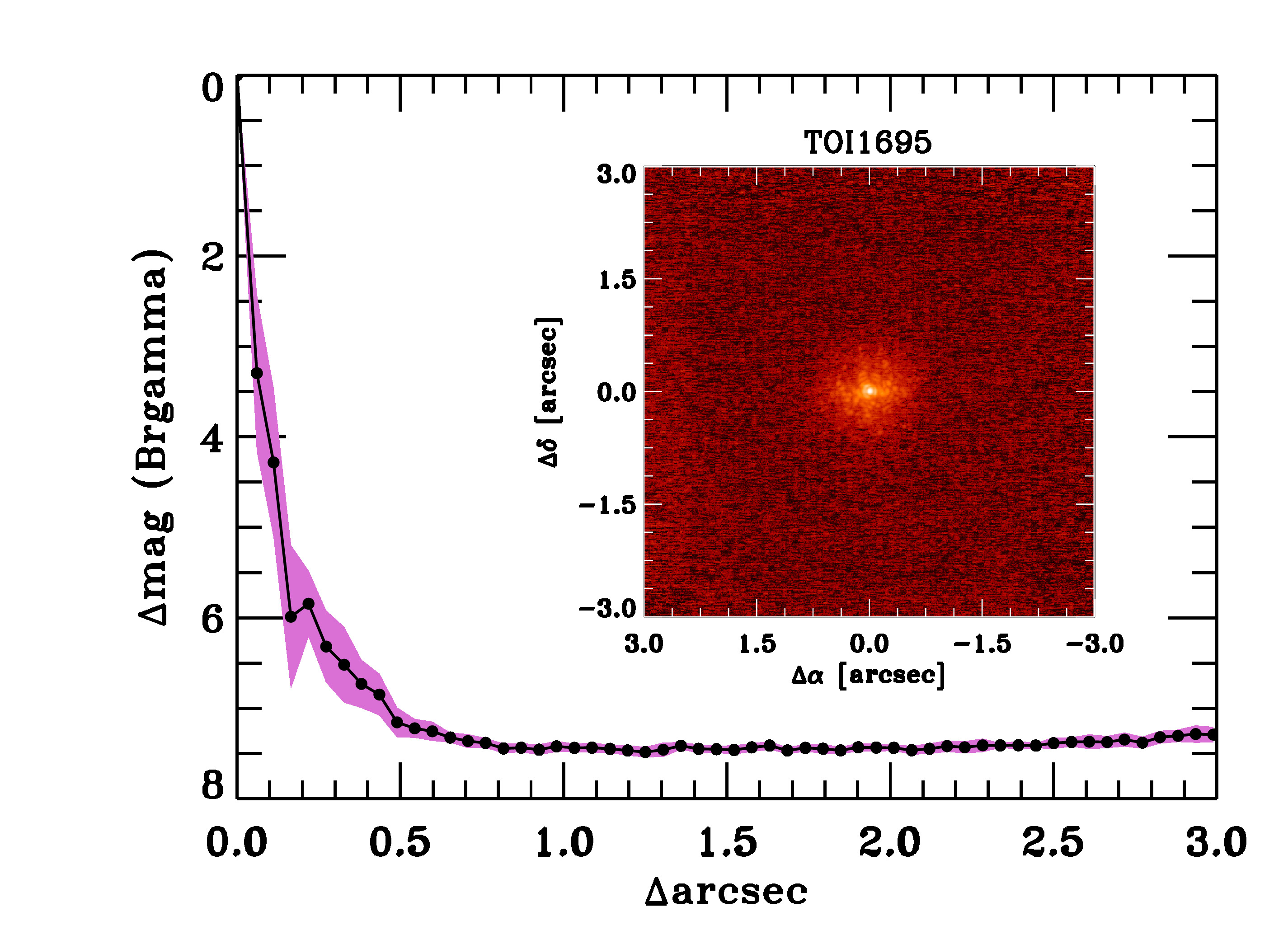}
    \caption{Companion sensitivity for the Keck NIR adaptive optics imaging.  The black points represent the 5$\sigma$ limits and are separated in steps of 1 FWHM ($\sim$0.054$\arcsec$); the purple represents the azimuthal dispersion (1$\sigma$) of the contrast determinations (see text). The inset image is of the primary target showing no additional companions to within 3" of the target.  \label{fig:ao_contrast}}
  
\end{figure}

No additional companions to within the limits of the data were detected (see Fig.~\ref{fig:ao_contrast}).  With contrast sensitivities of $\sim3.5$ mag at 0.06$\arcsec$\ (2.7\,au) and $\sim7$ mag at 0.5$\arcsec$\ (22\,au), the NIR AO observations indicate that there are likely no stellar companions down to $\sim$ M6 - L9 (see E.\ Mamajek's compilation of Mean Dwarf Stellar Colors version 2021.03.02 \footnote{
https://www.pas.rochester.edu/$\sim$emamajek/EEM\_dwarf\_UBVIJHK\\\_colors\_Teff.txt}).

\section{Stellar characterization}
\label{sec:carac}
\subsection{Stellar parameters}

We use four different methods to measure the stellar parameters of the star, including the effective temperature $T_\text{eff}$, the surface gravity $\log g$, the metallicity [M/H], the micro-turbulent $v_\text{micro}$, the macro-turbulent velocity $v_\text{macro}$, the bolometric flux $F_\text{bol}$, the bolometric luminosity ${\mathcal L}_\text{bol}$, the stellar mass $M_\star$, and the stellar radius $R_\star$. The methods of spectral Energy Distribution (SED) and of the TESS-exoFOP are summarized in Section~\ref{sec:SED}. The methods of comparison to synthetic spectra, based on SPIRou spectra, are explained in Section~\ref{sec:spectro}. All results are summarized and compared in Table~\ref{tab:star_param}.

\subsubsection{Spectral Energy Distribution}
\label{sec:SED}
\begin{figure}[hbt]
    \centering
    \includegraphics[width=89.3mm, clip=true]{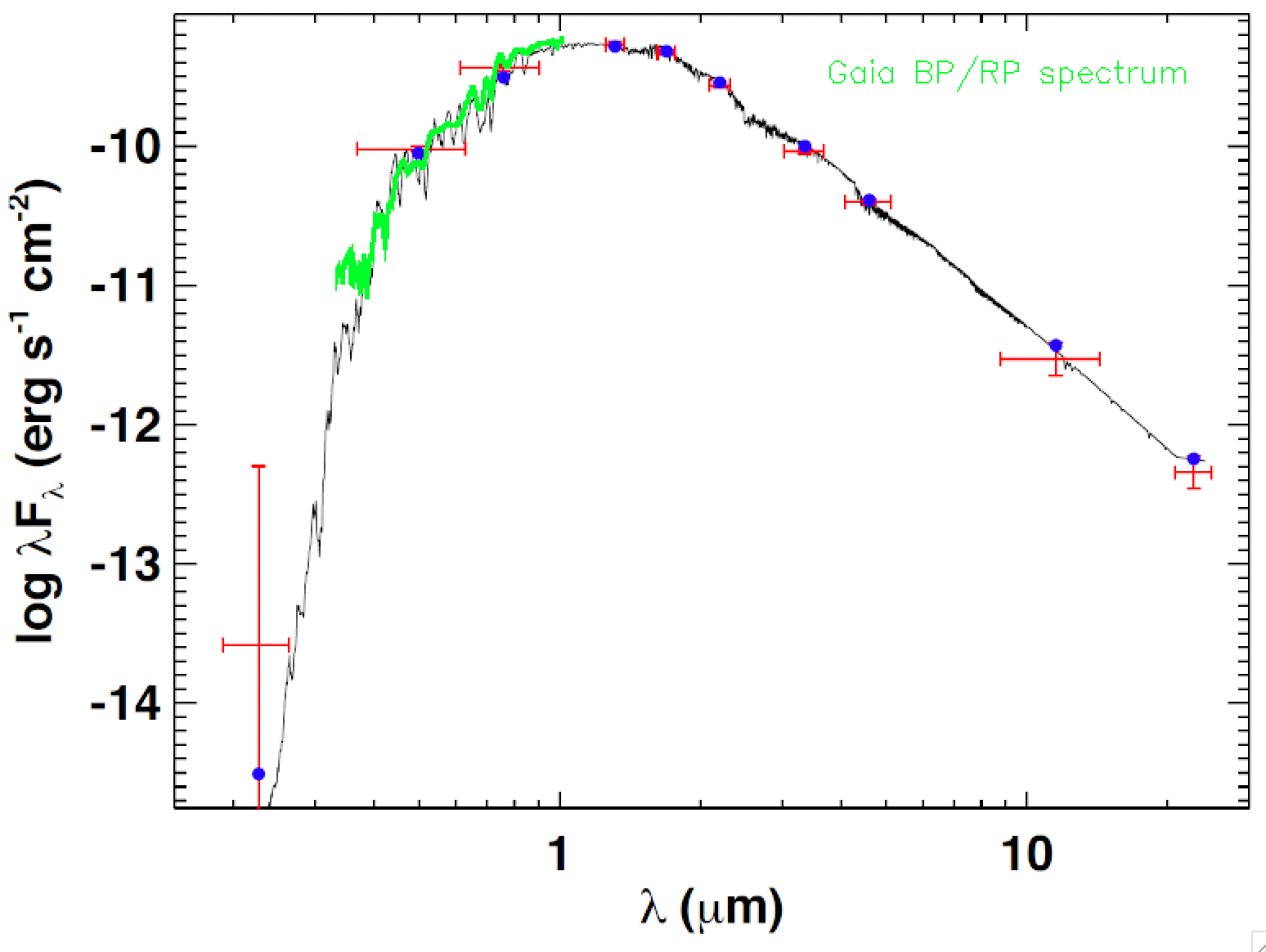}
\caption{Spectral energy distribution of TOI-1695. Red symbols represent the observed photometric measurements, where the horizontal bars represent the effective width of the passband. Blue symbols are the model fluxes from the best-fit NextGen atmosphere model (black). The green curve is the published Gaia DR3 spectrum for this star. \label{fig:SED}}
   
\end{figure}

We performed an analysis of the broadband spectral energy distribution (SED) of the star together with the {\it Gaia\/} EDR3 parallax \citep[with no systematic offset applied; see in particular][]{StassunTorres:2021}, in order to determine an empirical measurement of the stellar radius, following the procedures described in \citet{Stassun:2016,Stassun:2017,Stassun:2018}. We retrieved the $JHK_S$ magnitudes from {\it 2MASS}, the W1--W4 magnitudes from {\it WISE}, and the $G_{\rm BP} G_{\rm RP}$ magnitudes from {\it Gaia}, as well as the NUV flux from {\it GALEX}. Together, the available photometry spans the stellar SED over the wavelength range 0.2--22~$\mu$m (see Fig.~\ref{fig:SED}). All photometric data are summarized in Table~\ref{tab:SED_carac}.

\begin{table}
\caption{Main astrometric and photometric data of TOI-1695.\label{tab:SED_carac}}
\centering
\resizebox{\columnwidth}{!}{%
\begin{tabular}{lccc}
    Parameter & Unit & Value & Ref.  \\
    \hline
    TIC & & 422756130 &  \\
    Gaia DR2/EDR3 & & 534988616816537728 \\
    2MASS & & J01274094+7217472 \\
    WISE & & J012741.12+721747.6 \\
    RA & ICRS (J2000) & 01:27:40.973 & 1  \\
    DEC & ICRS (J2000) & +72:17:47.186 & 1 \\
    PM RA $\mu_\alpha$ & mas\,yr$^{-1}$ & 71.628$\pm$0.013 & 1 \\
    PM DEC $\mu_\delta$ & mas\,yr$^{-1}$ & 40.450$\pm$0.019 & 1 \\
    Epoch & & 2016.0 & 1 \\
    Gaia DR3 parallax & (mas) & 22.226$\pm$0.014 & 1\\
    distance & (pc) & 45.131$\pm$0.068 &  1 \\ 
    GAIA $G_{BP}$ & mag & 13.3064$\pm$0.0040 &  1 \\
    GAIA $G$ & mag & 12.1275$\pm$0.0011 & 1\\
    GAIA $G_\text{RP}$ & mag & 11.0464$\pm$0.0022 & 1 \\
    TESS $T$ & mag & 11.0294$\pm$0.0074 & \\
    2MASS $J$ & mag & 9.640$\pm$0.024 & 2\\
    2MASS $H$ & mag & 8.984$\pm$0.028 & 2\\
    2MASS $K$ & mag & 8.818$\pm$0.021 & 2\\
    WISE $3.4\mu$m & mag & 8.684$\pm$0.024 & 3 \\
    WISE $4.6\mu$m & mag & 8.61$\pm$0.02 & 3 \\
    WISE $12\mu$m & mag & 8.511$\pm$0.027 & 3 \\
    WISE $22\mu$m & mag & 8.40$\pm$0.29 & 3 \\
    \hline
\end{tabular}}
\tablefoot{(1) Gaia Collaboration et al. (2021); (2) Cutri et al. (2003); (3) Wright et al. (2010).}
\end{table}

We performed a fit using NextGen stellar atmosphere models~\citep{Hauschildt1999}, with the free parameters being the effective temperature ($T_{\rm eff}$) and metallicity ([Fe/H]), as well as the extinction $A_V$, which we limited to maximum line-of-sight value from the Galactic dust maps of \citet{Schlegel:1998}. The resulting fit (Fig.~\ref{fig:SED}) has a best-fit $A_V = 0.02 \pm 0.02$, $T_{\rm eff} = 3630 \pm 50$~K, and [Fe/H] = $0.0 \pm 0.5$, with a reduced $\chi^2$ of 1.3. It also fit well with the calibrated mean Gaia BP/RP spectrum available for this star in Gaia DR3~\citep{Montegriffo2022} as shown in Fig.~\ref{fig:SED}. Integrating the (unreddened) model SED gives the bolometric flux at Earth, $F_{\rm bol} = 6.82 \pm 0.24 \times 10^{-10}$ erg~s$^{-1}$~cm$^{-2}$, which with the {\it Gaia\/} parallax gives directly the bolometric luminosity, $L_{\rm bol} = 0.0431 \pm 0.0015$~L$_\odot$. Taking the $F_{\rm bol}$ and $T_{\rm eff}$ together with the {\it Gaia\/} parallax gives the stellar radius, $R_\star = 0.525 \pm 0.017$~R$_\odot$. In addition, we can estimate the stellar mass from the empirical $M_K$ relations of \citet{Mann:2019}, giving $M_\star = 0.539 \pm 0.027$~M$_\odot$. These mass and radius lead to $\log g$=4.72$\pm$0.14\,in cgs units.

The stellar parameters from the TESS project can be found in the TESS-exoFOP and are also reported in Table~\ref{tab:star_param}. TOI-1695 belongs to the specially curated list of cool dwarfs of the TIC~\citep{Muirhead2018,Stassun2019}. As explained in~\citet{Stassun2019}, the effective temperature $T_\text{eff}$ is retrieved from the cool dwarfs catalog of TIC v7~\citep{Stassun:2018}. The radius is derived from $T_\text{eff}$, Gaia magnitudes and Gaia parallax using equation (4) of \citet{Stassun2019}. The mass is obtained from the spline-interpolation of an empirical mass-T$_\text{eff}$ relationship.
 
The parameters derived from both methods are consistent within uncertainties.

\subsubsection{Comparison to synthetic spectra}
\label{sec:spectro}
We also derive the stellar parameters for TOI-1695 from a high-resolution template spectrum built from the tens of individual spectra acquired with SPIRou. The method used consists in the comparison of a grid of synthetic spectra to the template spectrum. We use two independent methods, both based on MARCS models \citep{Gustafsson2008}. 

The first method used is based on a process briefly described below (more details in \citealt{Cristofari2022, Cristofari2022b}). This process was tested and calibrated on reference stars. The synthetic spectra used for this analysis were computed from the MARCS model atmospheres with the \verb+Turbospectrum+ radiative transfer code \citep{Plez2012}, for a wide range of effective temperatures ($T_\text{eff}$), surface gravities ($\log{g}$) and metallicities ($[$M/H$]$). The models and template spectrum are compared by computing a $\chi^2$ value in key regions, containing lines that are both well reproduced by the models and sensitive to the stellar parameters. We therefore retrieve a three-dimensional grid of $\chi^2$ values, and perform a parabolo\"id fit to identify the best fitting parameters and estimate error bars on these values.
For TOI-1695, the parameters estimated with this method are $T_\text{eff}$~=~3627~$\pm$~31~K, $\log{g}$~=~4.60~$\pm$0.05~dex and [M/H]~=~0.10~$\pm$~0.10~dex. A fit of the star's projected rotational velocity $v\sin i$, fixing the micro-turbulent velocity to $v_\text{micro}$=1\,km\,s$^{-1}$, the macro-turbulence velocity to $v_\text{macro}$=0\,km\,s$^{-1}$ and the resolution linewidth of SPIRou to FWHM$\sim$4.3\,km\,s$^{-1}$, leads to a $v\sin i$=1.9$\pm$0.2\,km\,s$^{-1}$. Given that $v_\text{macro}$ is fixed to 0, we can only interpret this measurement as an upper-limit on $v\sin i$, therefore $<$2.5\,km\,s$^{-1}$ at 1--sigma, i.e. at most on the order of the pixel-scale of the SPIRou detector ($\sim$2.3\,km\,s$^{-1}$;  \citealt{Donati2020}). Modeled and observed spectral lines are compared together in Fig.~\ref{fig:linefit_cristo}.

The second method used is described in greater detail in section 4.2 in \citet{Martioli2022}. We calculated a grid of 650,000 synthetic spectra generated with \verb+MOOG+ \citep{Sneden2012} covering T$_\text{eff}$ from 2900\,K to 4100\,K with 50\,K step, $\log g$ from 3.5 to 5.3 with 0.1 dex step, [M/H] from -0.6 to 0.5 with 0.11 dex step, alpha from -0.2 to 0.4 with 0.06 dex step and $v_\text{mic}$ from 0.5\,km\,s$^{-1}$ to 4\,km\,s$^{-1}$ with 0.35\,km\,s$^{-1}$ step. We modified the Apache Point Observatory Galactic Evolution Experiment (APOGEE) line list \citep{Smith2021} by selecting only the lines that are most sensitive to T$_\text{eff}$ and $\log g$ within the spectral range from 1.5~$\mu$m to 1.7~$\mu$m. We searched the grid for the best fit parameters, which are adopted as priors in the \verb+iSpec+ integrator \citep{Blanco2019} to calculate the synthetic spectrum that best fits our SPIRou data within the domain of the selected lines. We used the \verb+MOOG+ code with the MARCS GES model \citep{Gustafsson2008} and \citet{Asplund2009} solar abundances. In total, 78 spectra were compared. The parameters of TOI-1695 estimated with this method are $T_\text{eff}$~=~3711~$\pm$~47~K, $\log{g}$~=~4.53~$\pm$0.13~dex, [M/H]~=~0.00~$\pm$~0.08~dex, and $v_\text{micro}$~=~0.3~$\pm$~0.7\,km\,s$^{-1}$. 

The values derived with these two methods agree well together, with moreover good agreement with the SED fitting at 2-$\sigma$.
In the rest of the analysis, we thus adopt the values $T_\text{eff}$=3650$\pm$40\,K, $\log g$=4.72$\pm$0.14, and [Fe/H]=0.0$\pm$0.1\,dex.
We also adopt the $M_\star$=0.54$\pm$0.03\,$M_\odot$ and $R_\star$=0.53$\pm$0.02\,R$_\odot$ derived from the SED fitting (Section~\ref{sec:SED} above) that rely directly on the spectrophotometry of the star with robust empirical relations. 

\begin{table*}[hbt]
    \centering
    \caption{\label{tab:star_param} Spectral parameters of the star derived from spectral template matching with the method from \citep{Cristofari2022} and de Almeida et al. (in prep.),  with the SED fitting method \citep{Stassun:2016,Stassun:2017,Stassun:2018}, and with the TESS project stellar parameters derivation method \citep{Stassun:2018}.}
    \begin{tabular}{lccccc}
        Parameter & Unit & TurboSpectrum+MARCS & Moog+MARCS & SED & TESS Input Catalog v8\tablefootmark{f} \\
        \hline
        $T_\text{eff}$ & K & 3627$\pm$31 & 3711$\pm$47 & 3630$\pm$50 & 3575$\pm$157\\
        $\log(g)$ &  & 4.60$\pm$0.05 & 4.53$\pm$0.13 & 4.5$\pm$0.5 & 4.723$\pm$0.009 \\
        $[$M/H$]$ & dex &  0.10$\pm$0.10 & 0.00$\pm$0.08 & 0.0$\pm$0.5 &  \\
        $v\sin i$ & km\,s$^{-1}$ & $<$2.5\tablefootmark{a}  & \\
        $v_\text{micro}$ & km\,s$^{-1}$ & fixed to 1.0 & 0.3$\pm$0.7 & \\
        $F_\text{bol}$ & 10$^{-10}$ erg\,s$^{-1}$\,cm$^{-2}$ &   & & 6.82$\pm$0.24\tablefootmark{b} &  \\
        $L_\text{bol}$ & L$_\odot$ &   & & 0.0431$\pm$0.0015\tablefootmark{c} & 0.039$\pm$0.009\\
        M$_\star$ & M$_\odot$ & & & 0.54$\pm$0.03\tablefootmark{d} & 0.51$\pm$0.02 \\
        R$_\star$ & R$_\odot$ & & & 0.53$\pm$0.02\tablefootmark{e} & 0.52$\pm$0.02 \\
        \hline
    \end{tabular}
    \tablefoot{\\
    \tablefoottext{a}{At 3--sigma and fixing $v_\text{macro}$=0\,km\,s$^{-1}$ (more details in Section~\ref{sec:spectro}).}\\
    \tablefoottext{b}{From direct integration of SED.}\\
    \tablefoottext{c}{Luminosity derived from the bolometric flux and the parallax.}\\
    \tablefoottext{d}{Mass determined from \citet{Mann2019} empirical relations. }\\
    \tablefoottext{e}{Radius determined from the bolometric flux and the parallax.} \\
    \tablefoottext{f}{Taken from \citet{Stassun2019}. See text for details.} \\
    }
\end{table*}

\subsection{Polarimetry}

An independent polarimetric reduction and LSD analysis of TOI-1695 SPIRou data using the Libre-Esprit (LE) pipeline \citep{Donati1997,Donati2020}, leads to the polarimetric longitudinal field, $B_\ell$, for TOI-1695 shown in Fig.~\ref{fig:magnetic} and given in Table~\ref{tab:polar}. The $B_\ell$ measurements are consistent with those obtained in the classical APERO reduction (e.g. \citealt{Martioli2022}). At a given epoch, $B_\ell$ is directly determined from the Stokes V and Stokes I profiles \citep{Donati1997}. We find that the star has a weak polarimetric variability and magnetic activity. The Stokes V profiles have a $\chi^2_r$=0.8 for a constant model, while the "null polarization" N \citep{Donati1997} has a $\chi^2_r$=1. Any variability in the data is thus fully consistent with the error bars.

\begin{figure*}[hbt]
    \centering
    \includegraphics[height=178.6mm, angle=-90]{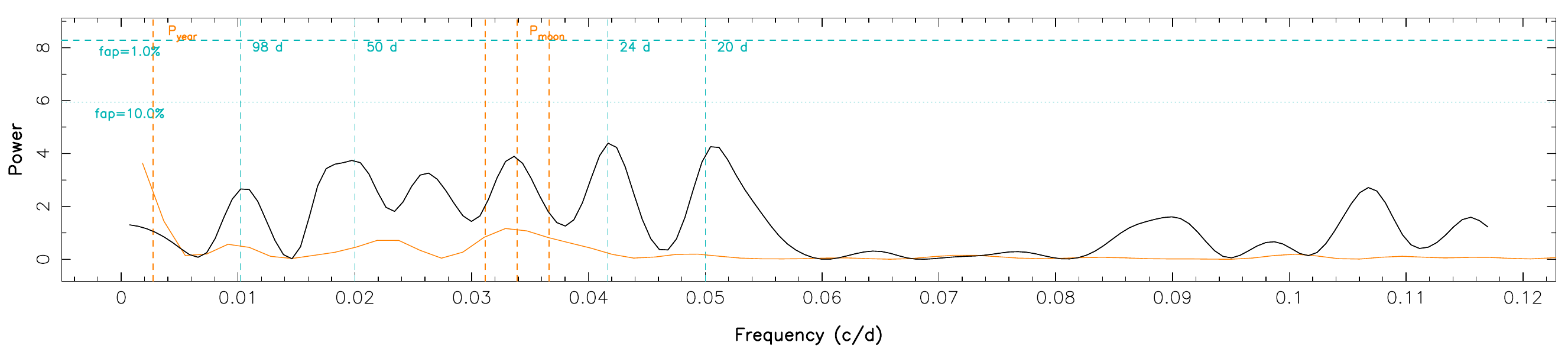}
    \caption{\label{fig:periodo_polar} Periodogram of the longitudinal magnetic field variations, with dominant frequencies highlighted in blue. The window function is shown in orange, with a peak close to the synodic orbital period of the Moon.  }
\end{figure*}

A power spectrum of the $B_\ell$ data (Fig.~\ref{fig:periodo_polar}) shows peaks beyond 20\,days, all with a false-alarm probability (FAP) $>$10\%. Thus no peculiar polarimetric signal is significantly detected for TOI-1695 given the current precision. The period of $\sim$48\,days seems nevertheless to dominate the power-spectrum at small frequencies. We fitted the data with a quasi-periodic Gaussian process (GP; see e.g. \citealt{Haywood2014,Aigrain2015}) using the same codes as in \citet{Donati2017}, at an initial period of 48\,days with typical decay-time $\tau_\text{decay}$=500\,days and smoothing factor $\gamma_\text{smooth}$=0.6. We obtain an almost flat GP model with a marginal likelihood difference $\Delta\ln{\mathcal L}$=2 compared to a constant model. The GP model is plotted over the longitudinal field on Fig.~\ref{fig:magnetic}.

We can thus conclude that (i) we do not see trace of a 15 days signal as reported in Section~\ref{sec:TESS}, (ii) the star was, if active, passing through a quiet phase during SPIRou observations, and (iii) more data are needed to confirm whether the hinted 48-days period could be a trace of the star's rotation signal.   

\subsection{RV variations and stellar activity effects}
\label{sec:activity}
We searched for correlations between RVs and the activity indicators, bisector span (BIS) and FWHM variations, obtained from the two APERO version v6 \& v7 studied in this work. The Pearson correlation coefficient for RV--BIS are $R_\text{v6}$=0.14 ($p$-value = 0.07) and $R_\text{v7}$=-0.22 ($p$-value = 0.004). The Pearson correlation coefficient for RV--FWHM is $R_\text{v6}$=-0.13 (p-value = 0.10) and $R_\text{v7}$=0.06 (p-value = 0.46). The only significant slope that we detect (1.7--$\sigma$) is for the FWHM and RV derived from the v6, of -0.62$\pm$0.37 m\,s$^{-1}$ per m\,s$^{-1}$. It disappears with the v7 suggesting that some instrumental signal was corrected from the SPIRou data with the last implementations in the pipeline. 

We thus find no significant correlations of astrophysical origin between RV and the activity indicators FWHM and BIS. This is compatible with the weak magnetic activity deduced from the almost flat longitudinal field $B_\ell$. 

\begin{figure}[hbt]
    \centering
    \includegraphics[width=89.3mm,clip=true,trim=70pt 49pt 30pt 32pt]{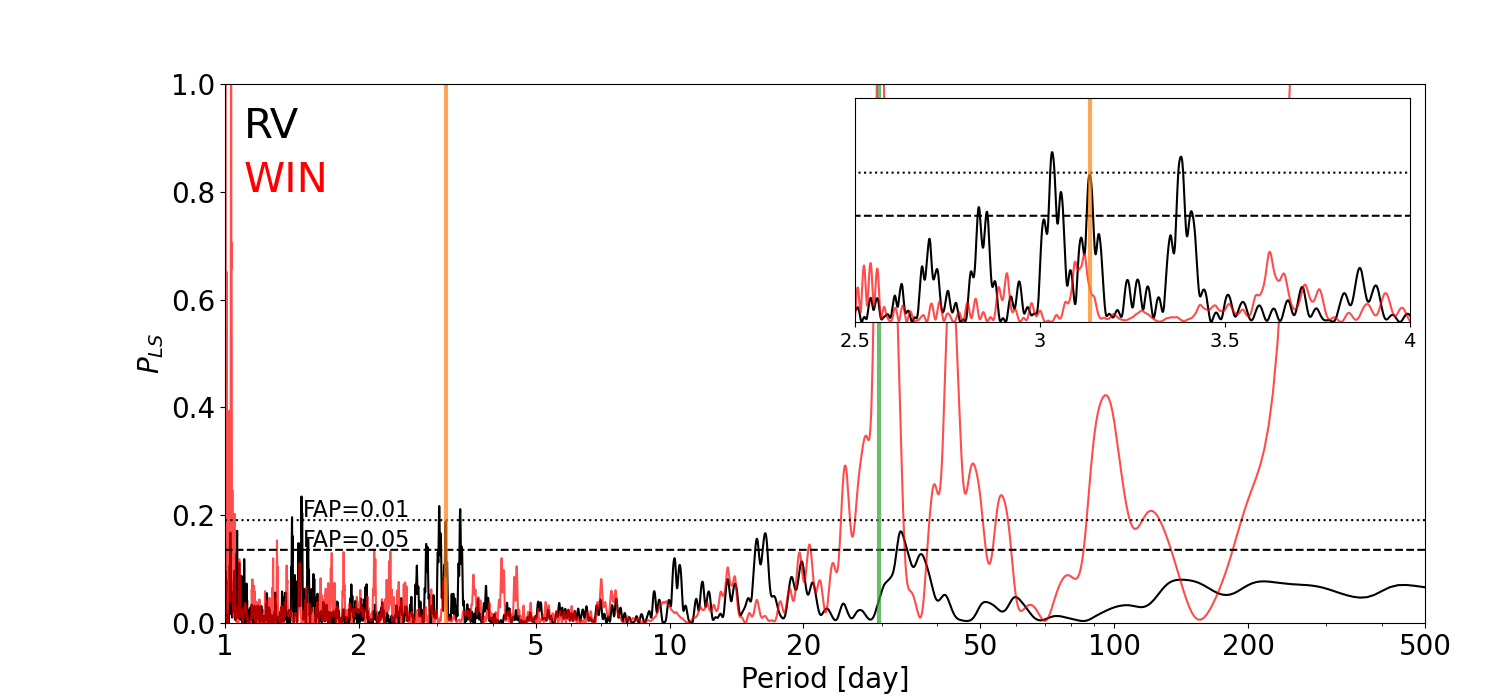}
    \includegraphics[width=89.3mm,clip=true,trim=70pt 50pt 30pt 32pt]{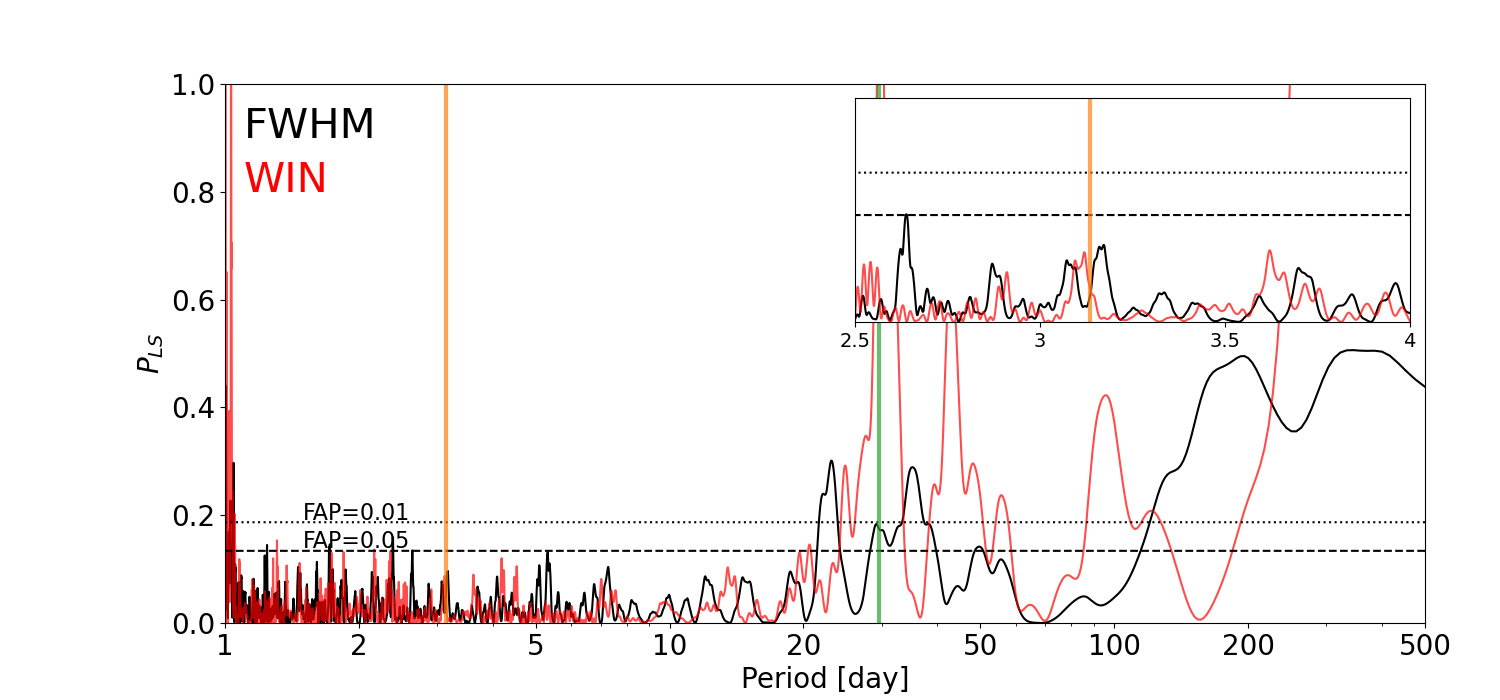}
    \includegraphics[width=89.3mm,clip=true,trim=70pt 0 30pt 32pt]{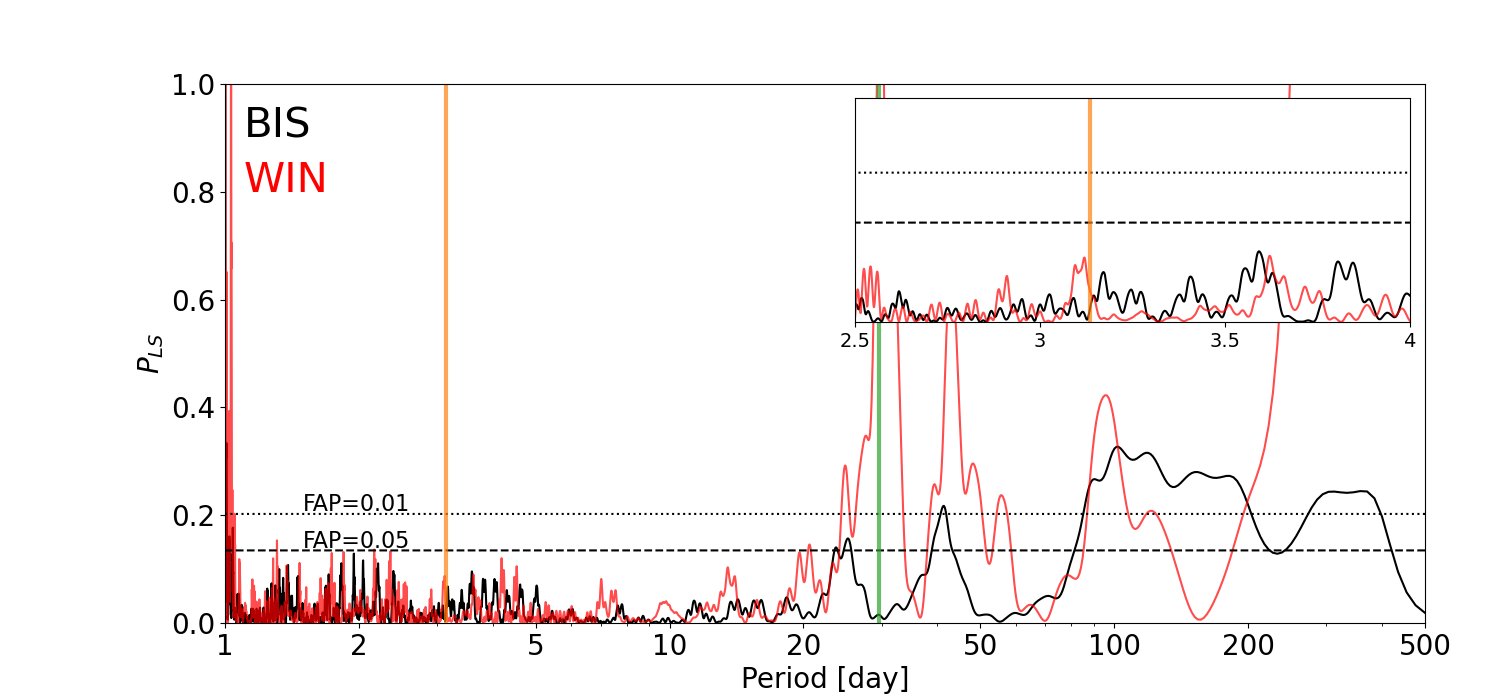}
    \caption{From top to bottom, RV, FWHM and BIS Lomb-Scargle periodograms with FAP levels 0.01 and 0.05 indicated as dotted and dashed lines for data extracted with the APERO v6 version. They are compared to the window function (red solid line). The vertical orange solid line indicates the 3.134-days period of the TESS transit signal. The moon synodic orbital period (29.53\,days) is also shown as a green solid lines. \label{fig:periodo_all}}
\end{figure}

The periodograms of the BIS and FWHM compared to the periodogram of the RVs in Fig.~\ref{fig:periodo_all} show indeed no significant peaks with FAP$<$5\% at periods $<$20\,days, especially none around 3.13 days, the orbital period of the candidate planet detected with TESS. Peaks with FAP$<$1\% appear at periods beyond 20\,days, some of which coincide with large peaks of the window-function.

\begin{figure}[hbt]
    \centering
    \includegraphics[width=89.3mm,clip=true,trim=70pt 0 50pt 32pt]{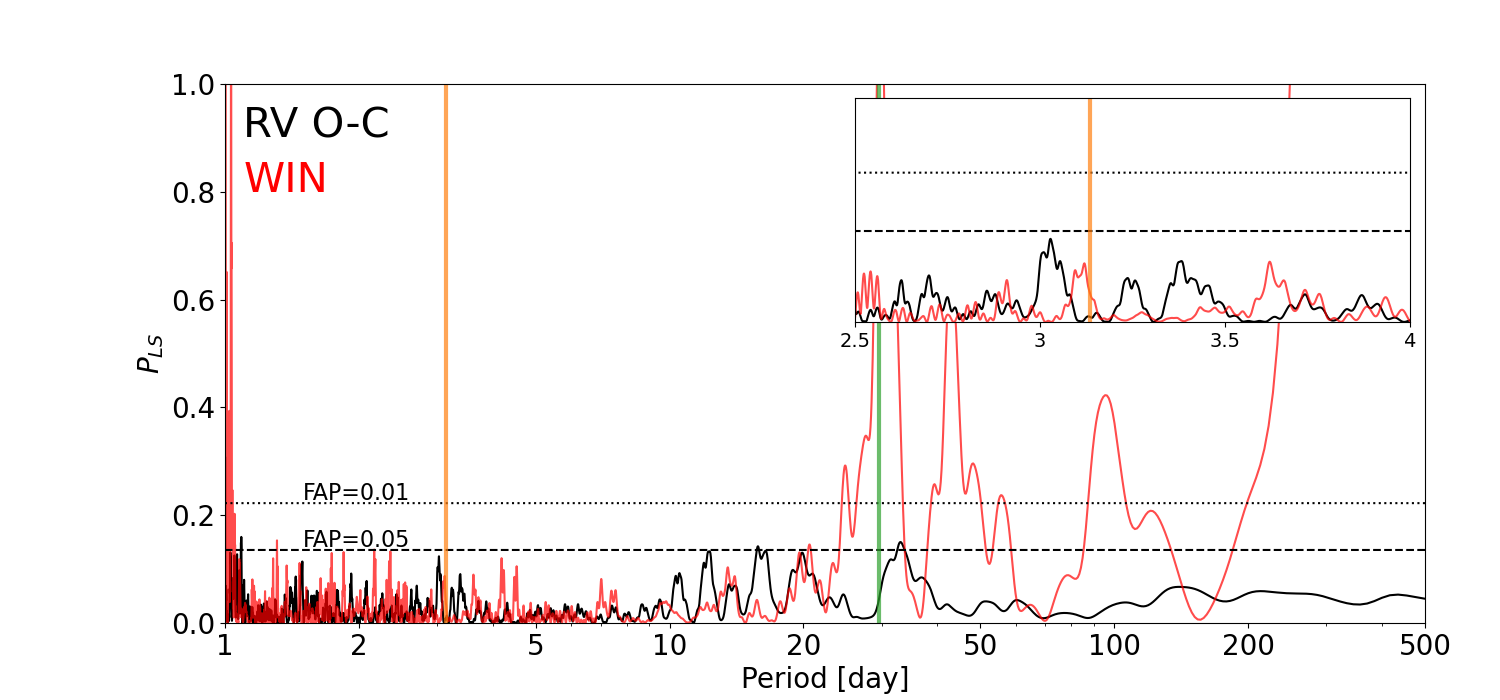}
    \caption{\label{fig:periodo_res} Periodogram of the residuals O-C after fitting a 3.134-days Keplerian to the RVs. Same color coding as in Fig.~\ref{fig:periodo_all}.}
\end{figure}

On the other hand, in the RV periodogram we have a positive detection of the candidate planet signal at 3.13\,days with a FAP$\sim$1\%. Two other strong peaks on both sides of this signal are due to aliases with the observing window frequencies. 
Indeed, most of the structures in the near periods of 3 days significantly weakens after fitting out a 3.13-days Keplerian to the RVs. As seen in the RV residuals periodogram in Fig.~\ref{fig:periodo_res}, residual RV variations then mainly concentrate around $\sim$16\,days. Interestingly, it is close to the 15-day period spotted in the lightcurve continuum variations of TESS sector 18 (Section~\ref{sec:TESS}). This would suggest an origin intrinsic to TOI-1695 itself or linked to its direct environment, even if no corresponding signal is detected in polarimetry. Studies of the TOI-1695's surrounding in Section~\ref{sec:imaging} exclude however that the RV variations could originate from background stars at less than 3\arcsec distance.  

Inspecting the periodogram of the LBL RV derived using the APERO v7 version (Fig.~\ref{fig:periodov7}) leads to similar conclusions. Only the power at low frequencies is significantly reduced from the RV, FWHM and the BIS variation, leaving only insignificant peaks, some of which are most likely bias of the Earth orbital period with the $\sim$20-days gap period. There is indeed some power at long period $\sim$365\,days which could be linked to yearly changes in the instrument resolution. 





\subsection{Concluding remark on the rotation period of TOI-1695}
\label{sec:rotation_discussion}
The TESS photometry and SPIRou RVs show traces of signals with periods within 14-19\,days. A rotation period of the star with a period in this range would be in agreement with the M2V spectral type found for TOI-1695 (see e.g. \citealt{Goulding2012,Goulding2013,Basri2011,Moutou2017}). There are few other observations that agree or not with such stellar rotation period.

Firstly, photometry and RV do not strictly agree with each other on the period of modulation. Photometry rather show a periodogram peak at 14.8 days. The LBL RV calculated from the APERO v6 data rather show a peak at 16 days. If they are calculated from the APERO v7 version the peak even shifts to $\sim$18 days.

Secondly, if the rotation period of the star were equal to 15\,days, following the hinted photometric variations in Section~\ref{sec:TESS}, the stellar radius of 0.53\,R$_\odot$ would imply a $v\sin i$ equal to 1.8 km\,s$^{-1}$, or smaller if the stellar spin and planet orbit are misaligned. This could agree with the $v\sin i$ of 1.9$\pm$0.2\,km\,s$^{-1}$ that was obtained in Section~\ref{sec:carac}, implying in this case that the equatorial plane of the star is aligned with the orbital plane of TOI-1695\,b. 

Thirdly, the SPIRou polarimetry of the star does not detect a 15-day periodicity of the star's magnetic field during the timespan of the SPIRou observations. If due to stellar rotation, we would have seen a stronger field and detected the rotational modulation in polarimetry. 

Finally, we retrieved available public photometry data to search for periodic signals indicative of stellar rotation.  As of 10 September 2022, TOI-1695 was observed by the All Sky-Automated Survey for SuperNovae \citep[ASAS-SN][]{Shappee2014,Kochanek2017} at 2043 epochs.  ASAS-SN images are obtained through a Sloan $g$ filter (and previously, for 879 epochs, in $V$-band).  A Lomb-Scargle periodogram analysis shows a signal at 52.4 days in $V$-band, but no significant signal in $g$-band.  We also retrieved 328 and 490 epochs of photometry in Sloan $g$ and $r$-band from the Zwicky Transient Factory archive \citep{Bellm2019,Masci2019}.  We find a FAP$<$1\% signal at 27 days in the $r$-band data but not the $g$-band data.  In principle, the 27-day signal could be an upper harmonic of the $\approx$52-day signal obtained from ASAS-SN. But no hint for a 15-day modulation. The absence of confirmatory detections in the other bands does not allow us to draw a firm conclusion about the rotation period of TOI-1695 from these data.  

We conclude that the 15-day signal seen in TESS sector 18, as well as the 16 (or 18) days signal seen in the SPIRou LBL RVs, are likely not related to the rotation of the star. Polarimetry, spectroscopy, photometry and RV agree nevertheless on a star rotation period larger than 15\,days. Moreover polarimetry, $V$-band and $r$-band photometry suggest a rotation period in the range 20--52\,days. Such long rotation period is found for M dwarfs with a stellar age $>$1\,Gyr \citep{Engle2018}. We thus infer that TOI-1695 is more than 1\,Gyr old.

\section{Joint analysis of TESS and SPIRou data}
\label{sec:planet}
We use here the same recipe as \citet{Martioli2022} to fit the combined photometric and velocimetric data from respectively TESS and SPIRou recorded for TOI-1695. We run a Markov-Chain Monte-Carlo (MCMC) algorithm with the \verb+emcee+ \citep{Foreman2013} routine with prior distribution shown in Table~\ref{tab:priors}. We fix the quadratic limb darkening parameters $u_0$ \& $u_1$ with respect to \citet{Claret2017} and \citet{Claret2022}, for a $T_\text{eff}$=3600\,K. We initialize 32 walkers, sample parameters space on 20,000 steps, and burn the 10,000 first samples. The posterior distributions of parameters are listed in Table~\ref{tab:MCMC_results}.

\begin{table}[hbt]
    \centering
    \caption{Prior distribution of fitted or fixed parameters entering the MCMC sampling. }
    \label{tab:priors}
    \begin{tabular}{lccc}
      Parameter  & Unit & Prior type & Prior distribution  \\
    \hline
    \multicolumn{4}{l}{Stellar parameters} \\ \\
      $M_\star$ & M$_\odot$ & Normal & $\mathcal N (0.54,0.03)$\\
      $R_\star$ & R$_\odot$ & Normal & $\mathcal N (0.53,0.02)$ \\ 
      $u_0$ & &	Fixed &	0.23\tablefootmark{$\dagger$} \\
      $u_1$ & &	Fixed &	0.38\tablefootmark{$\dagger$} \\
      \hline
    \multicolumn{4}{l}{Keplerian parameters} \\ \\
      $K$ & km\,s$^{-1}$ & Uniform & $\mathcal U$(0,10) \\
      $T_0$ & BTJD & Uniform & 	1791+$\mathcal U$(0.2,0.8) \\
      $P$	& day & Uniform	& 	$\mathcal U$(2.51,3.76) \\
      $a$	& R$_\star$ & Uniform & 	$\mathcal U$(10,23) \\
      $R_p$ & R$_\text{J}$ &	Uniform & 	$\mathcal U$(0.02,0.04) \\
      $I_p$ & $^\circ$ &	Uniform	& 	$\mathcal U$(70,110) \\
      $e$	& & Uniform or Fixed	& 	$\mathcal U$(0,1) or 0.0 \\
      $\omega$ & $^\circ$ & Uniform or Fixed	& 	$\mathcal U$(0,360) or 90 \\
      \hline
    \multicolumn{4}{l}{Gaussian process parameters} \\ \\
    $\sigma_\text{white,phot}$  & ppm       &  Uniform & $\mathcal U$($-10^{100}$,$10^{100}$) \\
    $\tau_\text{decay,phot}$ & day          &  Uniform & $\mathcal U$(0.25,1000)\\
    $\gamma_\text{smooth,phot}$ &           &  Uniform & $\mathcal U$(0.1,1.5)\\
    $P_\text{phot}$ & days                  &  Uniform & $\mathcal U$(10,20)     \\
    $A_\text{phot}$ & ppm                   &  Uniform & $\mathcal U$($10^{-4}$,2000)\\
    $\sigma_\text{white,RV}$  & m\,s$^{-1}$ &  Uniform & $\mathcal U$($-10^{100}$,$10^{100}$)\\
    $\tau_\text{decay,RV}$ & day            &  Uniform & $\mathcal U$(6,1000)   \\
    $\gamma_\text{smooth,RV}$ &             &  Uniform & $\mathcal U$(0.1,1.5)  \\
    $P_\text{RV}$ & days                    &  Uniform & $\mathcal U$(10,20)   \\
    $A_\text{RV}$ & m\,s$^{-1}$             &  Uniform & $\mathcal U$(0,100)   \\
    \hline
    \end{tabular}
    \tablefoot{\\
    \tablefoottext{$\dagger$}{Based on \citet{Claret2017,Claret2022}.}}
\end{table}

We apply the fit on different analysis schemes of the LBL RV, using both v6 and v7 reductions (RV$_\text{v6}$ and RV$_\text{v7}$ hereafter). This allows comparing the quality of the RV series extracted from both APERO versions and qualifying the improvements brought by the new pipeline. For both versions, fits without drift correction nor zero-point correction were poor in terms of $\chi^2$ and the O-C residuals. So we performed fits assuming:
\begin{itemize}
   \item DC+ZPC: accounting for the long-term calibration drift of the RV and a zero-point correction (see Section~\ref{sec:obs} for details);
   \item DC+ZPC+GP: adding a quasi-periodic (QP) Gaussian Process (GP) to the fit, as explained in more details below.
\end{itemize}
The result of the joint Keplerian fit of both lightcurve and RVs are all compared in Table~\ref{tab:MCMC_results}, assuming circular orbit. 

In the DC+ZPC scheme, the 3.134-days RV signal is found with a S/N larger than 5--$\sigma$. This allows us to confirm the detection of the planet signal with a corresponding companion mass in the super-Earth/sub-Neptune regime $\sim$6\,M$_\oplus$. Although compatible at 0.7--$\sigma$, the semi-amplitude of the Keplerian model fitted to the RV$_\text{v7}$ (3.6$\pm$0.7\,m\,s$^{-1}$) is smaller than that of the RV$_\text{v6}$ (4.1$\pm$0.7\,m\,s$^{-1}$). On the other hand, the RV-residuals dispersion is slightly smaller for RV$_\text{v6}$ than for RV$_\text{v7}$ with respective reduced $\chi^2$ of 0.8 and 1.1, and absolute O-C dispersion of 6\,m,s$^{-1}$ and  7\,m\,s$^{-1}$. We also noted a difference of $\sim$460\,m\,s$^{-1}$ in systemic RV which we identified to a different treatment of the calibration between the two versions. 
The difference in the Keplerian solutions and residuals, implying systematic variations unaccounted for in both, motivated us into digging further with the GP analysis of those datasets.

\begin{figure}
    \centering
    \begin{picture}(89.3,90)
    \put(0,0){\includegraphics[width=89.3mm]{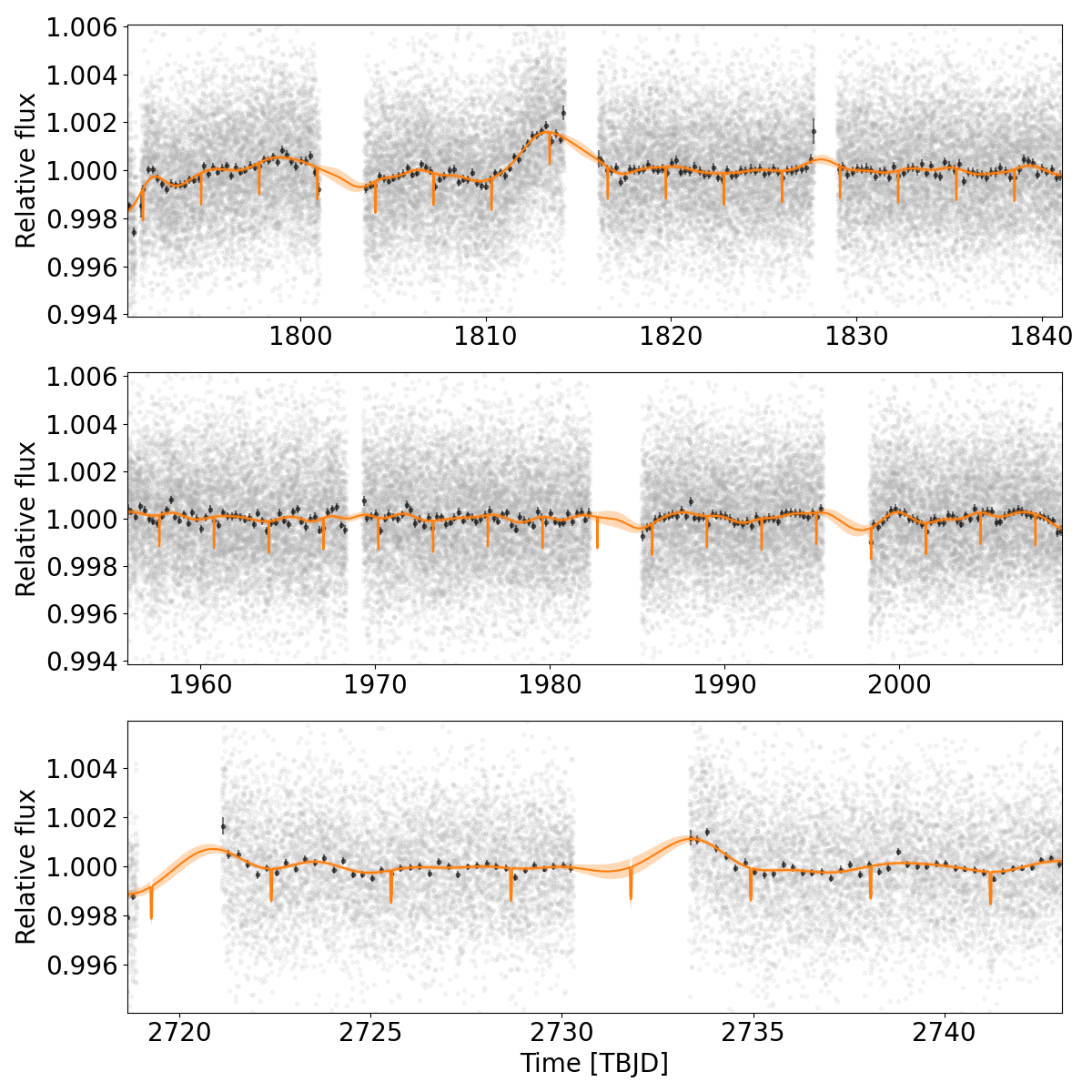}}
    \put(15,83){Sector 18}
    \put(50,83){Sector 19}
    \put(15,54){Sector 23}
    \put(55,54){Sector 24}
    \put(30,25){Sector 52}
    \end{picture}
    \caption{Gaussian process fit of the lightcurve with a quasi-periodic kernel. Grey points are raw PDCSAP flux TESS data, and black points are the binned data with $\Delta t_\text{bin}$=0.25\,days.}
    \label{fig:GP_LC}
\end{figure}
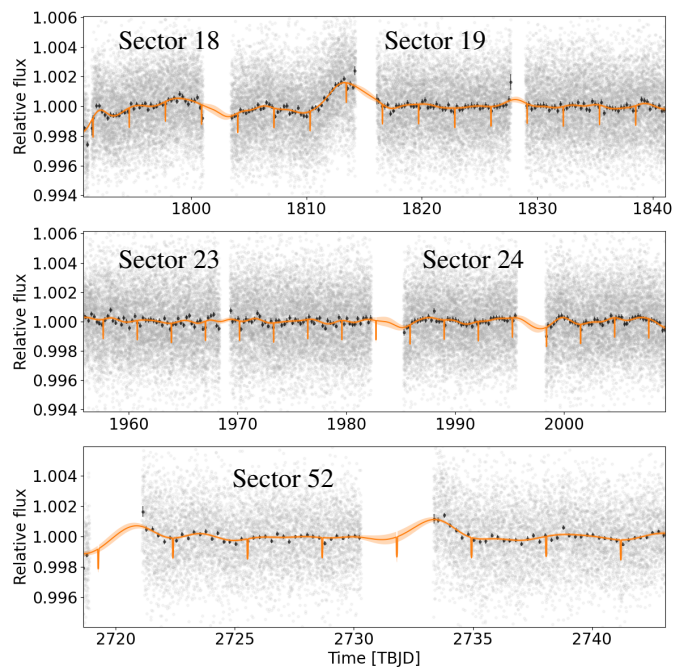

\begin{figure}
    \centering
    \begin{picture}(89.3,50)
    \put(0,23){\includegraphics[width=89.3mm]{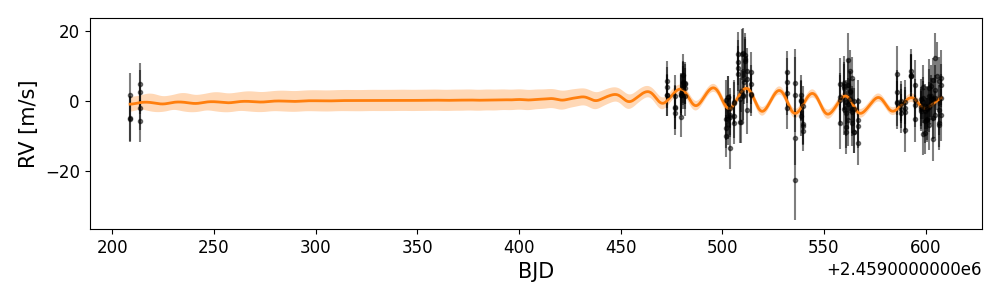}}
    \put(10,30){APERO v6}
    \put(0,0){\includegraphics[width=89.3mm]{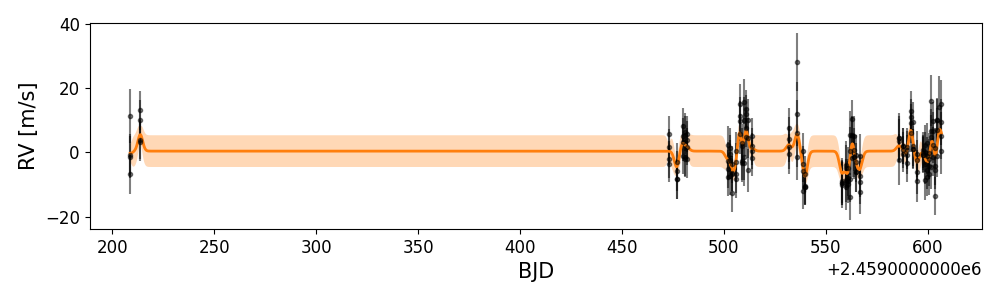}}
    \put(10,7){APERO v7}
    \end{picture}
    \caption{Gaussian process fit of the SPIRou RV in the case of the APERO v6 reduction (top) and the v7 reduction (bottom) with a quasi-periodic kernel.}
    \label{fig:GP_RV}
\end{figure}

The quasi-periodic Gaussian process that we used combines a squared exponential (SE) and a periodic kernel with the code \verb+george+ \citep{Ambikasaran2015} to remove the periodic pattern in the RV and the TESS-LC identified in Sections~\ref{sec:TESS} and \ref{sec:activity}. The GP is trained separately on the residuals RVs and on the photometric time series excluding the transits. The priors used on the GP parameters are added to Table~\ref{tab:priors}. We were careful not to allow the decay-time of the SE kernel to be smaller than twice the orbital period of TOI-1695\,b, thus preventing the GP to overfit and absorb some of the orbital signal present in the RV, along the recommendations drawn in \citet{Angus2018}. For optimising the GP hyper-parameters more efficiently using an MCMC, the photometric lightcurves were rebinned to time steps of 0.25\,days. Fig.~\ref{fig:GP_LC} and~\ref{fig:GP_RV} show the GP modeling of the TESS lightcurve and the SPIRou RV. 

The QP GP fit of the lightcurve converged to a period of 13.0$^{+3.4}_{-0.7}$\,days. For the RV$_\text{v6}$ it converged to a period of 16.2$^{+0.6}_{-3.9}$\,days and for the RV$_\text{v7}$ to 18.3$^{+0.5}_{-4.1}$\,days.
Once applying and then removing the GP from the RV and photometric time series, we fit again the orbital and transit signals from the RV and LC. 
The DC+ZPC+GP analysis leads to the smallest RV residual dispersion with reduced $\chi^2$ of 0.6 and 0.5 respectively of RV$_\text{v6}$ and RV$_\text{v7}$. The flux dispersion $\sim$2008\,ppm is constant over all schemes, which is not surprising given that the lightcurve continuum is always detrended around the transits epochs. The bayesian information criterion (BIC) among all schemes is clearly in favour of the DC+ZPC+GP scheme with the smallest BIC. The posterior distributions although different are compatible with the DC+ZPC solutions. As expected, the model semi-amplitudes of RV$_\text{v6}$ (3.95$\pm$0.66\,m\,s$^{-1}$) and RV$_\text{v7}$ (3.63$\pm$0.68\,m\,s$^{-1}$) are smaller (only slightly for RV$_\text{v7}$) than without modeling the systematic variations. 
The Keplerian fit of the RV$_\text{v7}$-GP residuals leads to the best residual dispersion among all analysed datasets, $\sigma_{O-C}$=4.8\,m\,s$^{-1}$, but with similar planetary parameters than without fitting the GP. 

As an additional check, we run a fit of the data in the DC+ZPC scheme, but considering a non-zero eccentricity. The eccentricity obtained when letting it freely vary is consistent with 0 at 1.3-$\sigma$ with $e$$<$$0.2$. More importantly, at maximum likelihood, the BIC is larger than for the zero-eccentricity case. This validates that fixing the eccentricity to zero in the other fits is a correct, most parsimonious approach, consistent with a tidally-circularized orbit of the planet, and moreover not generating significant systematic errors on the other orbital parameters. 

Varying the limb darkening coefficients led to wide posterior distributions of $u_0$ and $u_1$ with unrealistic values $u_0$$\sim$$0.5$$\pm$$0.4$ and $u_1$$\sim$$0.9$$\pm$$0.7$. Although compatible  within errors with the common values assumed in the rest of the analysis, they lead to less accurate planet radius estimation with highly biased posteriors on semi-major axis and radius. Notably, the median of the radius posterior distribution is 1.77$\pm$0.21\,R$_\oplus$, but the maximum a posteriori point estimate is $\sim$1.99\,R$_\oplus$. This shows that the present dataset is not able on its own to refine the limb darkening coefficients. It is more reasonable to rely on the known values of the limb darkening coefficients of an M2 star. 

\begin{figure}[hbt]
    \centering
    \includegraphics[width=89.3mm]{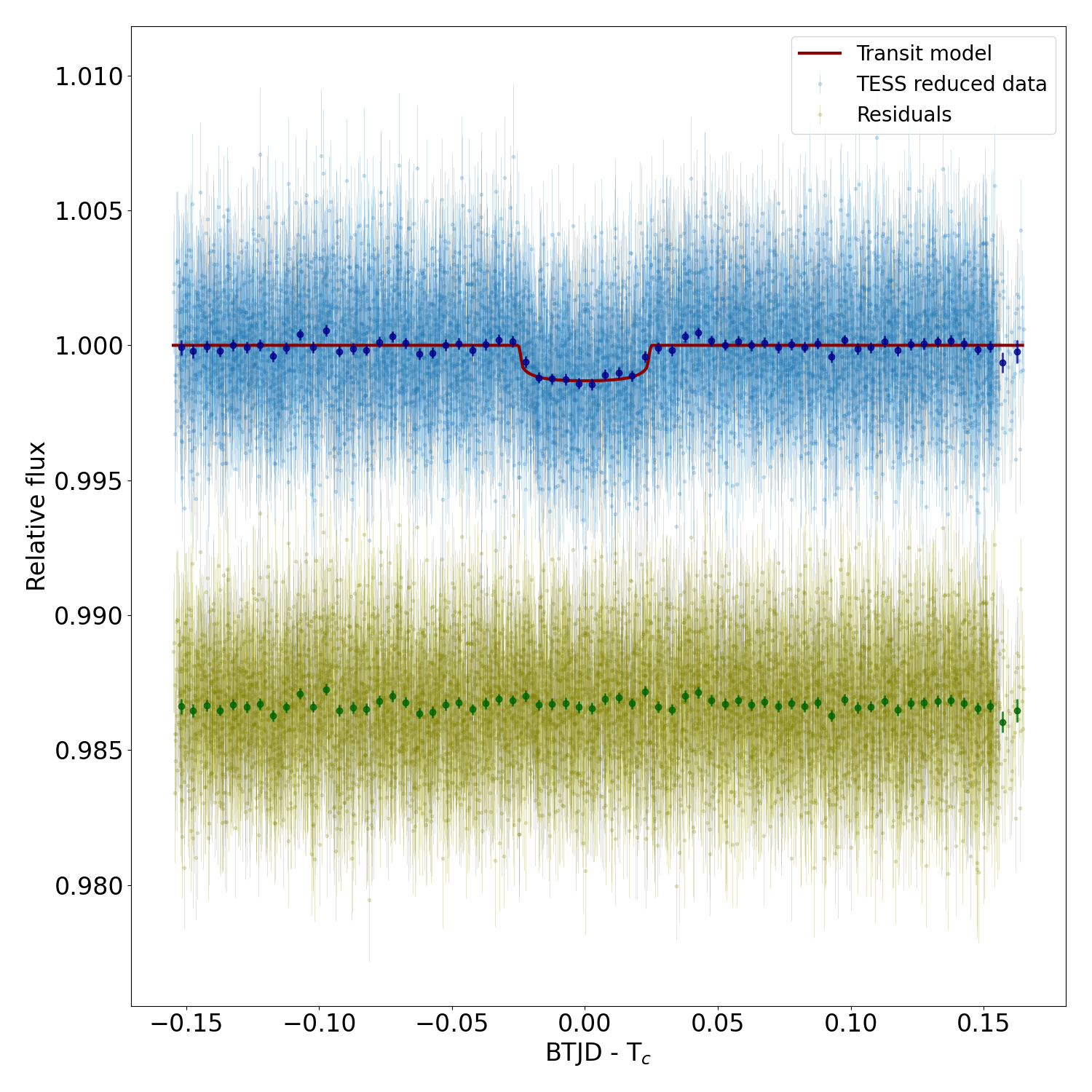}
    \caption{Results from fits to the TESS light curve, with all the observed transits stacked, with the model taken from the maximum a posteriori point estimation, based on maximizing the posterior probability distributions for each parameter. At this point, $T_0$=1791.5206\,BTJD, $P$=3.1342799\,days, $a$=21.1\,R$_\star$, $R_p$=0.034\,R$_\star$ \& $I_p$=89.4$^\circ$.  \label{fig:transit_modeling}}
\end{figure}

\begin{figure}[hbt]
\centering
\includegraphics[width=89.3mm,clip=true,trim=10pt 0pt 0pt 0pt]{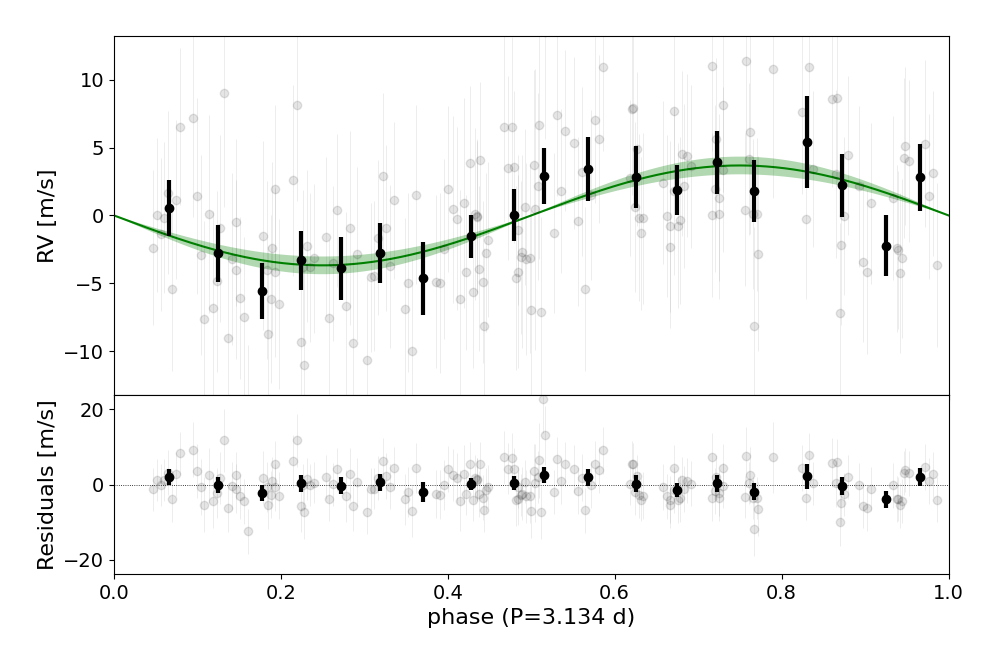}
    \caption{Phase-folded modelisation of the RV variation seen with SPIRou of the star TOI-1695 with a 3.13 days circular orbit. The data used here are derived from the APERO v7 version. Grey dots gather all the RV data and black dots are binned with a time step of 0.05\,days.  \label{fig:RV_modeling_v7}}
\end{figure}

Comparing all the different reduction/analysis schemes, we retain the solution with the smallest BIC. Finally, it is the DC+ZPC+GP derived from the APERO v7 version that leads to the best solution. It can be trusted that the GP did not overfit the RV signal as the planetary parameters are similar than if not fitting a GP. The corner plot of the posterior distribution of all varied and derived Keplerian parameters is shown in Fig.~\ref{fig:corner}. The best fitting model of the transit curve is shown in Fig.~\ref{fig:transit_modeling}, and the best fitting models of the phase-folded LBL RV is shown in Fig.~\ref{fig:RV_modeling_v7}. 

It results in a companion on an orbit of 3.1343\,days with a planetary radius of 2.03$\pm$0.18\,R$_\oplus$ and a planet mass of 5.5$\pm$1.0\,M$_\oplus$. This implies a sub-Earth planet density of 3.6$\pm$1.1\,g\,cm$^{-3}$, and an equilibrium temperature of 590$\pm$90\,K.

\begin{table*}[hbt]
    \centering
    \caption{MCMC results of the photometry (TESS) + spectroscopy (SPIRou) joint fit in the different RV-reduction schemes using APERO v6 and v7 versions, which are presented in Section~\ref{sec:planet}. For all parameters, except those of GPs, the main value is the median and the uncertainty is the average distance of the median to the 16th and 84th percentiles. For the GP hyper-parameters, since the posteriors are asymmetrical, the main value is given by the mode, and the uncertainties are the distance between this mode and the 16th and 84th percentiles. \label{tab:MCMC_results}}
    
\resizebox{\textwidth}{!}{%
    \begin{tabular}{@{}lc|cc|cc@{}}
    
    Param. & Unit & \multicolumn{2}{c|}{with APERO v6} & \multicolumn{2}{c}{with APERO v7}\\ 
    &  & DC + ZPC  & DC + ZPC + GP & DC + ZPC & DC + ZPC + GP \\
    \hline
\multicolumn{4}{l}{Fitted parameters} \\ \multicolumn{4}{l|}{}\\
$K$             & m/s               & 4.12$\pm$0.69              & 3.95$\pm$0.66         & 3.69$\pm$0.65 & 3.63$\pm$0.68\\
$t_c$           & BTJD              & 1791.5205$\pm$0.0013       & 1791.5204$\pm$0.0013  & 1791.5204$\pm$0.0013 & 1791.5203$\pm$0.0012\\
$P$             & day               & 3.1342788$\pm$0.0000077    & 3.1342793$\pm$0.0000077 & 3.1342788$\pm$0.0000078 & 3.1342790$\pm$0.0000072\\
$a_p$           & R$_\star$         & 	16.3$\pm$5.2             & 16.7$\pm$5.0          & 15.6$\pm$5.3 & 16.3$\pm$5.1\\
$R_p$/$R_\star$ &                   & 0.0349$\pm$0.0026          & 0.0344$\pm$0.0025     & 0.0351$\pm$0.0028 & 0.0344$\pm$0.0027\\
$I_c$           & $^\circ$          & 	89.9$\pm$3.5             & 89.8$\pm$3.3          & 90.0$\pm$3.8 & 89.8$\pm$3.5\\
$u_0$           &                   & 	0.22                     & 0.22                  & 0.22 & 0.22\\
$u_1$           &                   & 	0.39                     & 0.39                  & 0.39 & 0.39 \\
$\gamma$        & m\,s$^{-1}$       & -59921.62$\pm$0.47         & -59921.64$\pm$0.47    & -59461.93$\pm$0.47 & -59461.96$\pm$0.45\\
$e$             &                   & 0.0                        & 0.0                   & 0.0 & 0.0 \\
$\omega$        &  $^\circ$         & 90                         & 90                    & 90 & 90 \\
\hline
\multicolumn{4}{l}{Planet intrinsic parameters} \\\multicolumn{4}{l|}{}\\
$M_p$           & M$_\oplus$     & 6.2$\pm$1.1             & 6.0$\pm$1.0           & 5.6$\pm$1.0  & 5.5$\pm$1.0\\
$R_p$           & R$_\oplus$     & 2.06$\pm$0.17           & 2.03$\pm$0.17         & 2.07$\pm$0.18 & 2.03$\pm$0.18\\
$\rho_p$        & $\rho_\oplus$  & 0.71$\pm$0.22           & 0.71$\pm$0.21         & 0.63$\pm$0.20 &  0.66$\pm$0.21\\
$T_\text{eq}$   & K              & 580$\pm$90              & 575$\pm$86            & 590$\pm$100 & 590$\pm$90\\
$T_\text{irr}$  & K              & 900$\pm$140             & 890$\pm$130           & 920$\pm$160 &  900$\pm$140\\ 
$S_p$           & $S_\oplus$     & 27$\pm$17               & 26$\pm$16             & 30$\pm$17 & 27$\pm$17 \\ 
\hline
\multicolumn{4}{l}{Goodness of fit diagnostics} \\ \multicolumn{4}{l|}{}\\
$\chi^2_\text{r, RV}$   &                & 0.79            & 0.66                  & 1.16     & 0.49\\
O-C$_\text{RV}$         & m\,s$^{-1}$    & 5.9             & 5.39                  & 7.09     & 4.75 \\ 
O-C$_\text{phot}$       & ppm            & 2007            & 2006                  & 2010     & 2008 \\
BIC                     &                & -70137          & -70160                & -70049   & -70179\\
\tablefootmark{a}$\Delta$BIC &           & 28              & 25                    & 21       & 20\\
\hline
\multicolumn{4}{l}{GP hyper-parameters} \\\multicolumn{4}{l|}{}\\
$\sigma_\text{white,phot}$  & ppm       &                    & 98$\pm$19             &  & 100$\pm$18\\
$\tau_\text{decay,phot}$ & day          &                    & 7.2$^{+2.3}_{-5.1}$   &  & 6.1$^{+3.2}_{-4.7}$ \\
$\gamma_\text{smooth,phot}$ &           &                    & 0.34$^{+0.11}_{-0.06}$&  & 0.36$^{+0.15}_{-0.08}$\\
$P_\text{phot}$ & days                  &                    & 13.0$^{+3.4}_{-0.7}$  &  & 13.0$^{+4.1}_{-0.8}$\\ 
$A_\text{phot}$ & ppm                   &                    & 407$^{+46}_{-33}$     &  &  405$^{+48}_{-37}$\\
$\sigma_\text{white,RV}$  & m\,s$^{-1}$ &                    & 0.52$^{+0.53}_{-0.37}$&  & 0.58$^{+0.61}_{-0.39}$\\
$\tau_\text{decay,RV}$ & day            &                    & 370$^{+420}_{-316}$   &  & 32$^{+95}_{-18}$  \\
$\gamma_\text{smooth,RV}$ &             &                    & 0.91$^{+0.42}_{-0.50}$&  & 0.38$^{+0.38}_{-0.16}$\\
$P_\text{RV}$ & days                    &                    & 16.2$^{+0.6}_{-3.9}$  &  & 18.3$^{+0.5}_{-4.1}$\\
$A_\text{RV}$ & m\,s$^{-1}$             &                    & 3.5$^{+2.6}_{-1.3}$   &  & 5.0$^{+1.8}_{- 1.1}$ \\
\hline
    \end{tabular}
    }
    \tablefoot{\\
    \tablefoottext{a}{The $\Delta$BIC is calculated between the model without any signal (planet or GP) and the model considered in the given column header.}
    }
\end{table*}

\section{Discussion on TOI-1695 b mass and radius}
\label{sec:discussion}
\begin{figure}[hbt]
    \centering
    \includegraphics[width=89.3mm,clip=true]{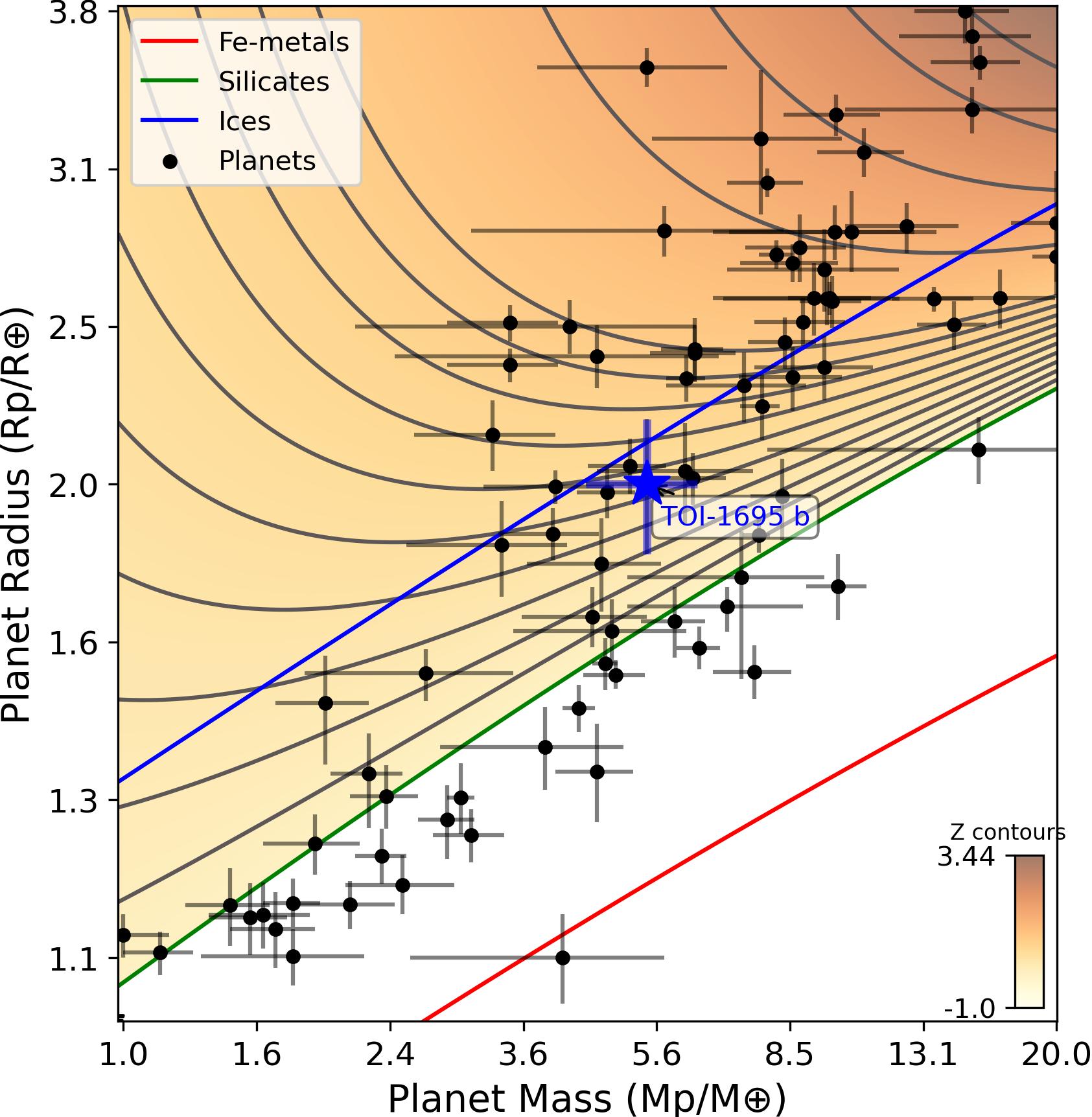}
    \caption{Mass-radius plot obtained based on the {\tt pyExoRaMa} code \citep{Zeng2021,Francesco2022}. The red, green and blue solid lines represent the mass-radius relation for respectively pure Fe, silicate, and H$_2$O core. The yellow/brown-colored $z$-contours (see \citealt{Zeng2021} for details) with black solid lines represent the radius inflation due to gas in an envelope surrounding the planet core corresponding to pure silicate. The comparison planets (black circles) are taken from the TepCat database \citep{Southworth2011}, selecting only exoplanets around M and K host-stars. TOI-1695\,b is shown as a blue star. \label{fig:MRplot}}
\end{figure}

We find that TOI-1695\,b is a sub-Neptune planet with an average density $\sim$3.6\,g\,cm$^{-3}$ smaller than the Earth density. The mass and radius of TOI-1695\,b are compared to other super-Earth/sub-Neptunes in Fig.~\ref{fig:MRplot}, as obtained thanks to the \verb+pyexorama+ package~\citep{Zeng2021,Francesco2022}. TOI-1695\,b sits in-between two similar sub-Neptunes, Wolf-503\,b \citep{Peterson2018, Polanski2021} and TOI-270\,d \citep{vaneylen2021,gunther2019} in terms of mass, radius and temperature, at the border of the radius valley but on the sub-Neptune side of it. It also has similar mass and radius as K2-3\,b \citep{Kosiarek2019} although warmer (463\,K for K2-3\,b). 

\subsection{The possible composition of the planet}
Assuming a pure silicate core leads to a planet with a significant gas envelope in the formalism developed by \citet{Zeng2021}. An M-R diagram comparing TOI-1695\,b to other super-Earths and sub-Neptunes is shown in Fig.~\ref{fig:MRplot}. Using the \verb+smint+ code \citep{Piaulet2021} based on the formalism of \citet{Lopez2014,Zeng2016, Aguichine2021}, we considered three different scenarios: 

\begin{itemize}
    \item a planet with H/He envelope and a solid interior leads to $f_\text{H/He}$=0.28$^{+0.46}_{-0.23}$\%;
    \item a 100\% silicate interior with pure water on top of it, we derive a water envelope with a mass fraction  of $f_{\text{H}_2\text{O}}$=55$^{+29}_{-30}\%$;
    \item a water-world with unfixed iron content in the core of the planet leads to an iron-to-silicate ratio $f'_\text{core}$=47$\pm$30\% and a water envelope with a mass fraction of $f_{\text{H}_2\text{O}}$=23$\pm$12\%.
\end{itemize} 

All of these scenarios agree on the evidence that, as for other sub-Neptune exoplanets of similar radius and mass, as in particular K2-3\,b, a fraction of the planet mass should be in the form of a gaseous envelope. With $T_\text{eq}$=590$\pm$90\,K TOI-1695\,b has an equilibrium temperature just below the water critical point, theoretically allowing water to be liquid if a surface pressure of about 200\,atm can be sustained. However, if a significant atmospheric layer is present, the temperature will also certainly increase with pressure. Thus water cannot be in a liquid phase on the surface of this planet, as was shown for GJ\,3470\,b whose irradiation temperature is similar \citep{Piette2020}.

\subsection{Atmospheric mass loss or gas-depleted formation?}
The worlds with radius $\sim$2\,R$_\oplus$ are reputed to be underabundant around Solar-type stars due to atmospheric loss forming rather $R_p$$<$$1.7$\,R$_\oplus$ planets \citep{Kite2021,Rogers2021}. Yet, TOI-1695\,b, as well as few others, exhibit such characteristics. Because it is located around an M-dwarf, TOI-1695\,b has undergone different irradiation conditions than planets around Solar-type stars and followed a different evolution with respect to photoevaporation. 

Using the equation 15 from \citet{Lecavelier2007}, we measure an order of magnitude for the evaporation rate for TOI-1695\,b of $dm/dt$$\sim$$10^{10}$\,g\,s$^{-1}$. This evaporation rate indicates that if a H/He envelope is still present today in the atmosphere of this planet (less than 1\% of the planet mass), it has to be, in this scenario, the remains of the evaporation of a much more massive envelope. If the gas represented about 10\% of the mass of the initial planet, this primitive atmosphere would have had a lifetime shorter than 1.5\,Gyr. This is about the minimum age of the system that we inferred from spectroscopic, photometric and activity analysis of TOI-1695 in Section~\ref{sec:carac}.

A lifetime of 10\,Gyr is obtained for an envelope with 70\% of the mass of the core, almost doubling the initial planet mass. Although possible, in this case the large mass loss rate requires a lot of fine tuning on the envelope mass and the age of the planet to explain our observations. It would make this detection fortuitous and thus unlikely if photoevaporation is responsible for the actual values of the mass and radius of TOI-1695\,b.

Therefore, if the age of this system is closer to $\sim$1\,Gyr, then photoevaporation could be at the origin of a shallow H/He atmosphere; but if the age of TOI-1695 is rather beyond 5\,Gyr, then this planet was more likely formed as is: with a small atmosphere, and survived because it formed far away and migrated inward. In this case the scenario proposed by gas-depleted formation~\citep{Lee2014} is preferred. 

\begin{figure}
    \centering
    \includegraphics[width=89.3mm]{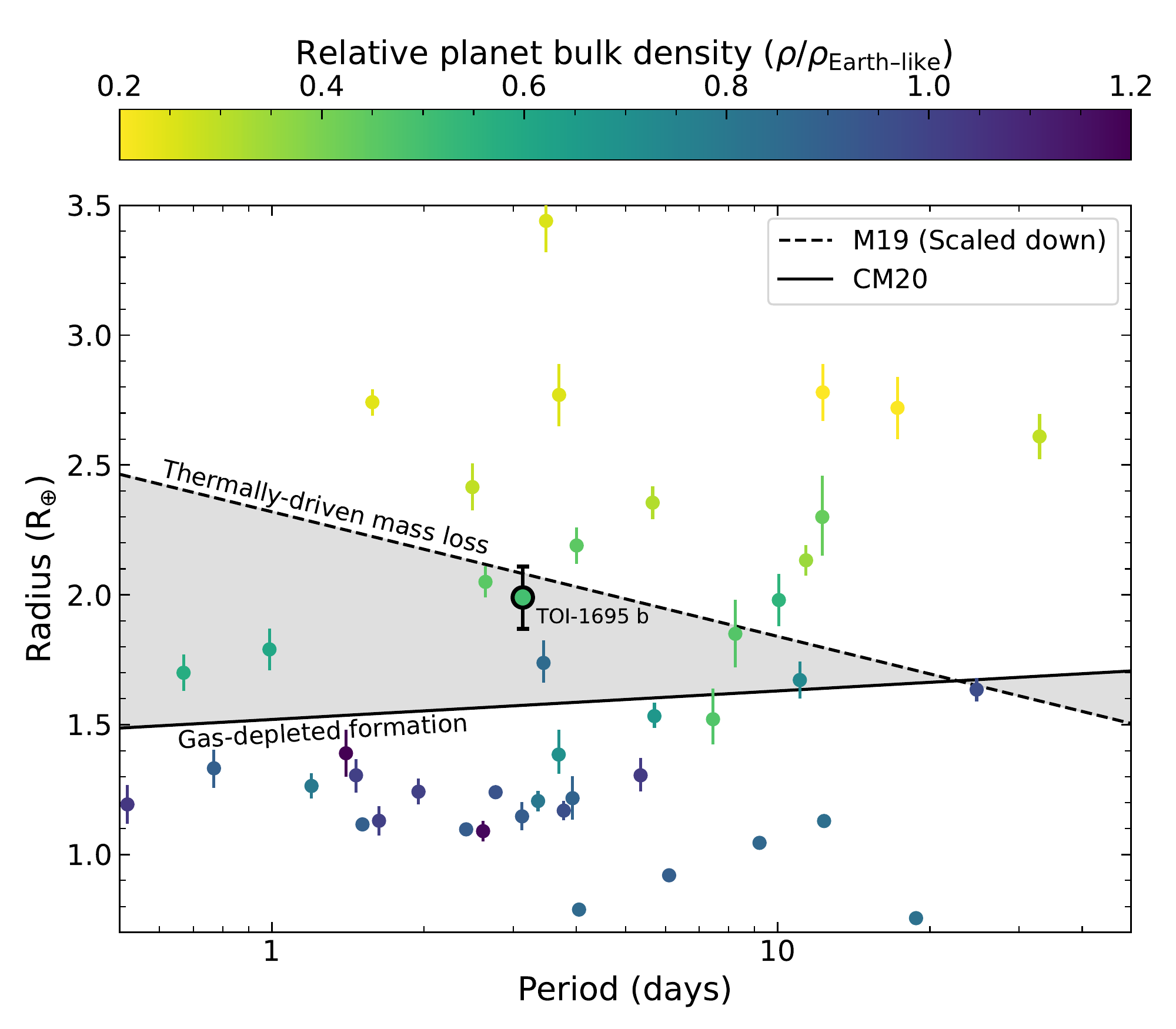}
    \caption{The $R_p$-$P_\text{orb}$ diagram for exoplanets around cool stars ($T_\text{eff}$$<$$4000$\,K) taken from the NASA exoplanet archive. The solid and dash lines represent respectively the \citet{CloutierMenou2020} and \citet{Martinez2019} empirical radius valley (see text). The shaded region shows exoplanets that from their gas content could be able to test gas-depleted formation and thermally-driven mass loss  scenarios. TOI-1695\,b is highlighted as a thick-lined black circle.}
    \label{fig:RpP_valley}
\end{figure}

This is illustrated in Fig.~\ref{fig:RpP_valley}. It compares TOI-1695\,b to other similar exoplanets around cool stars ($T_\text{eff}$$<$$4000$\,K) in an $R_p$-$P_\text{orb}$ diagram and confront it to rocky to non-rocky empirical transitions consistent with gas-depleted formation~\citep{CloutierMenou2020} and thermally-driven atmospheric escape~\citep{Martinez2019}. The latter includes photoevaporation \citep{Lopez2018}, as well as core-powered mass loss \citep{Gupta2022} mechanisms, each predicting a similar negative slope of the radius valley in $R_p$-$P_\text{orb}$ space for low-mass stars.  In contrast, the gas-depleted formation scenario can produce larger rocky super-Earths at longer orbital periods, resulting in a positive slope of the rocky/non rocky transition. Exoplanets in the shaded region of Fig.~\ref{fig:RpP_valley} are interesting candidates to test those formation/evolution scenarios. 

Determining the rocky/gaseous nature of exoplanets in the region enclosed by those two empirical $R_p$-$P_\text{orb}$ relations is therefore of key importance to determine which scenario dominates over the other around M dwarfs. Here, TOI-1695\,b as non-rocky tends to favour the gas-depleted formation scenario. However it is located close to the thermally-driven mass loss limit, and given the age uncertainty on the host star, the photoevaporation or core-powered mass loss scenarios cannot be rejected. 

\subsection{Prospect for atmospheric observations}

With a J magnitude of 9.6 for the star TOI-1695, the transits of TOI-1695 b 
can be observed in a search for atmospheric signatures. 
To characterize the potential for the detection of atmospheric absorption lines, 
we can evaluate the ratio of the atmospheric absorption depth to the noise 
in the transit light curve. Of course, the noise level will depend on the wavelength, telescope,
instrument, etc., and the depth of the atmospheric absorption 
depends on the species, abundances, oscillator strength and number
of the searched lines, etc. 
Therefore, now only a relative S/N ratio of detection 
can be calculated for an exoplanet atmosphere to be compared with other planets to 
be observed for the search of the same signature with the same instrument.  

Here, we have made the calculation in the J band, which is currently available from space 
and ground-based facilities, and which is close to the H$_2$O band observed 
in a large number of exoplanets \citep[{\it e.g.},][]{Fraine_2014,Benneke_2019,Tsiarias_2019,Mikal-Evans_2021}.
The atmospheric absorption depth, that is the fraction 
of the stellar flux that is absorbed by the atmosphere during the transit, is proportional to 
the area of the absorbing layer, which is given by the scale height of the atmosphere ($H$) 
multiplied by $2\pi R_p$, where $R_p$ is the planet radius, and inversely proportional to 
the stellar disk area ($\pi R_{star}^2$). 
The noise is assumed to be 
proportional to the square root of the stellar flux given by $F_J \propto 10^{-0.4 m_J}$, 
where $m_J$ is the star J magnitude.
The atmospheric scale height is given by $kT/\mu g$, where $T$ is the atmosphere temperature, $k$ the Boltzmann constant, 
$\mu$ the mean molar molecular mass, 
and $g$ is the planet gravity. 
Finally, with $g\propto M_p/R_p^2$, 
where $M_p$ is the mass of the planet, the signal-to-noise ratio is given by $S/N \propto H R_p / R_{star}^2 \sqrt{F_J}$,
in agreement with the TSM (transmission spectroscopic metric) 
as defined by \cite{Kempton_2018} \citep[see also,][]{Cointepas_2021}.

We calculated the S/N ratio expected for all known exoplanets transiting an M star, 
and normalized them to the value of 100 for AU Mic b as done in \cite{Martioli2022}.
We used the catalog of exoplanets published in the Exoplanets Encyclopedia 
on the 18$^{\mathrm th}$ July, 2022 \citep{Schneider_2011}. For the J magnitudes, 
we used the tabulated values when available, or calculated theoretical values 
from the V magnitudes and the stars effective temperatures assuming a black-body spectrum.
We considered only the M-type stars of the catalogue ; a star is considered to be an M-type star 
if it is tabulated with this stellar type or if no stellar type is given in the catalogue and its effective temperature is between 2200\,K and 3900\,K.  

\begin{figure}
   \centering
   \includegraphics[width=1\hsize]{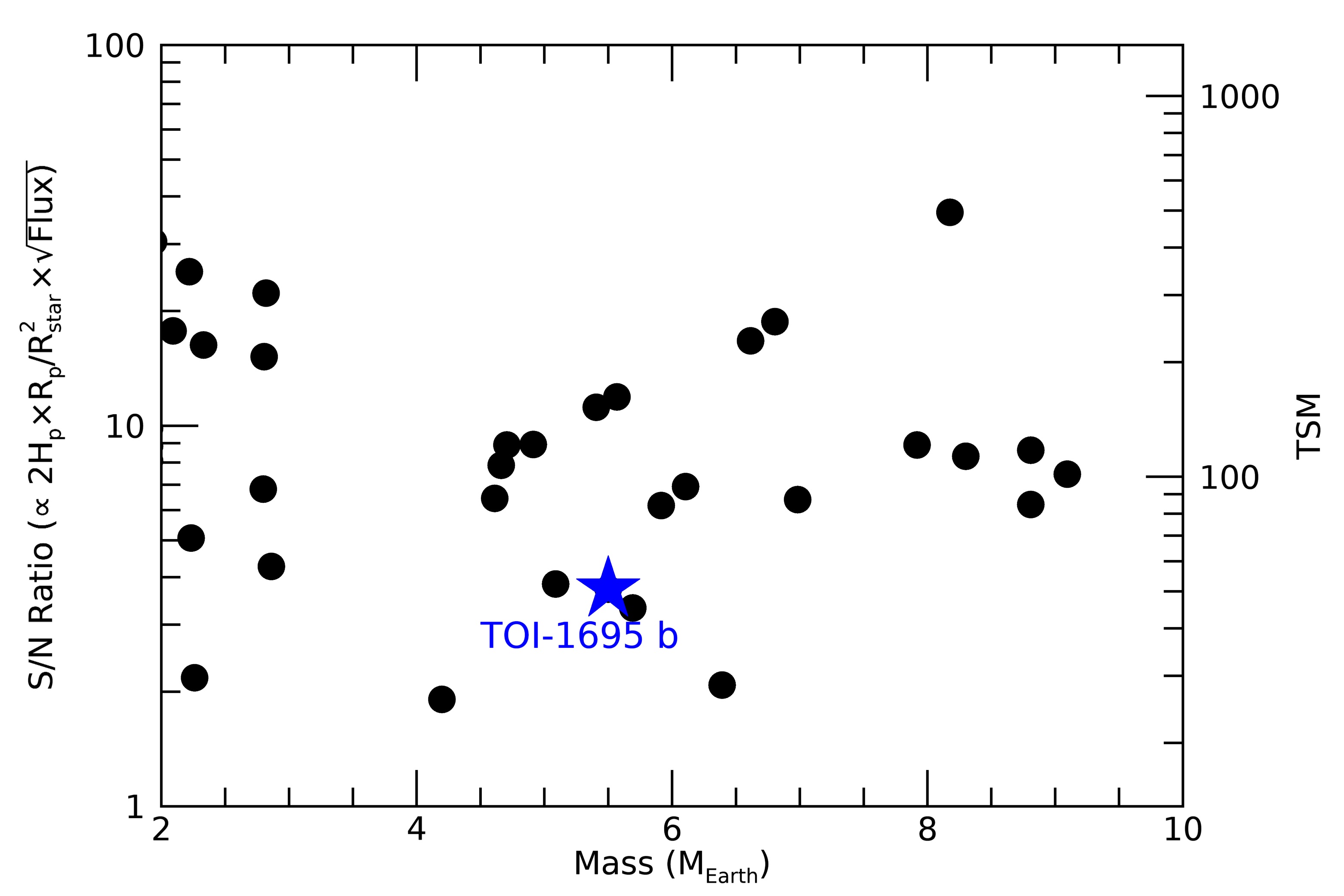}
      \caption{Signal-to-noise ratio (left axis) of the atmospheric signatures and TSM (right axis) in the J band as a function of the planetary mass for exoplanets orbiting M~type stars with masses between 2 and 10~Earth mass. The S/N ratios have been normalized to a reference S/N of 100 for AU Mic b as done in \cite{Martioli2022}.}
        \label{fig:SN_J_vs_Mass}
  \end{figure}

The result is shown in Fig.~\ref{fig:SN_J_vs_Mass} where we plotted the S/N ratio of the atmospheric signatures and the TSM in the J band as a function of the planetary mass for known exoplanets orbiting M~type stars with masses between 2 and 10\,M$_\oplus$. With an S/N of $4$ of TOI-1695\,b, in the 2 to 10~Earth mass range, several exoplanets yield a much better S/N. Even if we calculate the same plot for the S/N in the V band, with a stellar type M2V and a V magnitude of 13.0, the situation is barely improved and the conclusion remains the same. Nevertheless, TOI-1695\,b has a TSM of 48, a decent S/N ratio to be reached in only 10\,h of observation with JWST. 

We determined also the emission signature metric (ESM; eq. 4 in \citealt{Kempton_2018}) of TOI-1695\,b in the K-band, assuming a day-side temperature of the planet of 1.1$\times$\,\!$T_\text{eq}$. We found an ESM of 3, implying an S/N=3 for a JWST detection of the planet in the K-band during a secondary eclipse. This is again significantly below the threshold for defining the best "emission" targets, but still challenging in term of detectability.  

Thus, even though TOI-1695\,b is not the best target in terms of precision, its atmosphere could still be characterized with a limited observation time on large telescopes such as JWST.

\section{Summary and conclusion}
\label{sec:conclusion}

Thanks to RV follow-up observations with SPIRou at CFHT, we have established the planetary nature of a companion discovered by TESS around the star TOI-1695 with an orbital period of 3.134\,days. 

We have shown that TOI-1695\,b is a sub-Neptune planet, with a mass of 5.5$\pm$1.0\,M$_\oplus$ and a radius of 2.03$\pm$0.18\,R$_\oplus$. We found hints of a supplementary weak variability in both photometry ($P$$\sim$$12$-15\,days) and RV ($P$$\sim$$16$-19\,days), but not present in activity indicators (polarimetry, FWHM, bisector span). These activity indicators moreover do not show signs of any clear modulation below a period of $\sim$20\,days. It implies that TOI-1695 is likely to be a slow rotator, older than 1\,Gyr. The origin of the supplementary variability in RV and photometry is still uncertain.

The density of TOI-1695\,b is 3.6$\pm$1.1\,g\,cm$^{-3}$ suggesting the existence of an atmospheric layer on top of the solid bulk of the planet. If made of H/He, the gas envelope would have a low mass fraction $f_{H/He}$=0.28$_{-0.12}^{+0.46}$\%. If containing water and varying its core composition, the planet may have a water mass fraction of 23$\pm$12\%. This confirms that TOI-1695\,b stands in the sub-Neptune domain of the radius valley.

The detection prospects of the atmosphere of TOI-1695\,b with current and future instruments, such as the JWST and ELTs, with a TSM of 48 are relatively encouraging. Its atmosphere has a S/N ratio of detection lower than several other similar targets, and its characterization is thus not among the easiest. Nevertheless, its radius and orbital period, compared to the rocky/gaseous bounds of the different formation/evolution scenarios of super-Earth and sub-Neptune, make it an interesting case for testing those scenarios.

\begin{acknowledgements}
The authors gratefully thank the anonymous referee for his corrections and suggestions that led to improve the content of this article. The authors wish to recognize and acknowledge the very significant cultural role and reverence that the summit of MaunaKea has always had within the indigenous Hawaiian community. We are most fortunate to have the opportunity to conduct observations from this mountain. 
This paper includes data collected by the TESS mission,
which are publicly available from the Mikulski Archive for
Space Telescopes (MAST). Funding for the TESS mission is
provided by NASA’s Science Mission directorate. STScI is operated by the Association of Universities for Research in Astronomy, Inc., under NASA contract NAS 5-26555.
We acknowledge funding from the French National Research Agency (ANR) under contract number ANR\-18\-CE31\-0019 (SPlaSH). F.K. acknowledges support from the Université Paris Sciences et Lettres under the DIM-ACAV program Origines et conditions d'apparition de la vie. E. M. acknowledge funding from the Funda\c{c}\~{a}o de Amparo \`{a} Pesquisa do Estado de Minas Gerais (FAPEMIG) under the project number APQ-02493-22. J.-F.D. acknowledges funding from the European Research Council (ERC) under the H2020 research \& innovation programme (grant agreement \#740651 NewWorlds). This work is partly supported by the French National Research Agency in the framework of the Investissements d'Avenir program (ANR-15-IDEX-02), through the funding of the "Origin of Life" project of the Grenoble-Alpes University. J.H.C.M. is supported in the form of a work contract funded by Funda\c c\~ao para a Ci\^encia e Tecnologia (FCT) with the reference DL 57/2016/CP1364/CT0007; and also supported from FCT through national funds and by FEDER-Fundo Europeu de Desenvolvimento Regional through COMPETE2020-Programa Operacional Competitividade e Internacionaliza\c c\~ao for these grants UIDB/04434/2020 \& UIDP/04434/2020, PTDC/FIS-AST/32113/2017 \& POCI-01-0145-FEDER-032113, PTDC/FIS-AST/28953/2017 \& POCI-01-0145-FEDER-028953, PTDC/FIS-AST/29942/2017. J.H.C.M. also acknoleges the support from FCT - Funda\c c\~ao para a Ci\^encia e a Tecnologia through national funds and by FEDER through COMPETE2020 - Programa Operacional Competitividade e Internacionaliza\c c\~ao by these grants: UID/FIS/04434/2019; UIDB/04434/2020; UIDP/04434/2020; PTDC/FIS-AST/32113/2017 \& POCI-01-0145-FEDER-032113; PTDC/FIS-AST/28953/2017 \& POCI-01-0145-FEDER-028953.\\
\end{acknowledgements}

\onecolumn
\bibliographystyle{aa}
\bibliography{biblio}

\begin{appendix}
\onecolumn

\section{TESS aperture and field-of-view}
\begin{figure*}[h]
    \centering
    \includegraphics[width=178.6mm]{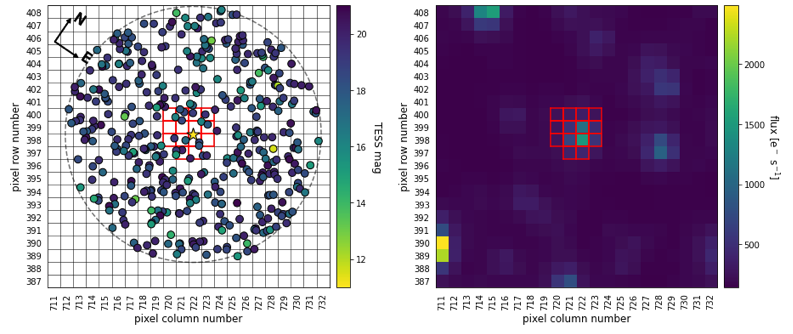}
    \caption{Aperture used to retrieve the photometry of TOI-1695 from the SPOC in sector 18, compared to the background sources identified by Gaia within 200\arcsec (left), and the background resolved sources on the full-frame image (right). This image is an output from TRICERATOPS (see text).}
    \label{fig:aperture}
\end{figure*}

\section{Spectral modelisation of the star}
\begin{figure*}[h]
    \centering
    \includegraphics[width=178.6mm,trim=170pt 0pt 10pt 30pt, clip=true]{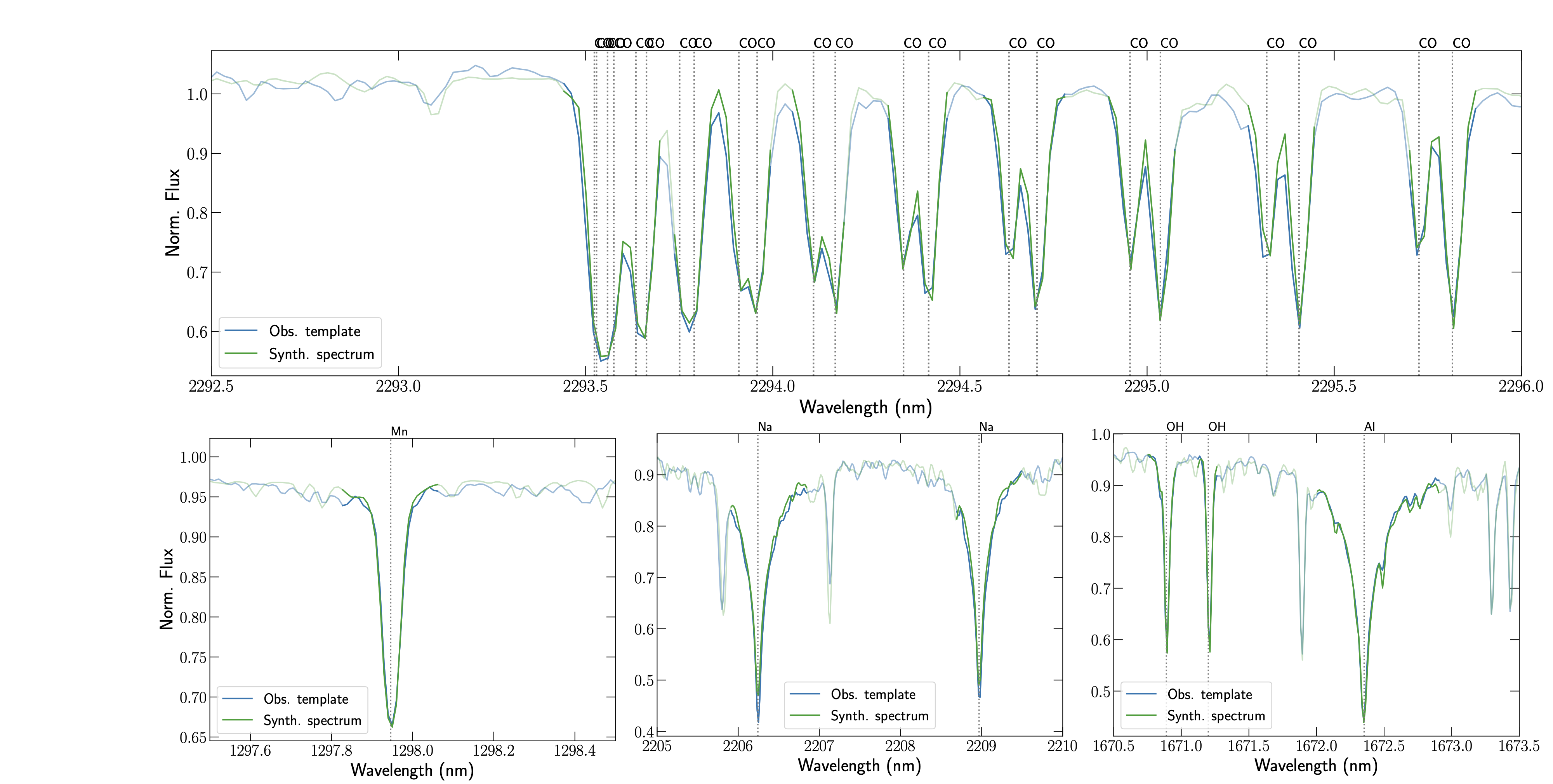}
    \caption{\label{fig:linefit_cristo} Stellar lines spectral fitting with MARCS model (see Section~\ref{sec:carac} for explanations), with CO (top), Mn (bottom-left), Na (bottom-middle) and OH (bottom-right) lines. The best fitting model is plotted in green solid line.}
\end{figure*}

\clearpage
\section{SPIRou polarimetry time series}
\begin{figure*}[h]
    \centering
    \includegraphics[height=178.6mm, angle=-90]{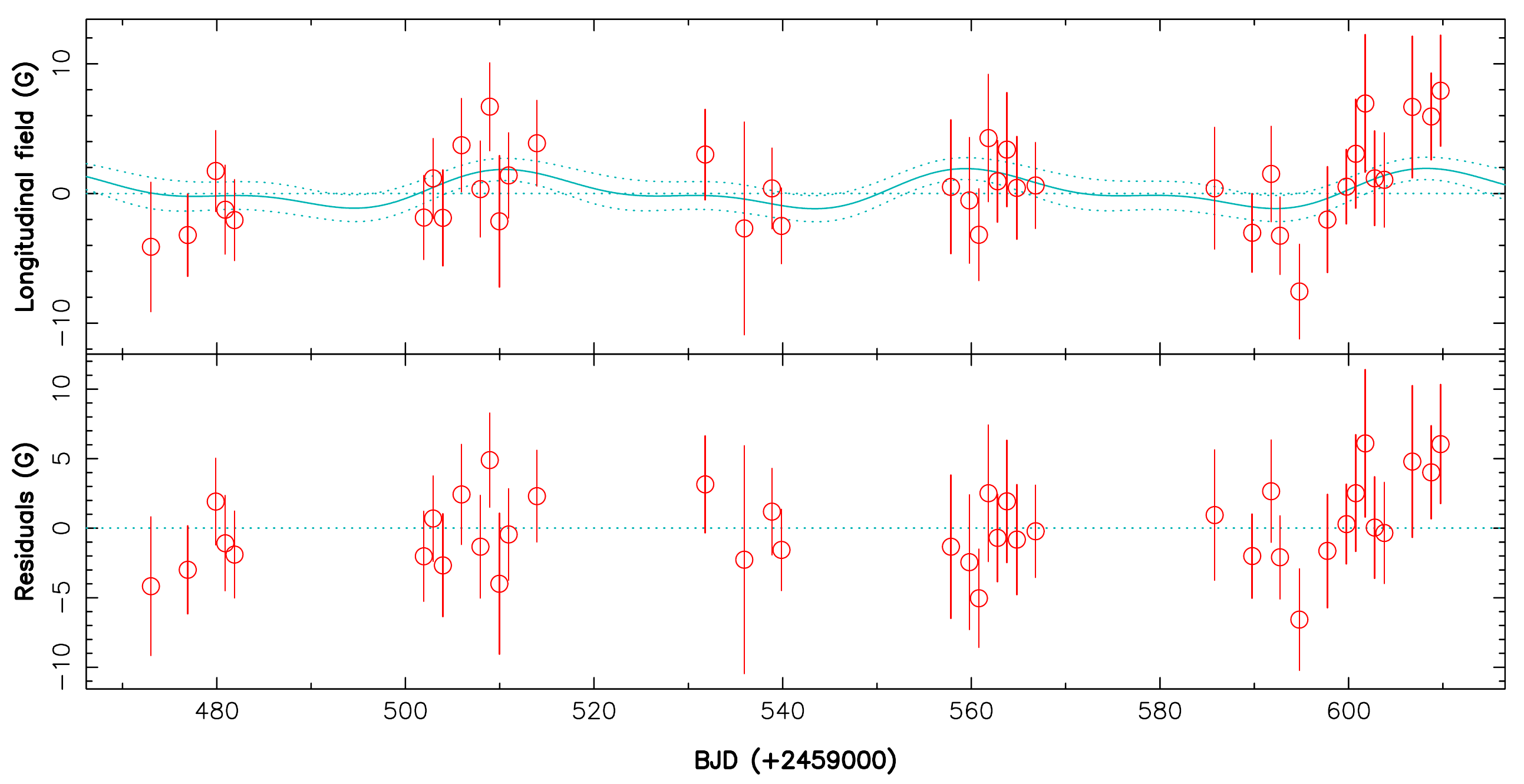}
    \caption{\label{fig:magnetic} Longitudinal magnetic field variations (red points). A Gaussian process best fit with a period of 48~d with a decay time of 500\,days and a smoothing factor commonly fixed to 0.6\,day is also shown in blue. }
\end{figure*}

\clearpage
\section{Periodograms of RV, FWHM and BIS obtained with the APERO v7 version}
\begin{figure}[h]
    \centering
    \includegraphics[width=89.3mm,clip=true]{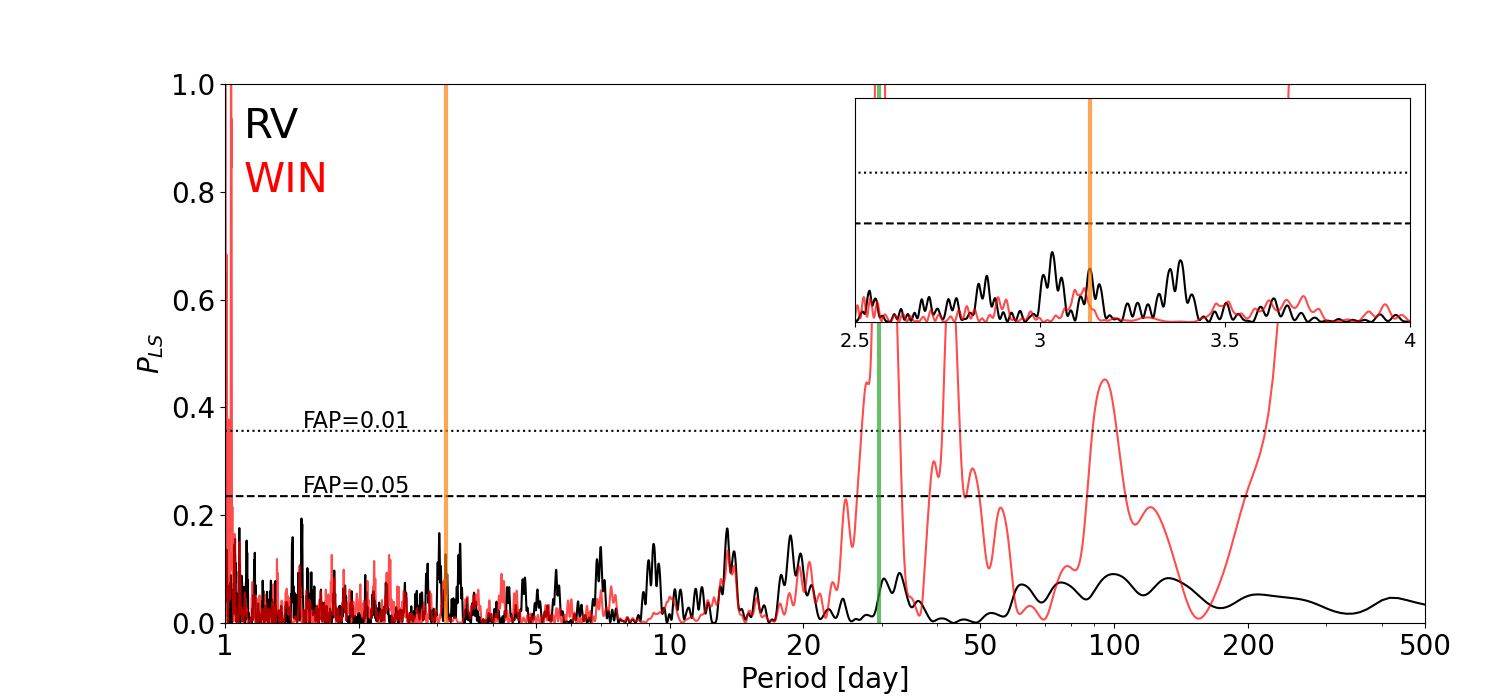}
    \includegraphics[width=89.3mm,clip=true]{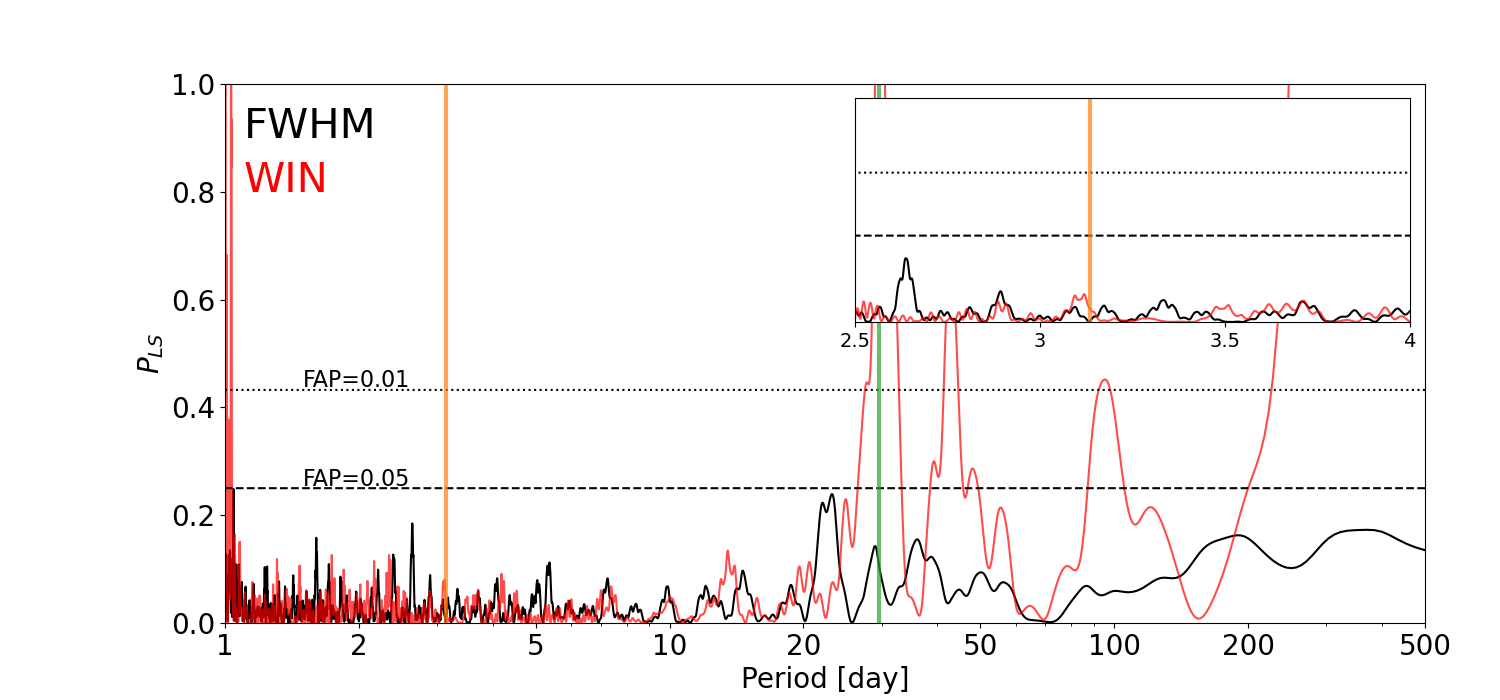}
    \includegraphics[width=89.3mm,clip=true]{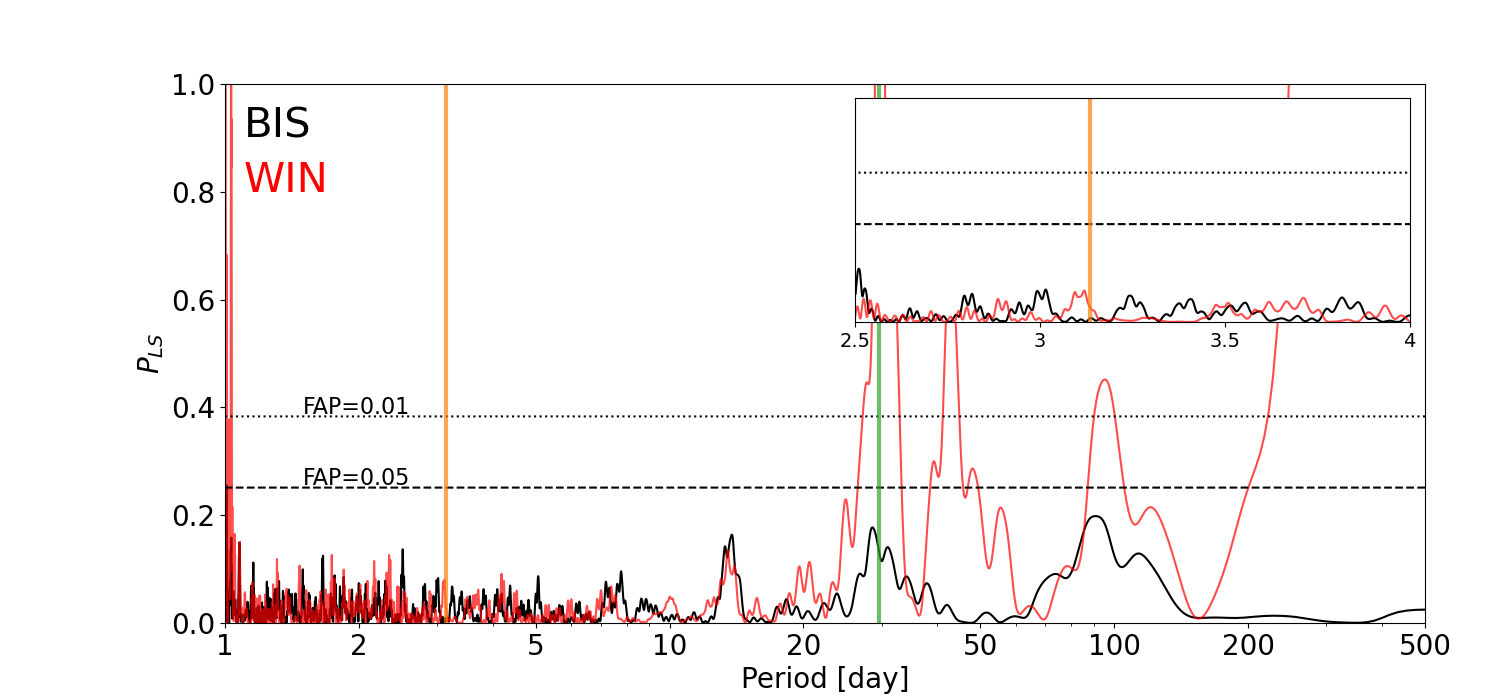}
    \caption{Same as Fig.~\ref{fig:periodo_all} but obtained with the APERO v7. \label{fig:periodov7}}
\end{figure}

\clearpage
\section{Corner plot of the MCMC posterior distributions for the transit and RV fit}
\begin{figure*}[h]
    \centering
    \includegraphics[width=170mm]{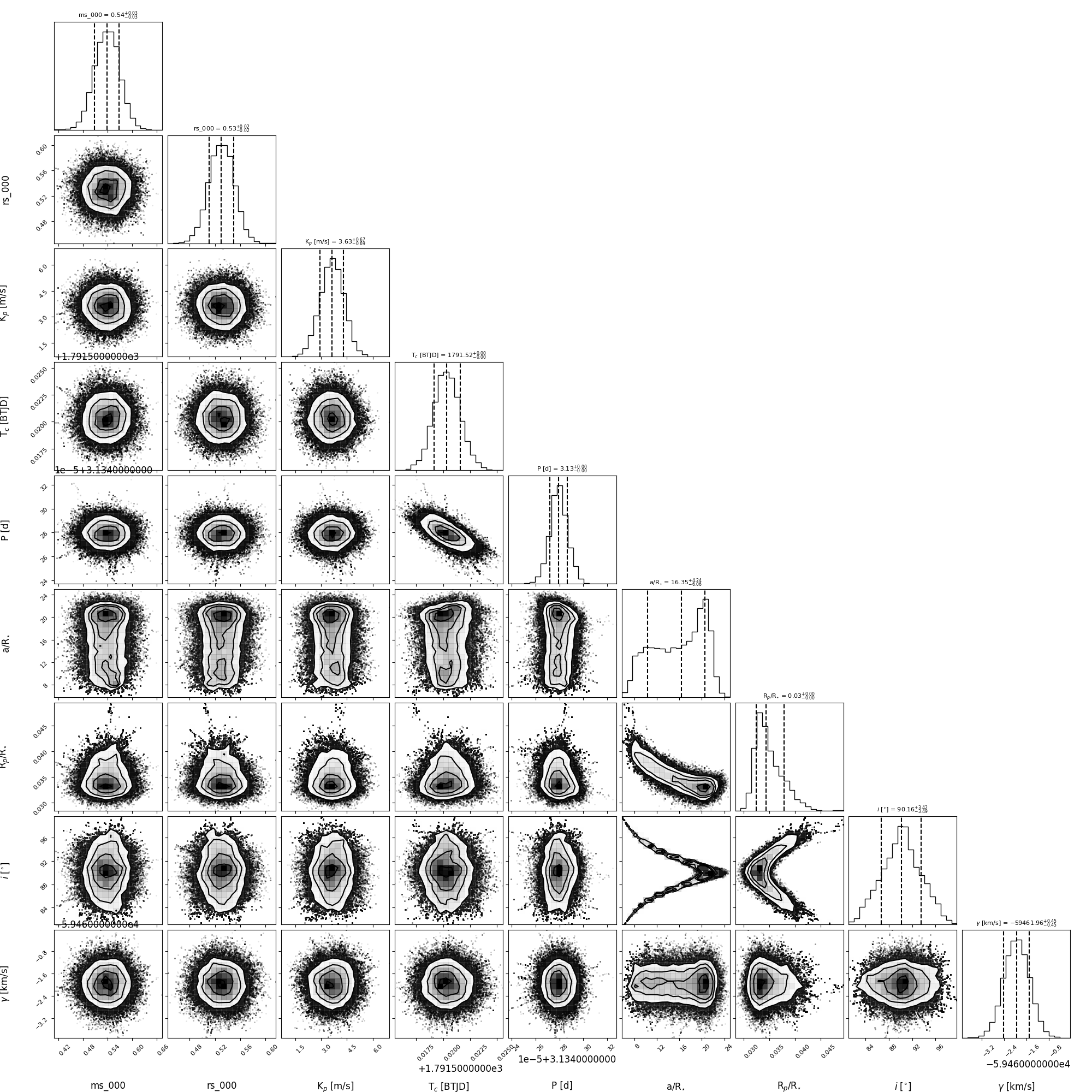}
    \caption{MCMC fit posteriors of the SPIRou and TESS data, in the scheme where RVs are drift corrected (DC), zero-point corrected (ZPC) and a GP with $P_\text{GP}$=14.8\,days is fitted out from the RVs and the lightcurve. The blue solid line spots the median of the posterior probability density function (pdf), and the black dotted lines locate the 16 and 84 percentiles of the posterior pdf.\label{fig:corner}}
\end{figure*}

\clearpage
\section{Observation logs. }
\begin{center}
\begin{longtable}{@{}c@{~~~}c@{~~~}c@{~~~}c@{~~~}c@{~~~}c@{~~~}c@{~~~}c@{~~~}|c@{~~~}c@{~~~}c@{~~~}|c@{}}
    \caption{\label{tab:log}Observations log of SPIRou observations of TOI-1695. The $\dagger$ symbol indicates data that were excluded from the drift corrected data reduced with the APERO v6 version. The $\star$ symbol indicates data removed from the drift corrected and zero point corrected data reduced with the APERO v7 version. This is discussed in Section~\ref{sec:obs}.}\\
Epoch	& UT Date	& BJD	& BERV	& Exp. time	& Airmass	& S/N	& H$_2$O	& \multicolumn{3}{|c|}{Moon}	& Pol. seq.\\
	& 	& -2,459,000 	& (km\,s$^{-1}$)	& (sec)	& 	& at 1670\,nm	& rel. abs.	& up? & angle ($^\circ$)	& phase	& number\\
\hline
    \endfirsthead
    \multicolumn{11}{c}{{\bfseries \tablename\ \thetable{}}: continued from previous page} \\
Epoch	& UT Date	& BJD	& BERV	& Exp. time	& Airmass	& S/N	& H$_2$O	& \multicolumn{3}{|c|}{Moon}	& Pol. seq.\\
	& 	& 	& (km\,s$^{-1}$)	& (sec)	& 	& at 1670\,nm	& rel. abs.	& on-sky &  ($^\circ$)	& phase	& number\\
\hline
    \endhead
1	& 2020-12-25T07:09:48	& 208.794938	& -9.5505	& 1204	& 1.69	& 142	& 4.17	& T	& 62	& 0.80	& 1\\
1	& 2020-12-25T07:30:20	& 208.809194	& -9.5649	& 1204	& 1.71	& 140	& 4.36	& T	& 62	& 0.80	& 2\\
1	& 2020-12-25T07:50:52	& 208.823449	& -9.5787	& 1204	& 1.75	& 145	& 4.54	& T	& 62	& 0.80	& 3\\
1	& 2020-12-25T08:11:23	& 208.837704	& -9.5920	& 1204	& 1.79	& 141	& 4.62	& T	& 62	& 0.80	& 4\\
2	& 2020-12-30T06:17:47	& 213.758646	& -10.7086	& 1204	& 1.66	& 135	& 2.10	& T	& 64	& -1.00	& 1\\
2	& 2020-12-30T06:38:19	& 213.772901	& -10.7233	& 1204	& 1.68	& 135	& 2.20	& T	& 64	& -1.00	& 2\\
2	& 2020-12-30T06:58:50	& 213.787157	& -10.7376	& 1204	& 1.70	& 137	& 2.25	& T	& 64	& -1.00	& 3\\
2	& 2020-12-30T07:19:22	& 213.801412	& -10.7516	& 1204	& 1.73	& 133	& 2.40	& T	& 64	& -1.00	& 4\\
3	& 2021-09-15T11:39:36	& 472.980910	& 15.1838	& 1204	& 1.67	& 142	& 2.10	& F	& 117	& 0.77	& 1\\
3	& 2021-09-15T12:00:07	& 472.995168	& 15.1704	& 1204	& 1.65	& 145	& 2.11	& F	& 117	& 0.77	& 2\\
3	& 2021-09-15T12:20:33	& 473.009361	& 15.1570	& 1204	& 1.65	& 145	& 2.12	& F	& 117	& 0.77	& 3\\
3	& 2021-09-15T12:40:59	& 473.023551	& 15.1436	& 1204	& 1.65	& 147	& 2.09	& F	& 117	& 0.77	& 4\\
4	& 2021-09-19T09:19:11	& 476.883601	& 14.7815	& 1204	& 1.90	& 148	& 3.08	& T	& 90	& 0.97	& 1\\
4	& 2021-09-19T09:39:37	& 476.897791	& 14.7706	& 1204	& 1.84	& 152	& 3.09	& T	& 91	& 0.97	& 2\\
4	& 2021-09-19T10:00:09	& 476.912048	& 14.7589	& 1204	& 1.79	& 150	& 3.04	& T	& 89	& 0.97	& 3\\
4	& 2021-09-19T10:20:35	& 476.926239	& 14.7468	& 1204	& 1.75	& 152	& 2.80	& T	& 89	& 0.97	& 4\\
5	& 2021-09-22T08:51:52	& 479.864766	& 14.3903	& 1204	& 1.95	& 155	& 1.22	& T	& 70	& -0.98	& 1\\
5	& 2021-09-22T09:12:23	& 479.879022	& 14.3797	& 1204	& 1.89	& 156	& 1.20	& T	& 70	& -0.98	& 2\\
5	& 2021-09-22T09:32:49	& 479.893215	& 14.3684	& 1204	& 1.83	& 153	& 1.12	& T	& 70	& -0.98	& 3\\
5	& 2021-09-22T09:53:15	& 479.907407	& 14.3565	& 1204	& 1.78	& 153	& 1.06	& T	& 70	& -0.98	& 4\\
6	& 2021-09-23T09:02:08	& 480.871949	& 14.2419	& 1204	& 1.91	& 150	& 4.95	& T	& 65	& -0.94	& 1\\
6	& 2021-09-23T09:22:40	& 480.886206	& 14.2308	& 1204	& 1.84	& 148	& 4.74	& T	& 65	& -0.94	& 2\\
6	& 2021-09-23T09:43:06	& 480.900398	& 14.2190	& 1204	& 1.79	& 146	& 3.13	& T	& 65	& -0.94	& 3\\
6	& 2021-09-23T10:03:32	& 480.914590	& 14.2067	& 1204	& 1.75	& 143	& 4.74	& T	& 64	& -0.94	& 4\\
7	& 2021-09-24T08:40:33	& 481.857005	& 14.1059	& 1204	& 1.97	& 154	& 1.61	& T	& 61	& -0.88	& 1\\
7	& 2021-09-24T09:00:59	& 481.871199	& 14.0953	& 1204	& 1.90	& 150	& 1.58	& T	& 60	& -0.88	& 2\\
7	& 2021-09-24T09:21:25	& 481.885388	& 14.0840	& 1204	& 1.84	& 153	& 1.54	& T	& 60	& -0.88	& 3\\
7	& 2021-09-24T09:41:51	& 481.899581	& 14.0721	& 1204	& 1.79	& 153	& 1.46	& T	& 60	& -0.88	& 4\\
8	& 2021-10-14T10:23:45	& 501.929496	& 10.2267	& 1204	& 1.65	& 152	& 3.84	& T	& 106	& 0.73	& 1\\
8	& 2021-10-14T10:44:17	& 501.943753	& 10.2116	& 1204	& 1.65	& 149	& 3.85	& T	& 106	& 0.73	& 2\\
8	& 2021-10-14T11:04:49	& 501.958010	& 10.1966	& 1204	& 1.65	& 148	& 4.00	& T	& 106	& 0.73	& 3\\
8	& 2021-10-14T11:25:15	& 501.972201	& 10.1818	& 1204	& 1.66	& 153	& 4.05	& F	& 106	& 0.73	& 4\\
9	& 2021-10-15T10:03:28	& 502.915441	& 10.0123	& 1204	& 1.65	& 156	& 2.77	& T	& 98	& 0.73	& 1\\
9	& 2021-10-15T10:24:00	& 502.929698	& 9.9972	& 1204	& 1.65	& 155	& 2.75	& T	& 98	& 0.82	& 2\\
9	& 2021-10-15T10:44:26	& 502.943889	& 9.9821	& 1204	& 1.65	& 156	& 2.59	& T	& 98	& 0.82	& 3\\
9	& 2021-10-15T11:04:52	& 502.958081	& 9.9672	& 1204	& 1.65	& 152	& 2.54	& T	& 98	& 0.82	& 4\\
10	& 2021-10-16T10:56:00	& 503.951960	& 9.7415	& 1204	& 1.65	& 130	& 0.85	& T	& 92	& 0.89	& 1\\
10	& 2021-10-16T11:16:32	& 503.966216	& 9.7265	& 1204	& 1.66	& 145	& 0.85	& T	& 92	& 0.89	& 2\\
10	& 2021-10-16T11:37:04	& 503.980473	& 9.7118	& 1204	& 1.68	& 141	& 0.85	& T	& 90	& 0.89	& 3\\
10	& 2021-10-16T11:57:30	& 503.994665	& 9.6976	& 1204	& 1.70	& 134	& 0.88	& T	& 90	& 0.89	& 4\\
11	& 2021-10-18T10:20:43	& 505.927524	& 9.2950	& 1204	& 1.65	& 137	& 3.05	& T	& 77	& 0.98	& 1\\
11	& 2021-10-18T10:41:15	& 505.941780	& 9.2798	& 1204	& 1.65	& 136	& 3.07	& T	& 77	& 0.98	& 2\\
11	& 2021-10-18T11:01:47	& 505.956036	& 9.2647	& 1204	& 1.66	& 142	& 3.17	& T	& 77	& 0.98	& 3\\
11	& 2021-10-18T11:22:13	& 505.970227	& 9.2499	& 1204	& 1.67	& 143	& 3.15	& T	& 77	& 0.98	& 4\\
12	& 2021-10-20T10:23:54	& 507.929789	& 8.8095	& 1204	& 1.65	& 145	& 5.63	& T	& 66	& -0.99	& 1\\
12	& 2021-10-20T10:44:20	& 507.943982	& 8.7943	& 1204	& 1.65	& 145	& 5.79	& T	& 66	& -0.99	& 2\\
12	& 2021-10-20T11:04:46	& 507.958173	& 8.7793	& 1204	& 1.66	& 142	& 5.82	& T	& 66	& -0.99	& 3\\
12	& 2021-10-20T11:25:12	& 507.972368	& 8.7646	& 1204	& 1.68	& 139	& 6.01	& T	& 66	& -0.99	& 4\\
13	& 2021-10-21T10:33:20	& 508.936373	& 8.5569	& 1204	& 1.65	& 148	& 3.10	& T	& 61	& -0.97	& 1\\
13	& 2021-10-21T10:53:46	& 508.950565	& 8.5418	& 1204	& 1.66	& 147	& 2.98	& T	& 61	& -0.97	& 2\\
13	& 2021-10-21T11:14:18	& 508.964820	& 8.5268	& 1204	& 1.68	& 144	& 2.88	& T	& 61	& -0.97	& 3\\
13	& 2021-10-21T11:34:44	& 508.979012	& 8.5123	& 1204	& 1.70	& 147	& 3.01	& T	& 61	& -0.97	& 4\\
14	& 2021-10-22T10:44:04	& 509.943857	& 8.3007	& 1204	& 1.66	& 111	& 2.15	& T	& 58	& -0.93	& 1\\
14	& 2021-10-22T11:04:36	& 509.958114	& 8.2856	& 1204	& 1.67	& 111	& 2.17	& T	& 58	& -0.93	& 2\\
14	& 2021-10-22T11:25:02	& 509.972306	& 8.2710	& 1204	& 1.69	& 110	& 2.26	& T	& 58	& -0.93	& 3\\
14	& 2021-10-22T11:45:28	& 509.986497	& 8.2567	& 1204	& 1.72	& 110	& 2.28	& T	& 58	& -0.93	& 4\\
15	& 2021-10-23T10:12:50	& 510.922184	& 8.0731	& 1204	& 1.65	& 149	& 3.08	& T	& 56	& -0.93	& 1\\
15	& 2021-10-23T10:33:16	& 510.936376	& 8.0579	& 1204	& 1.65	& 145	& 3.12	& T	& 56	& -0.87	& 2\\
15	& 2021-10-23T10:53:47	& 510.950632	& 8.0427	& 1204	& 1.66	& 143	& 3.27	& T	& 56	& -0.87	& 3\\
15	& 2021-10-23T11:14:14	& 510.964823	& 8.0279	& 1204	& 1.68	& 146	& 3.27	& T	& 56	& -0.87	& 4\\
16	& 2021-10-24T11:20:08	& 511.968953	& 7.7703	& 1204	& 1.69	& 120	& 1.76	& T	& 56	& -0.80	& 1\\
16	& 2021-10-24T11:40:34	& 511.983145	& 7.7560	& 1204	& 1.72	& 127	& 1.84	& T	& 56	& -0.80	& 2\\
16	& 2021-10-24T12:01:00	& 511.997337	& 7.7422	& 1204	& 1.76	& 118	& 1.84	& T	& 56	& -0.80	& 3\\
17	& 2021-10-26T09:50:42	& 513.906896	& 7.3218	& 1204	& 1.65	& 147	& 2.36	& T	& 61	& -0.72	& 1\\
17	& 2021-10-26T10:11:14	& 513.921151	& 7.3063	& 1204	& 1.65	& 147	& 2.33	& T	& 61	& -0.72	& 2\\
17	& 2021-10-26T10:31:40	& 513.935343	& 7.2910	& 1204	& 1.66	& 145	& 2.45	& T	& 61	& -0.63	& 3\\
17	& 2021-10-26T10:52:06	& 513.949535	& 7.2759	& 1204	& 1.67	& 139	& 2.57	& T	& 61	& -0.63	& 4\\
18	& 2021-11-13T06:36:44	& 531.772487	& 2.4459	& 1204	& 1.76	& 131	& 1.99	& T	& 88	& 0.69	& 1\\
18	& 2021-11-13T06:57:10	& 531.786678	& 2.4312	& 1204	& 1.73	& 138	& 2.15	& T	& 88	& 0.69	& 2\\
18	& 2021-11-13T07:17:42	& 531.800934	& 2.4160	& 1204	& 1.70	& 143	& 2.20	& T	& 88	& 0.69	& 3\\
18	& 2021-11-13T07:38:08	& 531.815126	& 2.4005	& 1204	& 1.68	& 144	& 2.26	& T	& 88	& 0.69	& 4\\
19	& 2021-11-17T09:58:59	& 535.912967	& 1.1234	& 1204	& 1.71	& 115	& 5.06	& T	& 62	& 0.96	& 1\\
19	& 2021-11-17T10:19:31	& 535.927223	& 1.1089	& 1204	& 1.75	& 79	& 5.31	& T	& 62	& 0.96	& 2\\
19	& 2021-11-17T10:40:03	& 535.941479	& 1.0948	& 1204	& 1.79	& 83	& 5.98	& T	& 62	& 0.99	& 3\\
19	& 2021-11-17T11:00:29	& 535.955670	& 1.0815	& 1204	& 1.84	& 91	& 6.18	& T	& 62	& 0.99	& 4\\
20	& 2021-11-20T08:16:08	& 538.841557	& 0.3204	& 1204	& 1.65	& 147	& 1.30	& T	& 56	& -0.99	& 1\\
20	& 2021-11-20T08:36:40	& 538.855808	& 0.3044	& 1204	& 1.65	& 147	& 1.29	& T	& 56	& -0.99	& 2\\
20	& 2021-11-20T08:57:12	& 538.870065	& 0.2886	& 1204	& 1.66	& 149	& 1.22	& T	& 56	& -0.99	& 3\\
20	& 2021-11-20T09:17:38	& 538.884257	& 0.2731	& 1204	& 1.68	& 148	& 1.19	& T	& 56	& -0.99	& 4\\
21	& 2021-11-21T08:28:22	& 539.850042	& 0.0164	& 1204	& 1.65	& 157	& 1.07	& T	& 57	& -0.96	& 1\\
21	& 2021-11-21T08:48:54	& 539.864298	& 0.0006	& 1204	& 1.66	& 157	& 1.08	& T	& 57	& -0.96	& 2\\
21	& 2021-11-21T09:09:25	& 539.878554	& -0.0151	& 1204	& 1.67	& 155	& 1.10	& T	& 57	& -0.96	& 3\\
21	& 2021-11-21T09:29:52	& 539.892747	& -0.0303	& 1204	& 1.69	& 152	& 1.15	& T	& 57	& -0.96	& 4\\
22	& 2021-12-09T07:21:22	& 557.803364	& -5.2062	& 1204	& 1.65	& 119	& 2.49	& T	& 97	& 0.32	& 1\\
22	& 2021-12-09T07:41:54	& 557.817620	& -5.2219	& 1204	& 1.66	& 116	& 2.48	& T	& 97	& 0.32	& 2\\
22	& 2021-12-09T08:02:26	& 557.831874	& -5.2373	& 1204	& 1.68	& 117	& 2.51	& T	& 97	& 0.32	& 3\\
22	& 2021-12-09T08:22:52	& 557.846067	& -5.2524	& 1204	& 1.70	& 119	& 2.48	& T	& 97	& 0.32	& 4\\
23	& 2021-12-11T06:42:23	& 559.776247	& -5.7413	& 1204	& 1.65	& 118	& 2.95	& T	& 82	& 0.53	& 1\\
23	& 2021-12-11T07:02:54	& 559.790503	& -5.7572	& 1204	& 1.65	& 120	& 2.92	& T	& 82	& 0.53	& 2\\
23	& 2021-12-11T07:23:26	& 559.804759	& -5.7729	& 1204	& 1.65	& 128	& 2.90	& T	& 82	& 0.53	& 3\\
23	& 2021-12-11T07:43:52	& 559.818950	& -5.7884	& 1204	& 1.67	& 123	& 2.98	& T	& 80	& 0.53	& 4\\
24	& 2021-12-12T06:41:13	& 560.775428	& -6.0201	& 1204	& 1.65	& 139	& 2.53	& T	& 74	& 0.63	& 1\\
24	& 2021-12-12T07:01:45	& 560.789685	& -6.0359	& 1204	& 1.65	& 139	& 2.57	& T	& 75	& 0.63	& 2\\
24	& 2021-12-12T07:22:17	& 560.803939	& -6.0516	& 1204	& 1.66	& 140	& 2.69	& T	& 75	& 0.63	& 3\\
24	& 2021-12-12T07:42:43	& 560.818130	& -6.0670	& 1204	& 1.67	& 140	& 2.77	& T	& 75	& 0.63	& 4\\
25	& 2021-12-13T07:08:52	& 561.794605	& -6.3189	& 1204	& 1.65	& 113	& 3.42	& T	& 69	& 0.72	& 1\\
25	& 2021-12-13T07:29:24	& 561.808859	& -6.3345	& 1204	& 1.66	& 117	& 3.72	& T	& 69	& 0.72	& 2\\
25	& 2021-12-13T07:49:55	& 561.823115	& -6.3497	& 1204	& 1.68	& 125	& 4.17	& T	& 69	& 0.72	& 3\\
25	& 2021-12-13T08:10:22	& 561.837306	& -6.3646	& 1204	& 1.70	& 120	& 4.21	& T	& 69	& 0.72	& 4\\
26	& 2021-12-14T06:15:31	& 562.757530	& -6.5535	& 1204	& 1.65	& 152	& 3.18	& T	& 64	& 0.81	& 1\\
26	& 2021-12-14T06:36:02	& 562.771786	& -6.5693	& 1204	& 1.65	& 154	& 2.98	& T	& 64	& 0.81	& 2\\
26	& 2021-12-14T06:56:29	& 562.785977	& -6.5849	& 1204	& 1.65	& 154	& 2.88	& T	& 64	& 0.81	& 3\\
26	& 2021-12-14T07:16:55	& 562.800168	& -6.6005	& 1204	& 1.66	& 155	& 2.85	& T	& 64	& 0.81	& 4\\
27	& 2021-12-15T06:25:47	& 563.764642	& -6.8348	& 1204	& 1.65	& 123	& 0.67	& T	& 60	& 0.88	& 1\\
27	& 2021-12-15T06:46:19	& 563.778898	& -6.8506	& 1204	& 1.65	& 119	& 0.70	& T	& 60	& 0.88	& 2\\
27	& 2021-12-15T07:06:50	& 563.793154	& -6.8662	& 1204	& 1.65	& 114	& 0.71	& T	& 60	& 0.88	& 3\\
27	& 2021-12-15T07:27:17	& 563.807345	& -6.8815	& 1204	& 1.67	& 127	& 0.72	& T	& 60	& 0.88	& 4\\
28	& 2021-12-16T07:14:59	& 564.798782	& -7.1435	& 1204	& 1.66	& 140	& 1.26	& T	& 57	& 0.93	& 1\\
28	& 2021-12-16T07:35:30	& 564.813036	& -7.1587	& 1204	& 1.68	& 144	& 1.28	& T	& 57	& 0.93	& 2\\
28	& 2021-12-16T07:56:02	& 564.827292	& -7.1736	& 1204	& 1.70	& 135	& 1.34	& T	& 57	& 0.93	& 3\\
28	& 2021-12-16T08:16:28	& 564.841483	& -7.1880	& 1204	& 1.73	& 138	& 1.42	& T	& 57	& 0.93	& 4\\
29	& 2021-12-18T06:47:02	& 566.779331	& -7.6582	& 1204	& 1.65	& 150	& 3.30	& T	& 56	& 0.99	& 1\\
29	& 2021-12-18T07:07:34	& 566.793582	& -7.6736	& 1204	& 1.66	& 160	& 3.32	& T	& 56	& 0.99	& 2\\
29	& 2021-12-18T07:28:06	& 566.807838	& -7.6887	& 1204	& 1.68	& 157	& 3.44	& T	& 56	& 0.99	& 3\\
29	& 2021-12-18T07:48:32	& 566.822028	& -7.7035	& 1204	& 1.70	& 161	& 3.50	& T	& 56	& 0.99	& 4\\
30	& 2022-01-06T06:29:30	& 585.766520	& -12.2163	& 1204	& 1.70	& 137	& 2.39	& T	& 92	& 0.17	& 1\\
30	& 2022-01-06T06:49:57	& 585.780715	& -12.2299	& 1204	& 1.72	& 138	& 2.69	& T	& 92	& 0.17	& 2\\
30	& 2022-01-06T07:10:28	& 585.794967	& -12.2431	& 1204	& 1.76	& 107	& 2.66	& T	& 92	& 0.17	& 3\\
30	& 2022-01-06T07:30:54	& 585.809157	& -12.2556	& 1204	& 1.80	& 110	& 2.67	& T	& 92	& 0.17	& 4\\
31	& 2022-01-08T06:41:25	& 587.774707	& -12.6328	& 1204	& 1.72	& 148	& 0.96	& T	& 77	& 0.36	& 1\\
31	& 2022-01-08T07:01:57	& 587.788963	& -12.6458	& 1204	& 1.76	& 151	& 1.01	& T	& 77	& 0.36	& 2\\
31	& 2022-01-08T07:22:28	& 587.803217	& -12.6583	& 1204	& 1.80	& 151	& 1.06	& T	& 77	& 0.36	& 3\\
31	& 2022-01-08T07:43:00	& 587.817472	& -12.6702	& 1204	& 1.85	& 148	& 1.13	& T	& 78	& 0.36	& 4\\
32	& 2022-01-10T06:10:19	& 589.753025	& -13.0040	& 1204	& 1.69	& 152	& 2.61	& T	& 66	& 0.56	& 1\\
32	& 2022-01-10T06:30:51	& 589.767280	& -13.0175	& 1204	& 1.72	& 154	& 2.68	& T	& 66	& 0.56	& 2\\
32	& 2022-01-10T06:51:22	& 589.781535	& -13.0305	& 1204	& 1.75	& 153	& 2.69	& T	& 65	& 0.56	& 3\\
32	& 2022-01-10T07:11:48	& 589.795727	& -13.0429	& 1204	& 1.80	& 154	& 2.67	& T	& 65	& 0.56	& 4\\
33$\dagger$	& 2022-01-12T06:36:43	& 591.771269	& -13.3958	& 1204	& 1.74	& 133	& 0.99	& T	& 58	& 0.74	& 1\\
33$\dagger$	& 2022-01-12T06:57:14	& 591.785525	& -13.4083	& 1204	& 1.78	& 129	& 1.05	& T	& 58	& 0.74	& 2\\
33$\dagger$	& 2022-01-12T07:17:46	& 591.799780	& -13.4202	& 1204	& 1.83	& 133	& 1.12	& T	& 58	& 0.74	& 3\\
33$\dagger$	& 2022-01-12T07:38:12	& 591.813970	& -13.4314	& 1204	& 1.89	& 139	& 1.15	& T	& 58	& 0.74	& 4\\
34	& 2022-01-13T05:02:35	& 592.705859	& -13.5142	& 1204	& 1.65	& 159	& 1.20	& T	& 56	& 0.82	& 1\\
34	& 2022-01-13T05:23:06	& 592.720111	& -13.5284	& 1204	& 1.66	& 154	& 1.26	& T	& 56	& 0.82	& 2\\
34	& 2022-01-13T05:43:38	& 592.734367	& -13.5423	& 1204	& 1.68	& 157	& 1.33	& T	& 56	& 0.82	& 3\\
34	& 2022-01-13T06:04:04	& 592.748557	& -13.5559	& 1204	& 1.70	& 158	& 1.34	& T	& 56	& 0.82	& 4\\
35	& 2022-01-15T06:31:35	& 594.767572	& -13.9225	& 1204	& 1.75	& 146	& 2.28	& T	& 58	& 0.94	& 1\\
35	& 2022-01-15T06:52:07	& 594.781827	& -13.9346	& 1204	& 1.80	& 136	& 2.24	& T	& 58	& 0.94	& 2\\
35	& 2022-01-15T07:12:33	& 594.796017	& -13.9461	& 1204	& 1.85	& 152	& 2.27	& T	& 58	& 0.94	& 3\\
35	& 2022-01-15T07:33:05	& 594.810273	& -13.9569	& 1204	& 1.91	& 159	& 2.30	& T	& 58	& 0.94	& 4\\
36	& 2022-01-18T05:51:37	& 597.739680	& -14.3897	& 1204	& 1.71	& 151	& 1.27	& T	& 71	& -1.00	& 1\\
36	& 2022-01-18T06:12:09	& 597.753935	& -14.4024	& 1204	& 1.74	& 151	& 1.30	& T	& 71	& -1.00	& 2\\
36	& 2022-01-18T06:32:35	& 597.768126	& -14.4146	& 1204	& 1.78	& 152	& 1.35	& T	& 71	& -1.00	& 3\\
36	& 2022-01-18T06:53:01	& 597.782317	& -14.4261	& 1204	& 1.83	& 152	& 1.40	& T	& 71	& -1.00	& 4\\
37	& 2022-01-19T06:18:59	& 598.758635	& -14.5620	& 1204	& 1.76	& 145	& 1.30	& T	& 77	& -0.98	& 1\\
37	& 2022-01-19T06:39:31	& 598.772889	& -14.5738	& 1204	& 1.80	& 149	& 1.37	& T	& 77	& -0.98	& 2\\
37	& 2022-01-19T07:00:03	& 598.787145	& -14.5849	& 1204	& 1.86	& 153	& 1.42	& T	& 77	& -0.98	& 3\\
37	& 2022-01-19T07:20:29	& 598.801335	& -14.5953	& 1204	& 1.92	& 153	& 1.45	& T	& 77	& -0.98	& 4\\
38	& 2022-01-20T05:59:54	& 599.745337	& -14.7018	& 1204	& 1.73	& 152	& 2.81	& F	& 84	& -0.94	& 1\\
38	& 2022-01-20T06:20:26	& 599.759591	& -14.7140	& 1204	& 1.77	& 149	& 3.02	& F	& 84	& -0.94	& 2\\
38	& 2022-01-20T06:40:58	& 599.773850	& -14.7256	& 1204	& 1.82	& 152	& 3.20	& T	& 84	& -0.94	& 3\\
38	& 2022-01-20T07:01:24	& 599.788038	& -14.7364	& 1204	& 1.87	& 151	& 3.39	& T	& 84	& -0.94	& 4\\
39	& 2022-01-21T05:56:31	& 600.742939	& -14.8467	& 1204	& 1.73	& 144	& 3.24	& F	& 91	& -0.88	& 1\\
39	& 2022-01-21T06:16:58	& 600.757130	& -14.8588	& 1204	& 1.77	& 142	& 3.33	& F	& 91	& -0.88	& 2\\
39	& 2022-01-21T06:37:24	& 600.771321	& -14.8703	& 1204	& 1.82	& 147	& 3.49	& F	& 91	& -0.88	& 3\\
39	& 2022-01-21T06:57:50	& 600.785511	& -14.8811	& 1204	& 1.87	& 146	& 3.59	& F	& 91	& -0.88	& 4\\
40	& 2022-01-22T05:55:23	& 601.742089	& -14.9888	& 1204	& 1.74	& 119	& 1.64	& F	& 98	& -0.81	& 1\\
40	& 2022-01-22T06:15:54	& 601.756344	& -15.0008	& 1204	& 1.78	& 123	& 1.77	& F	& 98	& -0.81	& 2\\
40	& 2022-01-22T06:36:26	& 601.770599	& -15.0121	& 1204	& 1.82	& 118	& 1.85	& F	& 98	& -0.81	& 3\\
40	& 2022-01-22T06:56:52	& 601.784791	& -15.0228	& 1204	& 1.88	& 104	& 1.92	& F	& 98	& -0.81	& 4\\
41	& 2022-01-23T06:03:58	& 602.748012	& -15.1322	& 1204	& 1.76	& 151	& 2.34	& F	& 105	& -0.72	& 1\\
41	& 2022-01-23T06:24:30	& 602.762267	& -15.1438	& 1204	& 1.80	& 146	& 2.28	& F	& 105	& -0.72	& 2\\
41	& 2022-01-23T06:45:02	& 602.776521	& -15.1547	& 1204	& 1.86	& 144	& 2.25	& F	& 105	& -0.72	& 3\\
41	& 2022-01-23T07:05:28	& 602.790713	& -15.1648	& 1204	& 1.92	& 147	& 2.33	& F	& 105	& -0.72	& 4\\
42	& 2022-01-24T06:19:05	& 603.758450	& -15.2746	& 1204	& 1.80	& 157	& 2.52	& F	& 112	& -0.62	& 1\\
42	& 2022-01-24T06:39:37	& 603.772704	& -15.2855	& 1204	& 1.85	& 156	& 2.70	& F	& 111	& -0.62	& 2\\
42	& 2022-01-24T07:00:08	& 603.786959	& -15.2956	& 1204	& 1.92	& 156	& 2.90	& F	& 111	& -0.62	& 3\\
42	& 2022-01-24T07:20:34	& 603.801150	& -15.3049	& 1204	& 1.99	& 155	& 3.20	& F	& 111	& -0.62	& 4\\
43	& 2022-01-25T06:29:43	& 604.765791	& -15.4096	& 1204	& 1.84	& 124	& 2.11	& F	& 116	& -0.51	& 1\\
43	& 2022-01-25T06:50:15	& 604.780045	& -15.4199	& 1204	& 1.90	& 124	& 2.18	& F	& 115	& -0.51	& 2\\
43	& 2022-01-25T07:10:47	& 604.794300	& -15.4294	& 1204	& 1.97	& 129	& 2.23	& F	& 117	& -0.51	& 3\\
43	& 2022-01-25T07:31:13	& 604.808491	& -15.4380	& 1204	& 2.05	& 124	& 2.32	& F	& 117	& -0.51	& 4\\
44$\dagger\star$	& 2022-01-26T06:37:52	& 605.774581	& -15.5406	& 652	& 1.90	& 75	& 4.28	& F	& 121	& -0.40	& 1\\
45	& 2022-01-27T05:39:40	& 606.730932	& -15.6273	& 1204	& 1.74	& 128	& 2.54	& F	& 122	& -0.29	& 1\\
45	& 2022-01-27T06:00:12	& 606.745188	& -15.6388	& 1204	& 1.78	& 124	& 2.42	& F	& 122	& -0.29	& 2\\
45	& 2022-01-27T06:20:44	& 606.759443	& -15.6497	& 1204	& 1.83	& 123	& 2.38	& F	& 122	& -0.29	& 3\\
45	& 2022-01-27T06:41:10	& 606.773634	& -15.6599	& 1204	& 1.89	& 122	& 2.42	& F	& 122	& -0.29	& 4\\
46	& 2022-01-28T05:44:06	& 607.733950	& -15.7455	& 1204	& 1.76	& 143	& 0.51	& F	& 123	& -0.19	& 1\\
46	& 2022-01-28T06:04:37	& 607.748206	& -15.7567	& 1204	& 1.80	& 116	& 0.50	& F	& 123	& -0.19	& 2\\
46	& 2022-01-28T06:25:09	& 607.762461	& -15.7673	& 1204	& 1.86	& 118	& 0.52	& F	& 123	& -0.19	& 3\\
46	& 2022-01-28T06:45:35	& 607.776651	& -15.7771	& 1204	& 1.92	& 107	& 0.55	& F	& 123	& -0.19	& 4\\
\hline
\end{longtable}
\end{center}

\section{RV and polarimetry data tables}
\begin{center}
\begin{longtable}{ccccccc}
    \caption{\label{tab:rvs_v6} SPIRou LBL RV observations of TOI-1695 obtained with the APERO v6 version. The FWHM and BIS proxies are derived from respectively the second and third derivatives of LBL profiles (see text for explanation).} \\
       BJD & $v_r$ & $\sigma_{v_r}$ & FWHM & eFWHM & BIS & eBIS \\
     -2,459,000       & (km\,s$^{-1}$) & (km\,s$^{-1}$) & (km\,s$^{-1}$) & (km\,s$^{-1}$) & (km\,s$^{-1}$) & (km\,s$^{-1}$) \\
       \hline
       \endfirsthead
       \multicolumn{5}{c}{{\bfseries \tablename\ \thetable{}}: continued from previous page} \\
       BJD & $v_r$ & $\sigma_{v_r}$ & FWHM & eFWHM & BIS & eBIS \\
     -2,459,000       & (km\,s$^{-1}$) & (km\,s$^{-1}$) & (km\,s$^{-1}$) & (km\,s$^{-1}$) & (km\,s$^{-1}$) & (km\,s$^{-1}$) \\
       \hline
       \endhead
208.79493816	& -59.9476	& 0.0066	& 5.527	&  0.014	&  -0.053	&  0.027 \\
208.80919391	& -59.9474	& 0.0064	& 5.526	&  0.014	&  -0.036	&  0.027 \\
208.82344915	& -59.9404	& 0.0063	& 5.524	&  0.014	&  -0.029	&  0.026 \\
208.83770450	& -59.9471	& 0.0064	& 5.529	&  0.014	&  -0.041	&  0.027 \\
213.75864561	& -59.9192	& 0.0060	& 5.533	&  0.013	&  -0.018	&  0.025 \\
213.77290070	& -59.9255	& 0.0061	& 5.512	&  0.013	&  -0.030	&  0.025 \\
213.78715749	& -59.9290	& 0.0060	& 5.532	&  0.013	&  -0.024	&  0.025 \\
213.80141198	& -59.9199	& 0.0063	& 5.533	&  0.013	&  -0.022	&  0.026 \\
472.98091030	& -59.9349	& 0.0059	& 5.485	&  0.013	&  -0.014	&  0.024 \\
472.99516782	& -59.9289	& 0.0057	& 5.486	&  0.012	&  -0.005	&  0.024 \\
473.00936133	& -59.9328	& 0.0057	& 5.478	&  0.012 	&   0.007	&  0.024 \\
473.02355075	& -59.9301	& 0.0057	& 5.475	&  0.012	&  -0.007	&  0.024 \\
476.88360128	& -59.9372	& 0.0063	& 5.474	&  0.014	&   0.018	&  0.026 \\
476.89779088	& -59.9340	& 0.0061	& 5.462	&  0.013	&  -0.011	&  0.025 \\
476.91204768	& -59.9288	& 0.0062	& 5.456	&  0.013	&  -0.029	&  0.025 \\
476.92623878	& -59.9308	& 0.0062	& 5.456	&  0.013	&  -0.003	&  0.025 \\
479.86476628	& -59.9310	& 0.0058	& 5.457	&  0.013	&  -0.012	&  0.024 \\
479.87902237	& -59.9284	& 0.0056	& 5.453	&  0.012	&  -0.011	&  0.024 \\
479.89321505	& -59.9282	& 0.0057	& 5.457	&  0.012	&  -0.015	&  0.023 \\
479.90740723	& -59.9338	& 0.0057	& 5.465	&  0.012	&  -0.020	&  0.023 \\
480.87194906	& -59.9313	& 0.0061	& 5.475	&  0.013	&  -0.031	&  0.025 \\
480.88620644	& -59.9303	& 0.0061	& 5.465	&  0.013	&  -0.049	&  0.026 \\
480.90039831	& -59.9250	& 0.0060	& 5.470	&  0.013	&  -0.035	&  0.025 \\
480.91458968	& -59.9281	& 0.0060	& 5.477	&  0.013	&  -0.012	&  0.025 \\
481.85700461	& -59.9358	& 0.0059	& 5.457	&  0.013	&  -0.018	&  0.025 \\
481.87119917	& -59.9356	& 0.0058	& 5.443	&  0.013	&  -0.032	&  0.024 \\
481.88538804	& -59.9370	& 0.0058	& 5.425	&  0.012	&  -0.011	&  0.024 \\
481.89958051	& -59.9333	& 0.0058	& 5.433	&  0.012	&  -0.024	&  0.024 \\
501.92949642	& -59.9339	& 0.0060	& 5.478	&  0.013	&  -0.026	&  0.025 \\
501.94375341	& -59.9382	& 0.0059	& 5.485	&  0.013	&  -0.022	&  0.025 \\
501.95800989	& -59.9363	& 0.0058	& 5.487	&  0.013	&  -0.008	&  0.024 \\
501.97220128	& -59.9305	& 0.0058	& 5.480	&  0.013	&   0.003	&  0.024 \\
502.91544136	& -59.9373	& 0.0059	& 5.476	&  0.013	&  -0.039	&  0.025 \\
502.92969804	& -59.9416	& 0.0059	& 5.474	&  0.013	&  -0.009	&  0.024 \\
502.94388941	& -59.9353	& 0.0059	& 5.477	&  0.013	&  -0.042	&  0.025 \\
502.95808148	& -59.9427	& 0.0060	& 5.490	&  0.013	&  -0.006	&  0.025 \\
503.95195991	& -59.9548	& 0.0061	& 5.492	&  0.013	&  -0.020	&  0.025 \\
503.96621637	& -59.9445	& 0.0056	& 5.502	&  0.012	&  -0.029	&  0.023 \\
503.98047314	& -59.9413	& 0.0057	& 5.487	&  0.012	&  -0.013	&  0.023 \\
503.99466510	& -59.9429	& 0.0058	& 5.498	&  0.013	&  -0.015	&  0.024 \\
505.92752409	& -59.9379	& 0.0062	& 5.508	&  0.013	&  -0.029	&  0.026 \\
505.94177953	& -59.9400	& 0.0061	& 5.498	&  0.013	&  -0.025	&  0.025 \\
505.95603647	& -59.9418	& 0.0061	& 5.473	&  0.013	&  -0.027	&  0.025 \\
505.97022731	& -59.9358	& 0.0060	& 5.514	&  0.013	&  -0.007	&  0.025 \\
507.92978917	& -59.9286	& 0.0064	& 5.469	&  0.014	&  -0.026	&  0.027 \\
507.94398159	& -59.9208	& 0.0063	& 5.472	&  0.014	&  -0.019	&  0.026 \\
507.95817331	& -59.9222	& 0.0063	& 5.471	&  0.014	&  -0.022	&  0.026 \\
507.97236802	& -59.9235	& 0.0064	& 5.469	&  0.014	&  -0.017	&  0.026 \\
508.93637316	& -59.9311	& 0.0060	& 5.495	&  0.013	&  -0.017	&  0.025 \\
508.95056467	& -59.9371	& 0.0059	& 5.496	&  0.013	&  -0.016	&  0.025 \\
508.96482027	& -59.9373	& 0.0060	& 5.499	&  0.013	&  -0.022	&  0.025 \\
508.97901207	& -59.9280	& 0.0060	& 5.519	&  0.013	&  -0.013	&  0.025 \\
509.94385745	& -59.9250	& 0.0073	& 5.490	&  0.016	&   0.000	&  0.030 \\
509.95811414	& -59.9231	& 0.0072	& 5.489	&  0.015	&  -0.036	&  0.030 \\
509.97230573	& -59.9348	& 0.0072	& 5.496	&  0.016	&  -0.034	&  0.030 \\
509.98649743	& -59.9347	& 0.0072	& 5.526	&  0.016	&  -0.013	&  0.030 \\
510.92218401	& -59.9222	& 0.0062	& 5.490	&  0.014	&  -0.023	&  0.026 \\
510.93637590	& -59.9131	& 0.0062	& 5.518	&  0.014	&  -0.017	&  0.026 \\
510.95063248	& -59.9140	& 0.0063	& 5.500	&  0.014	&  -0.029	&  0.027 \\
510.96482346	& -59.9139	& 0.0061	& 5.487	&  0.014	&  -0.020	&  0.026 \\
511.96895286	& -59.9232	& 0.0067	& 5.531	&  0.014	&  -0.021	&  0.027 \\
511.98314483	& -59.9198	& 0.0064	& 5.518	&  0.013	&  -0.027	&  0.026 \\
511.99733669	& -59.9306	& 0.0067	& 5.518	&  0.014	&  -0.016	&  0.027 \\
513.90689610	& -59.9232	& 0.0057	& 5.491	&  0.012	&  -0.026	&  0.024 \\
513.92115105	& -59.9252	& 0.0056	& 5.496	&  0.012	&  -0.015	&  0.023 \\
513.93534260	& -59.9270	& 0.0057	& 5.502	&  0.012	&  -0.039	&  0.024 \\
513.94953504	& -59.9273	& 0.0059	& 5.495	&  0.013	&  -0.040	&  0.025 \\
531.77248693	& -59.9328	& 0.0069	& 5.526	&  0.015	&  -0.034	&  0.029 \\
531.78667804	& -59.9338	& 0.0063	& 5.496	&  0.014	&  -0.031	&  0.026 \\
531.80093355	& -59.9376	& 0.0061	& 5.510	&  0.013	&  -0.062	&  0.025 \\
531.81512557	& -59.9268	& 0.0061	& 5.518	&  0.013	&  -0.048	&  0.025 \\
535.91296729	& -59.9410	& 0.0077	& 5.490	&  0.016	&  -0.036	&  0.031 \\
535.92722294	& -59.9531	& 0.0112	& 5.538	&  0.024	&  -0.070	&  0.045 \\
535.94147939	& -59.9292	& 0.0106	& 5.537	&  0.023	&  -0.101	&  0.044 \\
535.95567034	& -59.9257	& 0.0097	& 5.554	&  0.021	&  -0.073	&  0.040 \\
538.84155732	& -59.9325	& 0.0056	& 5.485	&  0.012	&  -0.038	&  0.024 \\
538.85580844	& -59.9317	& 0.0057	& 5.501	&  0.012	&  -0.050	&  0.024 \\
538.87006465	& -59.9355	& 0.0057	& 5.501	&  0.012	&  -0.036	&  0.024 \\
538.88425656	& -59.9297	& 0.0056	& 5.496	&  0.012	&  -0.027	&  0.024 \\
539.85004231	& -59.9332	& 0.0056	& 5.489	&  0.012	&  -0.040	&  0.024 \\
539.86429831	& -59.9276	& 0.0056	& 5.476	&  0.012	&  -0.042	&  0.023 \\
539.87855411	& -59.9345	& 0.0056	& 5.501	&  0.012	&  -0.048	&  0.023 \\
539.89274681	& -59.9326	& 0.0057	& 5.473	&  0.012	&  -0.045	&  0.023 \\
557.80336435	& -59.9221	& 0.0072	& 5.554	&  0.016	&  -0.026	&  0.030 \\
557.81761950	& -59.9161	& 0.0074	& 5.555	&  0.016	&  -0.035	&  0.030 \\
557.83187445	& -59.9272	& 0.0073	& 5.553	&  0.016	&  -0.028	&  0.030 \\
557.84606670	& -59.9196	& 0.0071	& 5.541	&  0.015	&  -0.040	&  0.029 \\
559.77624676	& -59.9350	& 0.0077	& 5.541	&  0.016	&  -0.060	&  0.031 \\
559.79050318	& -59.9408	& 0.0074	& 5.561	&  0.016	&  -0.036	&  0.030 \\
559.80475871	& -59.9346	& 0.0068	& 5.564	&  0.015	&  -0.054	&  0.028 \\
559.81894984	& -59.9353	& 0.0069	& 5.564	&  0.015	&  -0.039	&  0.029 \\
560.77542835	& -59.9428	& 0.0063	& 5.539	&  0.014	&  -0.033	&  0.026 \\
560.78968516	& -59.9381	& 0.0062	& 5.520	&  0.013	&  -0.013	&  0.025 \\
560.80393897	& -59.9466	& 0.0061	& 5.533	&  0.013	&  -0.032	&  0.025 \\
560.81812999	& -59.9446	& 0.0060	& 5.516	&  0.013	&  -0.034	&  0.025 \\
561.79460490	& -59.9219	& 0.0078	& 5.555	&  0.016	&  -0.032	&  0.031 \\
561.80885940	& -59.9272	& 0.0075	& 5.585	&  0.016	&  -0.040	&  0.031 \\
561.82311469	& -59.9326	& 0.0073	& 5.569	&  0.016	&  -0.039	&  0.030 \\
561.83730619	& -59.9371	& 0.0074	& 5.555	&  0.016	&  -0.028	&  0.031 \\
562.75753023	& -59.9328	& 0.0060	& 5.527	&  0.013	&  -0.027	&  0.025 \\
562.77178552	& -59.9345	& 0.0058	& 5.518	&  0.013	&  -0.030	&  0.025 \\
562.78597711	& -59.9320	& 0.0058	& 5.512	&  0.013	&  -0.041	&  0.024 \\
562.80016759	& -59.9267	& 0.0058	& 5.511	&  0.013	&  -0.027	&  0.024 \\
563.76464181	& -59.9453	& 0.0066	& 5.546	&  0.014	&  -0.005	&  0.026 \\
563.77889829	& -59.9400	& 0.0067	& 5.536	&  0.014	&  -0.050	&  0.027 \\
563.79315376	& -59.9405	& 0.0067	& 5.531	&  0.014	&  -0.023	&  0.027 \\
563.80734474	& -59.9301	& 0.0061	& 5.527	&  0.013	&  -0.049	&  0.025 \\
564.79878234	& -59.9410	& 0.0061	& 5.565	&  0.013	&  -0.035	&  0.025 \\
564.81303560	& -59.9329	& 0.0060	& 5.521	&  0.013	&  -0.018	&  0.024 \\
564.82729206	& -59.9388	& 0.0061	& 5.534	&  0.013	&  -0.035	&  0.025 \\
564.84148262	& -59.9299	& 0.0061	& 5.542	&  0.013	&  -0.049	&  0.025 \\
566.77933148	& -59.9446	& 0.0063	& 5.522	&  0.014	&  -0.055	&  0.026 \\
566.79358211	& -59.9428	& 0.0061	& 5.519	&  0.013	&  -0.032	&  0.025 \\
566.80783755	& -59.9503	& 0.0062	& 5.536	&  0.013	&  -0.024	&  0.026 \\
566.82202828	& -59.9385	& 0.0061	& 5.540	&  0.013	&  -0.046	&  0.025 \\
585.76652048	& -59.9380	& 0.0064	& 5.551	&  0.014	&  -0.013	&  0.027 \\
585.78071470	& -59.9383	& 0.0064	& 5.539	&  0.014	&  -0.010	&  0.026 \\
585.79496722	& -59.9291	& 0.0079	& 5.575	&  0.017	&  -0.037	&  0.033 \\
585.80915704	& -59.9367	& 0.0075	& 5.578	&  0.016	&  -0.035	&  0.032 \\
587.77470698	& -59.9397	& 0.0056	& 5.512	&  0.012	&  -0.011	&  0.023 \\
587.78896268	& -59.9365	& 0.0056	& 5.478	&  0.012	&  -0.007	&  0.023 \\
587.80321708	& -59.9375	& 0.0056	& 5.510	&  0.012	&  -0.012	&  0.023 \\
587.81747208	& -59.9402	& 0.0055	& 5.499	&  0.012	&  -0.000	&  0.023 \\
589.75302522	& -59.9403	& 0.0062	& 5.508	&  0.013	&  -0.007	&  0.026 \\
589.76728010	& -59.9335	& 0.0062	& 5.490	&  0.013	&  -0.039	&  0.025 \\
589.78153548	& -59.9306	& 0.0062	& 5.519	&  0.013	&  -0.026	&  0.025 \\
589.79572667	& -59.9318	& 0.0061	& 5.489	&  0.013	&  -0.007	&  0.025 \\
592.70585888	& -59.9231	& 0.0061	& 5.495	&  0.013	&  -0.009	&  0.025 \\
592.72011124	& -59.9234	& 0.0061	& 5.505	&  0.013	&  -0.022	&  0.025 \\
592.73436680	& -59.9242	& 0.0060	& 5.519	&  0.013	&  -0.035	&  0.025 \\
592.74855726	& -59.9241	& 0.0060	& 5.522	&  0.013	&  -0.021	&  0.024 \\
594.76757159	& -59.9434	& 0.0066	& 5.536	&  0.014	&  -0.018	&  0.027 \\
594.78182683	& -59.9326	& 0.0069	& 5.530	&  0.015	&  -0.035	&  0.028 \\
594.79601747	& -59.9382	& 0.0063	& 5.512	&  0.014	&  -0.014	&  0.026 \\
594.81027260	& -59.9363	& 0.0060	& 5.521	&  0.013	&  -0.031	&  0.025 \\
597.73968050	& -59.9398	& 0.0057	& 5.508	&  0.012	&  -0.001	&  0.023 \\
597.75393471	& -59.9394	& 0.0057	& 5.503	&  0.012	&  -0.001	&  0.023 \\
597.76812613	& -59.9385	& 0.0056	& 5.519	&  0.012	&  -0.018	&  0.023 \\
597.78231695	& -59.9362	& 0.0056	& 5.506	&  0.012	&  -0.019	&  0.023 \\
598.75863479	& -59.9294	& 0.0062	& 5.511	&  0.013	&  -0.033	&  0.025 \\
598.77288920	& -59.9320	& 0.0060	& 5.521	&  0.013	&  -0.019	&  0.025 \\
598.78714471	& -59.9302	& 0.0059	& 5.520	&  0.013	&  -0.016	&  0.025 \\
598.80133531	& -59.9404	& 0.0060	& 5.517	&  0.013	&  -0.026	&  0.025 \\
599.74533714	& -59.9339	& 0.0064	& 5.475	&  0.014	&  -0.027	&  0.026 \\
599.75959124	& -59.9358	& 0.0061	& 5.505	&  0.013	&   0.003	&  0.025 \\
599.77384994	& -59.9391	& 0.0060	& 5.506	&  0.013	&  -0.006	&  0.025 \\
599.78803765	& -59.9324	& 0.0060	& 5.508	&  0.013	&  -0.036	&  0.024 \\
600.74293878	& -59.9394	& 0.0063	& 5.538	&  0.014	&  -0.025	&  0.026 \\
600.75712977	& -59.9377	& 0.0061	& 5.523	&  0.013	&  -0.039	&  0.025 \\
600.77132087	& -59.9407	& 0.0061	& 5.536	&  0.013	&  -0.016	&  0.025 \\
600.78551106	& -59.9409	& 0.0062	& 5.526	&  0.013	&  -0.015	&  0.026 \\
601.74208946	& -59.9335	& 0.0076	& 5.549	&  0.016	&  -0.040	&  0.030 \\
601.75634424	& -59.9354	& 0.0071	& 5.541	&  0.015	&  -0.027	&  0.029 \\
601.77059913	& -59.9397	& 0.0073	& 5.561	&  0.015	&  -0.008	&  0.029 \\
601.78479052	& -59.9290	& 0.0083	& 5.598	&  0.018	&  -0.004	&  0.034 \\
602.74801173	& -59.9299	& 0.0060	& 5.518	&  0.013	&  -0.015	&  0.025 \\
602.76226731	& -59.9290	& 0.0060	& 5.533	&  0.013	&  -0.007	&  0.025 \\
602.77652129	& -59.9253	& 0.0060	& 5.495	&  0.013	&  -0.017	&  0.025 \\
602.79071298	& -59.9300	& 0.0058	& 5.500	&  0.013	&   0.000	&  0.024 \\
603.75844953	& -59.9340	& 0.0063	& 5.482	&  0.014	&  -0.018	&  0.026 \\
603.77270410	& -59.9382	& 0.0061	& 5.485	&  0.013	&  -0.026	&  0.025 \\
603.78695928	& -59.9429	& 0.0061	& 5.484	&  0.013	&  -0.002	&  0.025 \\
603.80115015	& -59.9489	& 0.0061	& 5.510	&  0.013	&  -0.025	&  0.025 \\
604.76579074	& -59.9210	& 0.0071	& 5.562	&  0.015	&  -0.022	&  0.029 \\
604.78004541	& -59.9281	& 0.0069	& 5.557	&  0.015	&  -0.029	&  0.028 \\
604.79430017	& -59.9295	& 0.0064	& 5.539	&  0.014	&  -0.025	&  0.027 \\
604.80849134	& -59.9340	& 0.0067	& 5.559	&  0.015	&  -0.037	&  0.028 \\
606.73093229	& -59.9328	& 0.0070	& 5.566	&  0.015	&  -0.007	&  0.029 \\
606.74518775	& -59.9433	& 0.0071	& 5.549	&  0.015	&  -0.025	&  0.029 \\
606.75944270	& -59.9392	& 0.0071	& 5.557	&  0.015	&  -0.032	&  0.029 \\
606.77363376	& -59.9428	& 0.0069	& 5.539	&  0.015	&  -0.026	&  0.029 \\
607.73395032	& -59.9287	& 0.0062	& 5.514	&  0.013	&  -0.039	&  0.025 \\
607.74820557	& -59.9363	& 0.0074	& 5.545	&  0.016	&  -0.034	&  0.030 \\
607.76246093	& -59.9318	& 0.0070	& 5.533	&  0.015	&  -0.021	&  0.028 \\
607.77665138	& -59.9256	& 0.0078	& 5.564	&  0.016	&  -0.024	&  0.030 \\
       \hline
\end{longtable}
\end{center}

\begin{center}
\begin{longtable}{ccccccc}
    \caption{\label{tab:rvs_v7} SPIRou LBL RV observations of TOI-1695 obtained with the APERO v7 version. } \\
       BJD & $v_r$ & $\sigma_{v_r}$ & FWHM & FWHM error & BIS & BIS error \\
     -2,459,000       & (km\,s$^{-1}$) & (km\,s$^{-1}$) & (km\,s$^{-1}$) & (km\,s$^{-1}$) & (km\,s$^{-1}$) & (km\,s$^{-1}$) \\
       \hline
       \endfirsthead
       \multicolumn{5}{c}{{\bfseries \tablename\ \thetable{}}: continued from previous page} \\
       BJD & $v_r$ & $\sigma_{v_r}$ & FWHM & FWHM error & BIS & BIS error \\
     -2,459,000       & (km\,s$^{-1}$) & (km\,s$^{-1}$) & (km\,s$^{-1}$) & (km\,s$^{-1}$) & (km\,s$^{-1}$) & (km\,s$^{-1}$) \\
       \hline
       \endhead
208.79493816	&  -59.4531	&  0.0084	&  5.8675	&  0.018	&  -0.002	&  0.033 \\
208.80919391	&  -59.4712	&  0.0064	&  5.7880	&  0.014	&  -0.011	&  0.027 \\
208.82344915	&  -59.4654	&  0.0063	&  5.7882	&  0.014	&  -0.004	&  0.027 \\
208.83770450	&  -59.4662	&  0.0065	&  5.7769	&  0.014	&  -0.016	&  0.028 \\
213.75864561	&  -59.4454	&  0.0062	&  5.7935	&  0.013	&   0.001	&  0.026 \\
213.77290070	&  -59.4544	&  0.0061	&  5.7817	&  0.013	&   0.010	&  0.026 \\
213.78715749	&  -59.4551	&  0.0061	&  5.7930	&  0.013	&  -0.005	&  0.026 \\
213.80141198	&  -59.4483	&  0.0063	&  5.7954	&  0.014	&   0.004	&  0.027 \\
472.98091030	&  -59.4657	&  0.0057	&  5.7911	&  0.012	&  -0.007	&  0.024 \\
472.99516782	&  -59.4577	&  0.0056	&  5.7730	&  0.012	&   0.006	&  0.024 \\
473.00936133	&  -59.4672	&  0.0056	&  5.7632	&  0.012	&   0.018	&  0.024 \\
473.02355075	&  -59.4615	&  0.0056	&  5.7823	&  0.012	&  -0.004	&  0.023 \\
476.88360128	&  -59.4672	&  0.0062	&  5.7740	&  0.014	&   0.018	&  0.027 \\
476.89779088	&  -59.4646	&  0.0061	&  5.7706	&  0.013	&  -0.005	&  0.026 \\
476.91204768	&  -59.4671	&  0.0060	&  5.7695	&  0.013	&  -0.006	&  0.026 \\
476.92623878	&  -59.4618	&  0.0060	&  5.7687	&  0.013	&  -0.007	&  0.026 \\
479.86476628	&  -59.4563	&  0.0057	&  5.7578	&  0.012	&  -0.008	&  0.024 \\
479.87902237	&  -59.4511	&  0.0057	&  5.7536	&  0.012	&  -0.009	&  0.024 \\
479.89321505	&  -59.4542	&  0.0056	&  5.7515	&  0.012	&  -0.005	&  0.024 \\
479.90740723	&  -59.4604	&  0.0056	&  5.7654	&  0.012	&  -0.003	&  0.024 \\
480.87194906	&  -59.4620	&  0.0058	&  5.7839	&  0.013	&  -0.009	&  0.025 \\
480.88620644	&  -59.4627	&  0.0058	&  5.7716	&  0.013	&  -0.029	&  0.025 \\
480.90039831	&  -59.4544	&  0.0058	&  5.7647	&  0.013	&  -0.028	&  0.025 \\
480.91458968	&  -59.4555	&  0.0059	&  5.7744	&  0.013	&   0.008	&  0.025 \\
481.85700461	&  -59.4616	&  0.0058	&  5.7705	&  0.013	&  -0.012	&  0.025 \\
481.87119917	&  -59.4676	&  0.0057	&  5.7683	&  0.012	&  -0.016	&  0.024 \\
481.88538804	&  -59.4636	&  0.0057	&  5.7367	&  0.012	&  -0.010	&  0.024 \\
481.89958051	&  -59.4597	&  0.0056	&  5.7474	&  0.012	&  -0.020	&  0.024 \\
501.92949642	&  -59.4617	&  0.0059	&  5.7822	&  0.013	&  -0.018	&  0.025 \\
501.94375341	&  -59.4641	&  0.0059	&  5.7733	&  0.013	&  -0.015	&  0.025 \\
501.95800989	&  -59.4665	&  0.0059	&  5.7847	&  0.013	&   0.007	&  0.025 \\
501.97220128	&  -59.4564	&  0.0059	&  5.7808	&  0.013	&   0.003	&  0.025 \\
502.91544136	&  -59.4598	&  0.0061	&  5.7711	&  0.013	&  -0.018	&  0.026 \\
502.92969804	&  -59.4637	&  0.0060	&  5.7617	&  0.013	&   0.003	&  0.026 \\
502.94388941	&  -59.4621	&  0.0060	&  5.7726	&  0.013	&  -0.012	&  0.026 \\
502.95808148	&  -59.4689	&  0.0060	&  5.7901	&  0.013	&   0.005	&  0.026 \\
503.95195991	&  -59.4779	&  0.0060	&  5.7833	&  0.013	&  -0.009	&  0.025 \\
503.96621637	&  -59.4718	&  0.0055	&  5.7844	&  0.012	&  -0.015	&  0.023 \\
503.98047314	&  -59.4717	&  0.0055	&  5.7966	&  0.012	&  -0.013	&  0.023 \\
503.99466510	&  -59.4689	&  0.0057	&  5.7794	&  0.012	&  -0.003	&  0.024 \\
505.92752409	&  -59.4635	&  0.0060	&  5.7910	&  0.013	&  -0.026	&  0.025 \\
505.94177953	&  -59.4672	&  0.0061	&  5.7887	&  0.013	&  -0.022	&  0.026 \\
505.95603647	&  -59.4689	&  0.0059	&  5.7718	&  0.013	&  -0.009	&  0.025 \\
505.97022731	&  -59.4604	&  0.0058	&  5.7931	&  0.013	&  -0.000	&  0.025 \\
507.92978917	&  -59.4545	&  0.0062	&  5.7577	&  0.014	&  -0.007	&  0.027 \\
507.94398159	&  -59.4490	&  0.0061	&  5.7651	&  0.013	&  -0.006	&  0.026 \\
507.95817331	&  -59.4504	&  0.0061	&  5.7491	&  0.013	&  -0.013	&  0.026 \\
507.97236802	&  -59.4451	&  0.0062	&  5.7546	&  0.013	&  -0.013	&  0.026 \\
508.93637316	&  -59.4568	&  0.0059	&  5.7890	&  0.013	&  -0.008	&  0.025 \\
508.95056467	&  -59.4624	&  0.0059	&  5.7848	&  0.013	&  -0.015	&  0.025 \\
508.96482027	&  -59.4631	&  0.0060	&  5.7795	&  0.013	&  -0.021	&  0.025 \\
508.97901207	&  -59.4581	&  0.0059	&  5.7913	&  0.013	&  -0.021	&  0.025 \\
509.94385745	&  -59.4552	&  0.0070	&  5.7879	&  0.015	&  -0.009	&  0.030 \\
509.95811414	&  -59.4496	&  0.0070	&  5.7608	&  0.015	&  -0.027	&  0.030 \\
509.97230573	&  -59.4670	&  0.0070	&  5.7817	&  0.015	&  -0.019	&  0.030 \\
509.98649743	&  -59.4686	&  0.0070	&  5.8111	&  0.015	&  -0.013	&  0.030 \\
510.92218401	&  -59.4565	&  0.0058	&  5.7855	&  0.013	&  -0.022	&  0.025 \\
510.93637590	&  -59.4478	&  0.0059	&  5.7871	&  0.013	&  -0.015	&  0.025 \\
510.95063248	&  -59.4534	&  0.0060	&  5.7824	&  0.013	&  -0.028	&  0.025 \\
510.96482346	&  -59.4489	&  0.0059	&  5.7599	&  0.013	&  -0.021	&  0.025 \\
511.96895286	&  -59.4491	&  0.0065	&  5.8208	&  0.014	&  -0.014	&  0.028 \\
511.98314483	&  -59.4547	&  0.0062	&  5.8100	&  0.013	&  -0.028	&  0.026 \\
511.99733669	&  -59.4650	&  0.0066	&  5.8064	&  0.014	&  -0.018	&  0.028 \\
513.90689610	&  -59.4572	&  0.0056	&  5.7917	&  0.012	&  -0.034	&  0.024 \\
513.92115105	&  -59.4618	&  0.0055	&  5.7801	&  0.012	&  -0.025	&  0.023 \\
513.93534260	&  -59.4638	&  0.0056	&  5.7952	&  0.012	&  -0.033	&  0.024 \\
513.94953504	&  -59.4601	&  0.0057	&  5.7967	&  0.012	&  -0.033	&  0.024 \\
531.77248693	&  -59.4610	&  0.0069	&  5.8188	&  0.015	&  -0.000	&  0.029 \\
531.78667804	&  -59.4636	&  0.0065	&  5.7800	&  0.014	&  -0.015	&  0.027 \\
531.80093355	&  -59.4657	&  0.0061	&  5.7979	&  0.013	&  -0.027	&  0.025 \\
531.81512557	&  -59.4576	&  0.0061	&  5.7894	&  0.013	&  -0.010	&  0.025 \\
535.91296729	&  -59.4633	&  0.0073	&  5.7752	&  0.016	&  -0.006	&  0.031 \\
535.92722294	&  -59.4559	&  0.0103	&  5.8492	&  0.023	&  -0.040	&  0.044 \\
535.94147939	&  -59.4496	&  0.0099	&  5.8508	&  0.022	&  -0.044	&  0.043 \\
535.95567034	&  -59.4333	&  0.0090	&  5.8318	&  0.020	&  -0.048	&  0.039 \\
538.84155732	&  -59.4670	&  0.0056	&  5.7769	&  0.012	&  -0.027	&  0.024 \\
538.85580844	&  -59.4628	&  0.0057	&  5.7716	&  0.012	&  -0.033	&  0.024 \\
538.87006465	&  -59.4751	&  0.0057	&  5.7695	&  0.012	&  -0.034	&  0.024 \\
538.88425656	&  -59.4688	&  0.0057	&  5.7742	&  0.012	&  -0.017	&  0.024 \\
539.85004231	&  -59.4689	&  0.0056	&  5.7647	&  0.012	&  -0.030	&  0.023 \\
539.86429831	&  -59.4651	&  0.0055	&  5.7558	&  0.012	&  -0.000	&  0.023 \\
539.87855411	&  -59.4692	&  0.0055	&  5.7668	&  0.012	&  -0.021	&  0.023 \\
539.89274681	&  -59.4692	&  0.0056	&  5.7679	&  0.012	&  -0.020	&  0.024 \\
557.80336435	&  -59.4724	&  0.0070	&  5.8099	&  0.015	&   0.011	&  0.030 \\
557.81761950	&  -59.4710	&  0.0071	&  5.8210	&  0.015	&   0.026	&  0.030 \\
557.83187445	&  -59.4715	&  0.0070	&  5.8075	&  0.015	&   0.021	&  0.030 \\
557.84606670	&  -59.4715	&  0.0069	&  5.8090	&  0.015	&   0.018	&  0.029 \\
559.77624676	&  -59.4681	&  0.0073	&  5.7912	&  0.016	&  -0.006	&  0.031 \\
559.79050318	&  -59.4750	&  0.0071	&  5.8125	&  0.015	&   0.005	&  0.030 \\
559.80475871	&  -59.4731	&  0.0066	&  5.8147	&  0.014	&  -0.006	&  0.028 \\
559.81894984	&  -59.4692	&  0.0069	&  5.8231	&  0.015	&  -0.004	&  0.029 \\
560.77542835	&  -59.4681	&  0.0063	&  5.7951	&  0.014	&  -0.010	&  0.027 \\
560.78968516	&  -59.4721	&  0.0060	&  5.7848	&  0.013	&   0.007	&  0.025 \\
560.80393897	&  -59.4731	&  0.0059	&  5.8008	&  0.013	&  -0.007	&  0.025 \\
560.81812999	&  -59.4709	&  0.0058	&  5.7846	&  0.013	&  -0.013	&  0.025 \\
561.79460490	&  -59.4529	&  0.0075	&  5.8266	&  0.017	&  -0.008	&  0.032 \\
561.80885940	&  -59.4580	&  0.0074	&  5.8299	&  0.016	&  -0.009	&  0.032 \\
561.82311469	&  -59.4722	&  0.0072	&  5.8262	&  0.016	&  -0.014	&  0.031 \\
561.83730619	&  -59.4668	&  0.0071	&  5.8188	&  0.016	&  -0.000	&  0.030 \\
562.75753023	&  -59.4580	&  0.0060	&  5.7945	&  0.013	&  -0.023	&  0.025 \\
562.77178552	&  -59.4651	&  0.0059	&  5.8029	&  0.013	&  -0.007	&  0.025 \\
562.78597711	&  -59.4584	&  0.0058	&  5.7839	&  0.013	&  -0.031	&  0.025 \\
562.80016759	&  -59.4530	&  0.0057	&  5.7840	&  0.013	&  -0.010	&  0.024 \\
563.76464181	&  -59.4655	&  0.0063	&  5.8150	&  0.014	&   0.009	&  0.026 \\
563.77889829	&  -59.4657	&  0.0066	&  5.7974	&  0.014	&  -0.016	&  0.027 \\
563.79315376	&  -59.4632	&  0.0066	&  5.7884	&  0.014	&  -0.009	&  0.027 \\
563.80734474	&  -59.4589	&  0.0061	&  5.7839	&  0.013	&  -0.027	&  0.025 \\
564.79878234	&  -59.4646	&  0.0060	&  5.8243	&  0.013	&  -0.004	&  0.025 \\
564.81303560	&  -59.4591	&  0.0059	&  5.8107	&  0.013	&   0.004	&  0.025 \\
564.82729206	&  -59.4647	&  0.0062	&  5.8055	&  0.013	&  -0.011	&  0.026 \\
564.84148262	&  -59.4614	&  0.0060	&  5.8159	&  0.013	&  -0.006	&  0.025 \\
566.77933148	&  -59.4741	&  0.0067	&  5.7850	&  0.015	&  -0.022	&  0.029 \\
566.79358211	&  -59.4721	&  0.0065	&  5.7605	&  0.014	&  -0.004	&  0.028 \\
566.80783755	&  -59.4771	&  0.0067	&  5.7811	&  0.015	&   0.007	&  0.029 \\
566.82202828	&  -59.4656	&  0.0065	&  5.7960	&  0.014	&  -0.033	&  0.028 \\
585.76652048	&  -59.4594	&  0.0068	&  5.8175	&  0.015	&  -0.001	&  0.029 \\
585.78071470	&  -59.4601	&  0.0067	&  5.7761	&  0.014	&  -0.006	&  0.028 \\
585.79496722	&  -59.4660	&  0.0080	&  5.8302	&  0.017	&   0.018	&  0.034 \\
585.80915704	&  -59.4590	&  0.0077	&  5.8197	&  0.017	&  -0.008	&  0.033 \\
587.77470698	&  -59.4634	&  0.0057	&  5.7748	&  0.012	&   0.005	&  0.024 \\
587.78896268	&  -59.4610	&  0.0056	&  5.7658	&  0.012	&  -0.001	&  0.023 \\
587.80321708	&  -59.4624	&  0.0056	&  5.7784	&  0.012	&  -0.000	&  0.023 \\
587.81747208	&  -59.4612	&  0.0055	&  5.7682	&  0.012	&   0.013	&  0.023 \\
589.75302522	&  -59.4619	&  0.0061	&  5.7736	&  0.013	&  -0.014	&  0.026 \\
589.76728010	&  -59.4594	&  0.0060	&  5.7631	&  0.013	&  -0.023	&  0.026 \\
589.78153548	&  -59.4571	&  0.0059	&  5.7856	&  0.013	&  -0.014	&  0.025 \\
589.79572667	&  -59.4578	&  0.0060	&  5.7794	&  0.013	&   0.005	&  0.025 \\
591.77126900	&  -59.4521	&  0.0060	&  5.8048	&  0.013	&  -0.036	&  0.025 \\
591.78552487	&  -59.4560	&  0.0060	&  5.7901	&  0.013	&  -0.032	&  0.025 \\
591.79978033	&  -59.4588	&  0.0060	&  5.7753	&  0.013	&  -0.023	&  0.025 \\
591.81397029	&  -59.4540	&  0.0057	&  5.7920	&  0.012	&  -0.010	&  0.024 \\
592.70585888	&  -59.4500	&  0.0063	&  5.7629	&  0.014	&  -0.020	&  0.026 \\
592.72011124	&  -59.4573	&  0.0062	&  5.7706	&  0.013	&  -0.017	&  0.026 \\
592.73436680	&  -59.4582	&  0.0061	&  5.7728	&  0.013	&  -0.030	&  0.026 \\
592.74855726	&  -59.4583	&  0.0063	&  5.7868	&  0.014	&  -0.001	&  0.026 \\
594.76757159	&  -59.4715	&  0.0069	&  5.7864	&  0.015	&  -0.002	&  0.029 \\
594.78182683	&  -59.4658	&  0.0071	&  5.7823	&  0.015	&  -0.019	&  0.030 \\
594.79601747	&  -59.4743	&  0.0065	&  5.7616	&  0.014	&   0.007	&  0.028 \\
594.81027260	&  -59.4677	&  0.0063	&  5.7637	&  0.014	&   0.003	&  0.027 \\
597.73968050	&  -59.4661	&  0.0056	&  5.7792	&  0.012	&  -0.004	&  0.024 \\
597.75393471	&  -59.4646	&  0.0056	&  5.7788	&  0.012	&  -0.007	&  0.023 \\
597.76812613	&  -59.4661	&  0.0054	&  5.7850	&  0.012	&  -0.009	&  0.023 \\
597.78231695	&  -59.4655	&  0.0054	&  5.7863	&  0.012	&  -0.011	&  0.023 \\
598.75863479	&  -59.4584	&  0.0063	&  5.7973	&  0.014	&  -0.017	&  0.026 \\
598.77288920	&  -59.4642	&  0.0062	&  5.8018	&  0.013	&  -0.015	&  0.026 \\
598.78714471	&  -59.4606	&  0.0062	&  5.7775	&  0.013	&  -0.002	&  0.026 \\
598.80133531	&  -59.4692	&  0.0060	&  5.7917	&  0.013	&  -0.017	&  0.025 \\
599.74533714	&  -59.4562	&  0.0061	&  5.7671	&  0.013	&  -0.015	&  0.026 \\
599.75959124	&  -59.4670	&  0.0059	&  5.7894	&  0.013	&   0.009	&  0.025 \\
599.77384994	&  -59.4649	&  0.0058	&  5.7783	&  0.013	&   0.000	&  0.025 \\
599.78803765	&  -59.4605	&  0.0058	&  5.7853	&  0.013	&  -0.025	&  0.025 \\
600.74293878	&  -59.4719	&  0.0064	&  5.8115	&  0.014	&  -0.029	&  0.027 \\
600.75712977	&  -59.4654	&  0.0063	&  5.7874	&  0.014	&  -0.045	&  0.027 \\
600.77132087	&  -59.4672	&  0.0062	&  5.7938	&  0.014	&  -0.013	&  0.026 \\
600.78551106	&  -59.4695	&  0.0062	&  5.7853	&  0.014	&  -0.011	&  0.026 \\
601.74208946	&  -59.4553	&  0.0070	&  5.8273	&  0.015	&  -0.025	&  0.029 \\
601.75634424	&  -59.4568	&  0.0068	&  5.8079	&  0.015	&  -0.027	&  0.029 \\
601.77059913	&  -59.4661	&  0.0070	&  5.8272	&  0.015	&  -0.015	&  0.029 \\
601.78479052	&  -59.4454	&  0.0078	&  5.8485	&  0.017	&  -0.004	&  0.033 \\
602.74801173	&  -59.4522	&  0.0060	&  5.7973	&  0.013	&  -0.012	&  0.025 \\
602.76226731	&  -59.4602	&  0.0058	&  5.8083	&  0.013	&  -0.001	&  0.024 \\
602.77652129	&  -59.4490	&  0.0058	&  5.7821	&  0.013	&  -0.000	&  0.024 \\
602.79071298	&  -59.4566	&  0.0057	&  5.7914	&  0.012	&  -0.006	&  0.024 \\
603.75844953	&  -59.4634	&  0.0061	&  5.7663	&  0.013	&  -0.009	&  0.026 \\
603.77270410	&  -59.4690	&  0.0059	&  5.7794	&  0.013	&  -0.016	&  0.025 \\
603.78695928	&  -59.4704	&  0.0059	&  5.7685	&  0.013	&   0.001	&  0.025 \\
603.80115015	&  -59.4784	&  0.0059	&  5.7862	&  0.013	&  -0.002	&  0.025 \\
604.76579074	&  -59.4525	&  0.0069	&  5.8172	&  0.015	&  -0.018	&  0.029 \\
604.78004541	&  -59.4554	&  0.0067	&  5.8201	&  0.015	&  -0.038	&  0.029 \\
604.79430017	&  -59.4523	&  0.0065	&  5.8048	&  0.014	&  -0.021	&  0.028 \\
604.80849134	&  -59.4634	&  0.0067	&  5.8235	&  0.015	&  -0.025	&  0.028 \\
606.73093229	&  -59.4486	&  0.0072	&  5.8098	&  0.016	&  -0.018	&  0.031 \\
606.74518775	&  -59.4544	&  0.0072	&  5.7984	&  0.016	&  -0.045	&  0.031 \\
606.75944270	&  -59.4588	&  0.0073	&  5.8186	&  0.016	&  -0.035	&  0.031 \\
606.77363376	&  -59.4635	&  0.0072	&  5.8029	&  0.016	&  -0.031	&  0.031 \\
       \hline
\end{longtable}
\end{center}

\begin{table*}[hbt]
    \centering
    \caption{$B_\ell$ and null polarization $N$ observations of TOI-1695.  \label{tab:polar}}
  
    \begin{tabular}{lcccccccc}
       BJD & $B_\ell$ & $\sigma_B$ & $N$ & $\sigma_N$ & EqW \\
        -2,459,000   & (G)& (G) & & & (\AA) \\
       \hline
473.0016398 & -4.12 & 4.99 &   5.00 & 4.87 & 1.651 \\
476.9045374 & -3.21 & 3.17 &   1.46 & 3.36 & 1.954 \\
479.8851662 &  1.73 & 3.12 &   1.11 & 3.26 & 1.944 \\
480.8918336 & -1.24 & 3.43 &  -2.74 & 3.55 & 1.930 \\
481.8775293 & -2.05 & 3.12 &  -6.80 & 3.40 & 1.942 \\
501.9502817 & -1.85 & 3.24 &  -1.71 & 3.30 & 1.813 \\
502.9355799 &  1.16 & 3.08 &   0.02 & 3.17 & 1.898 \\
503.9727203 & -1.88 & 3.70 &   3.94 & 3.71 & 1.802 \\
505.9486172 &  3.73 & 3.60 &   6.75 & 3.73 & 1.792 \\
507.9496392 &  0.35 & 3.70 &  -6.78 & 3.82 & 1.830 \\
508.9567142 &  6.69 & 3.39 &  -0.73 & 3.51 & 1.915 \\
509.9643374 & -2.14 & 5.07 &  -4.97 & 5.20 & 1.793 \\
510.9424300 &  1.40 & 3.29 &  -5.95 & 3.33 & 1.955 \\
513.9267436 &  3.88 & 3.30 &  -1.46 & 3.37 & 1.918 \\
531.7939318 &  3.00 & 3.49 &   2.10 & 3.66 & 1.930 \\
535.9308136 & -2.69 & 8.20 &  -2.33 & 7.68 & 1.738 \\
538.8621726 &  0.39 & 3.11 &   2.17 & 3.19 & 1.961 \\
539.8701053 & -2.49 & 2.92 &  -0.01 & 3.14 & 1.946 \\
557.8238866 &  0.52 & 5.16 &  -3.23 & 5.19 & 1.752 \\
559.7975022 & -0.53 & 4.85 &  -4.68 & 4.96 & 1.785 \\
560.7960801 & -3.17 & 3.54 &  -1.15 & 3.60 & 1.877 \\
561.8158711 &  4.28 & 4.91 &  -5.31 & 4.91 & 1.792 \\
562.7782783 &  0.94 & 3.15 &  -2.67 & 3.18 & 1.932 \\
563.7854884 &  3.39 & 4.40 &  -4.50 & 4.45 & 1.897 \\
564.8188796 &  0.44 & 3.96 &  -2.45 & 4.08 & 1.890 \\
566.8006265 &  0.62 & 3.32 &  -1.53 & 3.40 & 1.890 \\
585.7837539 &  0.41 & 4.69 &   0.73 & 6.50 & 1.776 \\
589.7735483 & -3.04 & 3.03 &  -1.98 & 3.38 & 1.975 \\
591.7921350 &  1.51 & 3.68 &   1.54 & 3.72 & 1.947 \\
592.7261387 & -3.25 & 2.99 &   2.83 & 3.11 & 1.967 \\
594.7893179 & -7.56 & 3.65 &  -9.50 & 4.00 & 1.613 \\
597.7601225 & -2.01 & 4.08 &  -0.30 & 4.21 & 1.586 \\
599.7656805 &  0.53 & 2.87 &  -4.19 & 2.98 & 1.895 \\
600.7635277 &  3.07 & 4.20 &  -1.51 & 4.24 & 1.655 \\
601.7610720 &  6.95 & 5.31 &   5.46 & 5.43 & 1.739 \\
602.7681757 &  1.18 & 3.65 &  -6.27 & 3.77 & 1.712 \\
603.7787482 &  1.05 & 3.64 &  -1.39 & 3.82 & 1.704 \\
606.7510680 &  6.67 & 5.46 &   4.70 & 5.63 & 1.662 \\
       \hline
    \end{tabular}
\end{table*}

\end{appendix}

\end{document}